\documentclass[preprint,aps,showpacs,preprint,floatfix]{revtex4-1} \usepackage{graphicx}
\usepackage{graphicx}
\usepackage{subfigure}
\usepackage{array}
\usepackage{color}
\usepackage{amsmath}
\usepackage{amsxtra}
\usepackage{tablists}
\usepackage{diagbox}
\usepackage{amstext}
\usepackage{amssymb}
\usepackage{latexsym}
\usepackage{dsfont}
\usepackage{verbatim}
\usepackage{mathrsfs}
\usepackage{graphicx}
\usepackage{afterpage}
\usepackage[position=t,singlelinecheck=off]{subfig}
\usepackage{caption}
\usepackage{multirow}
\usepackage{marvosym}
\usepackage{floatrow}
\usepackage[utf8]{inputenc}
\usepackage{caption}
\newfloatcommand{capbtabbox}{table}[][\FBwidth]

\newcommand{\beq}{\begin{equation}}
\newcommand{\eeq}{\end{equation}}
\newcommand{\bea}{\begin{eqnarray}}
\newcommand{\eea}{\end{eqnarray}}

\begin{document}
\title{Study of $HZZ$ anomalous couplings by angular differential  cross sections }
\author{Yi-Song Lu}
\email{yisong$_$lu@snnu.edu.cn}
\affiliation{School of Physics $\&$ Information Technology, Shaanxi Normal University, Xi'an 710119, China}
\author{You-Kai Wang}
\email{wangyk@snnu.edu.cn}
\affiliation{School of Physics $\&$ Information Technology, Shaanxi Normal University, Xi'an 710119, China}
\author{Xiang-Yuan You}
\affiliation{School of Physics $\&$ Information Technology, Shaanxi Normal University, Xi'an 710119, China}

\date{\today}

\begin{abstract}
Being one of the golden channels for precise measurement of Higgs properties, the process $gg \to H\to ZZ \rightarrow 4l$ provides an opportunity to detect the anomalous $HZZ$ couplings  in searching of new physics beyond the Standard Model. In this paper, we adopt the method of spinor helicity amplitudes to calculate the amplitudes of process  $gg \to H \to ZZ \rightarrow 2e2\mu$ at the LHC in the frame of the effective dimension-6 operators. Signal processes of CP-even and CP-odd anomalous $HVV$ couplings in Higgs on/off-shell region and their interference are simulated by using the package MCFM. The most sensitive angle $\phi$ to extract the CP violated $HZZ$ coupling coefficients, which is defined as the angle between two $Z$ decay planes in Higgs rest frame, is found for experimental measurements.  

\end{abstract}

\maketitle

\section{INTRODUCTION}
Precise test of the properties of the Higgs boson is one of the most important tasks in High energy physics after its discovery \cite{Barklow:2017awn,Bolognesi:2012mm}. Many proposals have been studied at the high luminosity LHC or future $e^+e^-$ colliders. In this paper, we focus on the couplings between Higgs and the $Z$ boson pair, especially its CP properties beyond the SM. Such possible CP violation couplings are widely believed to be connected with matter and antimatter asymmetry and play an important role in the evolution of the universe.

CP properties of the $HZZ$ coupling have been studied in many literatures. Theoretically, the New Physics CP violation terms can be described by adding new high dimensional terms and the so-called Warsaw bases~\cite{Grzadkowski:2010es} are the most widely used. The paper \cite{Gao:2010qx,Bolognesi:2012mm} made the MC simulation before the Higgs discovery at the LHC. Both total cross sections and angular distributions  have been calculated. Experimentally, CMS collaboration has measured the coupling strength between Higgs and gauge bosons~\cite{CMS:2019chr,CMS:2022ley}. Constraints of the parameter spaces of the new couplings beyond the SM have been given in reference~\cite{CMS:2019ekd}.

In view of the existed researches, we think three points should be highly concerned. First, the contribution from the interference effects between processes including the new $HZZ$ couplings SM background should be clearly estimated. Second, constraints from off shell Higgs energy regions should be emphasised. The anomalous couplings can be very sensitive to the small cross section in high energy region which may be mostly contributed from the above mentioned interference effects. Third, more accurate physics parameters can be subtracted from the differential cross sections rather than the total cross sections as LHC is accumulating more and more data. Therefore, it is worth estimating precise constraints on the anomalous couplings of $HZZ$ by its differential distributions.

The rest of this article is organized as follows. In Section~\ref{THEORETICAL CALCULATION}, the spinor helicity amplitudes with anomalous couplings are calculated based on dimension-6 operators. The formulas are summed in appendix~\ref{Appendix-formula}. In Section ~\ref{Kinematics analyze}, the kinematics analyze is introduced in order to get the angular distribution. It is embedded into the MCFM8.0~\cite{Campbell:2013una,Campbell:2014nqn} package with anomalous $HZZ$ couplings and the diagrams analysis are shown in Section~\ref{SIMULATION}. Section~\ref{summary} is the summary.

\section{THEORETICAL CALCULATION}\label{THEORETICAL CALCULATION}
The $HZZ$ anomalous couplings  have been written down explicitly in reference~\cite{He:2019kgh}. Here we just give a brief review. In the SM effective field theory \cite{Buchmuller:1985jz,Grzadkowski:2010es}, with the dimension-6 operators, the effective Lagrangian can be written as
\begin{equation}\label{Effective Lagrangian}
	{\cal L}=\frac{a_1}{v}M_Z^2 H Z^\mu Z_\mu-
	\frac{a_2}{v}HZ^{\mu\nu}Z_{\mu\nu}-
	\frac{a_3}{v}HZ^{\mu\nu}\Tilde{Z}_{\mu\nu},
\end{equation}
where $v=246~\mbox{GeV}$ is the electroweak vacuum expectation value. $Z_\mu$ is the $Z$ boson field, $Z_{\mu\nu}=\partial_\mu Z_\nu-\partial_\nu Z_\mu$ is the field strength tensor of the $Z$ boson, and $\Tilde{Z}_{\mu\nu}=\frac{1}{2}\epsilon_{\mu\nu\rho\sigma Z^{\rho\sigma}}$ represents its dual field strength.  The anomalous $HZZ$ interactions contain three dimensionless complex numbers $a_1,a_2,a_3$. In our case, the amplitude can be divided into three parts as follows,
\begin{equation}
	\mathcal{A}=a_1\mathcal{A}_{\rm SM}+a_2\mathcal{A}_{\rm CP\raisebox{0mm}{-}even}+a_3\mathcal{A}_{\rm CP\raisebox{0mm}{-}odd}.
\end{equation}
These three terms correspond to the amplitudes of SM, Beyond Standard Model~(BSM) CP-even and BSM CP-odd, respectively. The complete amplitude with a mediated Higgs boson can be expressed as
\begin{equation}
	\mathcal{A}_i=i\mathcal{A}^{gg\to H}P_H\mathcal{A}_i^{H\to ZZ\to2e2\mu},
\end{equation}
where $\mathcal{A}^{gg\to H}$ is the amplitude of Higgs production from gluon-gluon fusion\cite{He:2019kgh},  $P_H=\frac{1}{s-M^2_H+iM_H\Gamma}$ is the Higgs boson propagator, and the label $i$ represents SM, BSM CP-even, BSM CP-odd terms. The squared amplitude of the interference term can be donated as
\begin{equation}
2{\mbox{Re}}(\mathcal{A}_i\mathcal{A}_j^*)=|\mathcal{A}^{gg\to H}P_H|^2 ~2{\mbox{Re}}(\mathcal{A}_i^{H\to ZZ\to2e2\mu}\mathcal{A}_j^{*H\to ZZ\to2e2\mu}),
\end{equation}
where $j$ is the same index as $i$.
The results of interference including different terms are summed up in Appendix~\ref{Appendix-formula}.

One can do phase space integration to get a differential distribution. The cross section contributed from interference can be expressed in the following form
\begin{equation}
	\sigma_{ij}=\dfrac{1}{8E_1E_2}\int d\Pi_4 \left<\mathcal{A}_i\mathcal{A}_j^*+\mathcal{A}_i^*\mathcal{A}_j\right>,
\end{equation}
where $ \int d\Pi_4 $ is the phase space integral of the final state of the four bodies.

Helicity amplitude of the process $gg \rightarrow H \rightarrow ZZ \rightarrow 2e2\mu$ is written in the appendix~\ref{Appendix-formula}. We also consider its interference with the box process $gg \rightarrow ZZ \rightarrow 2e2\mu$. The full formula of the amplitude of the box process are complex and can be found in~\cite{Campbell:2013una}. They have been embedded in the package MCFM for numerical calculation. 

	\section{Kinematics analyze}\label{Kinematics analyze}
	
   In this section, to be consist with last part, we introduce the explicit angles definition. $ p_i $ with $ i=1\cdots6 $ is used to represent the momenta of the six external legs and the process $gg\rightarrow H \rightarrow Z_1 Z_2 \rightarrow \mu^- \mu^+ e^- e^+$ can be expressed as
	$g(p_1)g(p_2) \rightarrow H(p_{12}) \rightarrow Z_1(p_{56})Z_2(p_{34}) \rightarrow \mu^- (p_5) \mu^+(p_6) e^-(p_3) e^+(p_4)$,
	where $ p_{12}=p_1+p_2,\ p_{56}=p_5+p_6$ and $\ p_{34}=p_3+p_4 $. The eight independent variables of the final states of the four bodies can be constructed as two invariant masses $ m_{\mu^- \mu^+} $  and $m_{e^- e^+} $ and six angles $ \theta^*,\phi^*, \theta_1, \theta_2, \phi_1(\mbox{or} \ \phi_2) $ and $ \phi $, which will be defined in the followings. Firstly, we will illustrate the angular distribution in the Higgs production plane, and then in its decay plane.

\begin{figure}[htb]
	\centering
	\includegraphics[scale=0.5]{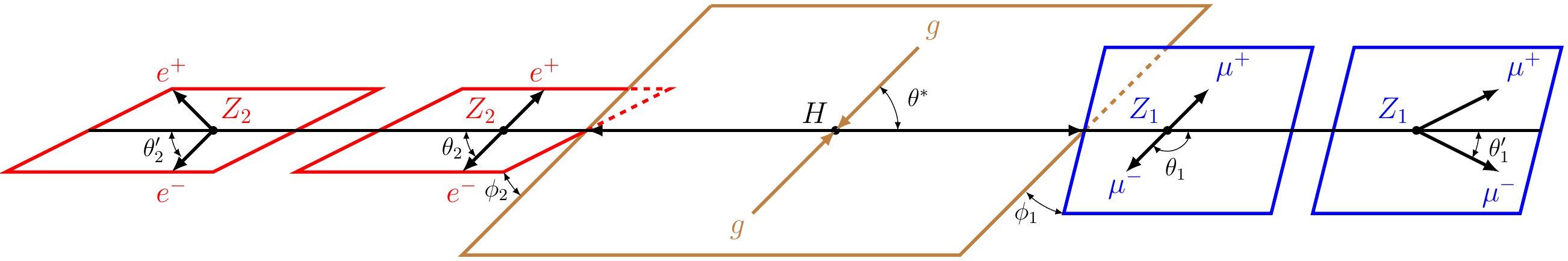}
	\caption{The kinematic distribution for $gg\rightarrow H \rightarrow Z_1 Z_2 \rightarrow \mu^+ \mu^- e^+ e^-$ process, where $\phi_1+\phi_2=\phi$. }
\label{ggHZZmumuee}
\end{figure}
Fig.~\ref{ggHZZmumuee} illustrates two angles of the Higgs production plane. $ \theta^*\in [0, \pi]$ is the angle between $ Z_1 $ boson momentum direction and z-axis in Higgs rest frame. The expression for $ \theta^* $ is
\begin{equation}
	\theta^*=\cos^{-1}\left(\dfrac{\vec{p}_{56} \cdot \hat{n}_z}{|\vec{p}_{56}||\hat{n}_z|}\right) \ , \ \
	\hat{n}_z=(0, 0, 1).
\end{equation}

$\phi^*$ is the azimuth angle of the two $Z$ bosons to the initial gluons momentum direction in the Higgs rest frame. It is trivial and can be integrated directly as the square of the amplitude has no dependence on this angle. 

$ \phi_1  \in [-\pi, \pi]$ is the angle between the $ Z_1 $ production and decay planes which can be defined either by the $ \mu $ lepton momentum in the $ Z_1 $ rest frame or by the $ \mu $ momentum in the Higgs rest frame. We define it in the Higgs rest frame. The expression for $ \phi_1 $ is
\begin{equation}
	\phi_1=\left[\cos^{-1}\left(\hat{n}_{prod} \cdot \hat{n}_{decay1}\right)\right] \cdot
\mbox{sign}[{\left(\hat{n}_{prod} \times \hat{n}_{decay1}\right) \cdot \vec{p}_{56}}],
\end{equation}
where
\begin{equation}
	\begin{split}
		&\hat{n}_{decay1}=\dfrac{\vec{p}_5 \times \vec{p}_6}{|\vec{p}_5 \times \vec{p}_6|}, \\
		&\hat{n}_{prod}=\dfrac{\hat{n}_{z} \times \vec{p}_{56}}{|\hat{n}_{z} \times \vec{p}_{56}|}.
	\end{split}
\end{equation}

	\begin{figure}[htb]
		\centering
		\includegraphics[scale=0.7]{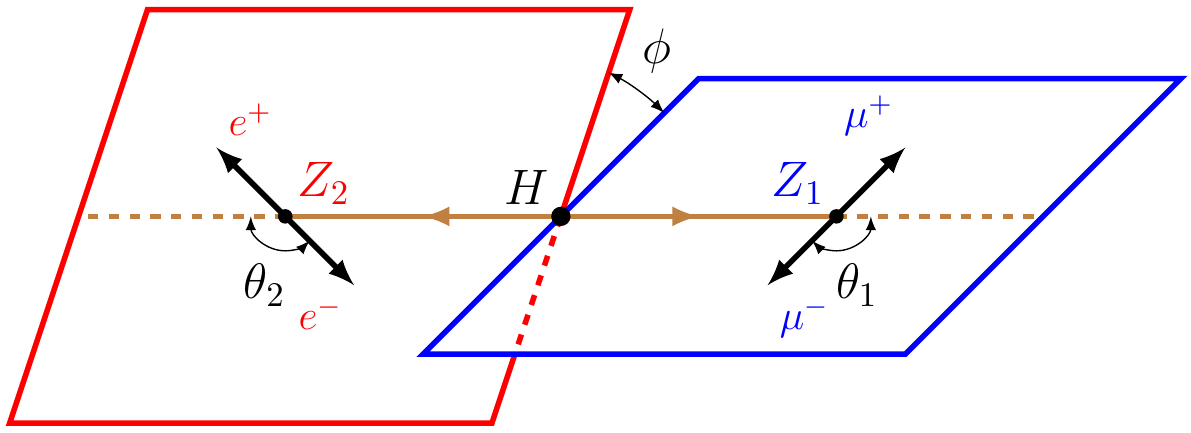}
		\caption{Angle definition between $Z_1$ and $Z_2$ decay planes in
			$gg\rightarrow H \rightarrow Z_1 Z_2 \rightarrow \mu^+ \mu^- e^+ e^-$ process. }
    \label{ggHZZ}
	\end{figure}
	$\phi_2$ can be defined similarly as $\phi_1$, with exchanging corresponding labels in the formulas above.

	In Fig.~\ref{ggHZZ}, $ \theta_1 / \theta_2 \in[0, \pi] $ is defined as the decay polar angle of the $ \mu^- / e^- $ in $ Z_1 / Z_2 $ rest frame and the direction of  $ Z_1 / Z_2 $ is the same as it is in $H$ rest frame.
	\begin{equation}
	\theta_1=\\\cos^{-1}\left(\dfrac{\vec{p}_{56} \cdot \vec{p}_5}{|\vec{p}_{56}||\vec{p}_5|}\right).
	\end{equation}
	\begin{equation}
	\theta_2=\cos^{-1}\left(\dfrac{\vec{p}_{34} \cdot \vec{p}_3}{|\vec{p}_{34}||\vec{p}_3|}\right).
	\end{equation}
	$ \phi \in [-\pi, \pi]$ is the angle between the $ Z_1 $ decay  and $ Z_2 $ decay planes. Since the boost along the Z direction does not change this angle, it can be defined in terms of either the lepton momentum in the Higgs rest frame or the lepton momentum in the $Z$ boson rest frame. We choose to define it in the Higgs rest frame. The expression for $ \phi $ is
	\begin{equation}
	\phi=\left[\cos^{-1}\left(\hat{n}_{decay1} \cdot \hat{n}_{decay2}\right)\right] \cdot
\mbox{sign}[{\left(\hat{n}_{decay2} \times \hat{n}_{decay1}\right) \cdot \vec{p}_{56}}],
	\end{equation}
	where $ \hat{n}_{decay1} $ and 	$ \hat{n}_{decay2} $ are unit vectors perpendicular to the decay planes of $ Z_1 $ and $ Z_2 $, respectively,
	\begin{equation}
	\begin{split}
	\hat{n}_{decay1}=\dfrac{\vec{p}_5 \times \vec{p}_6}{|\vec{p}_5 \times \vec{p}_6|}, \\
	\hat{n}_{decay2}=\dfrac{\vec{p_3} \times \vec{p}_4}{|\vec{p}_3 \times \vec{p}_4|}.
	\end{split}
	\end{equation}

The four-body phase space integral can be rewritten into a number of two-body phase space integrals through variable substitution
\begin{equation}
\begin{split}
	\int d\Pi_4&=\int\prod_{i=1}^{4}\dfrac{d^3p_i}{(2\pi)^32E_i}(2\pi)^4\delta^{(4)}(p_1+p_2-p_3-p_4-p_5-p_6)\\
	&=\frac{1}{4^4(2\pi)^{12}}\int\lambda ds\ dt\int d\cos\theta^*\ d\phi^* \int d\cos\theta_1\ d\phi_1 \int d\cos\theta_2\ d\phi_2 \label{eq:xkjjbl. 1},
	\end{split}
\end{equation}
where
\begin{equation*}
	s=\dfrac{k_1^2}{k^2}, \qquad t=\dfrac{k_2^2}{k^2}, \qquad\lambda=\sqrt{1+s^2+t^2-2s-2t-2st}.
\end{equation*}
$k_1=p_3+p_4$ and $k_2=p_5+p_6$ represent the momentum of the two $Z$ bosons, and $ k=p_1+p_2 $ represents the momentum of the Higgs boson. $\phi^*$ is the aforementioned trivial azimuth angle which can be integrated directly. $\phi_2$ can be replaced by $\phi$ as $\phi=\phi_1+\phi_2$.  
	
	\subsection*{The form of angular distribution}
	
	In order to cross-check the simulated angular distribution, we parameterize the amplitudes by the angular variables defined in the above section, and obtain the form of the angular distribution by integrating the phase space of the four bodies final states.
	
	For the angular distribution of signal processes, some papers\cite{Gao:2010qx,Anderson:2013afp,Ecker:2018ouo} have simulated these processes and analytic expressions have been given. Therefore, we will focus on calculating the angular distributions contributed from interference in the followings.

By redefining the coefficients, we can get the differential distribution of $\phi$. The angular distribution of the interference between BSM CP-even and BSM CP-odd process can be expressed as	
\begin{equation}
	\frac{d\sigma_{\rm even\, odd}}{d\phi}=A_1\sin 2\phi-A_2\sin\phi.
\label{dsigdphi-even-odd}
\end{equation}
	
The angular distribution of the interference between SM and BSM CP-odd process can be expressed as
\begin{equation}
	\frac{d\sigma_{\rm SM\, odd}}{d\phi}=B_1\sin 2\phi-B_2\sin\phi.
\label{dsigdphi-SM-odd}
\end{equation}
	
The angular distribution of the interference between SM and BSM CP-even process can be expressed as
\begin{equation}
	\frac{d\sigma_{\rm SM\, even}}{d\phi}=C_1\cos^2\phi-C_2\cos\phi+C_3,
\label{dsigdphi-SM-even}
\end{equation}
\begin{equation}
	\frac{d\sigma_{\rm SM\, even}}{d\cos\theta_1}=C_4,
\label{dsigdcosth-SM-even}
\end{equation}
where $ C_1, C_2, C_3$ and $C_4 $ are the unknown coefficients. Calculation details are shown in Appendix~\ref{Appendix-formula}.

	\section{SIMULATION}\label{SIMULATION}
	
	\subsection{Simulation parameters}
	We write amplitudes of the process of $gg\rightarrow H \rightarrow Z_1 Z_2 \rightarrow \mu^+ \mu^- e^+ e^-$ listed in Appendix~\ref{Appendix-formula} that with anomalous HZZ couplings into MCFM8.0. Information of total cross section and differential cross sections by integrating the phase space and parton distribution functions(PDF)\cite{Campbell:2015qma,Boughezal:2016wmq} are obtained. The proton-proton collisions are simulated at the center of mass energy of 13 TeV. Higgs mass is set to be 125 GeV. The renormalization and factorization scales are set to be the dynamic scale $ m_{4l} $/2. For PDF we choose MSTW08LO~\cite{Martin:2009iq}. The angular distribution of the differential cross section may be apparently changed by certain cut criteria. Here we perform the kinematic distribution with no cut and the presence of some basic CMS cuts respectively. The basic the CMS cuts are chosen to be \cite{CMS-PAS-HIG-13-002}	
	\begin{equation}
	\begin{split}
	&P_T(\mu)>5~\mathrm{GeV}, \quad |\eta_\mu|<2. 4, \\
	&P_T(e)>7~\mathrm{GeV}, \quad |\eta_e|<2.5, \\
	&m_{4l}>100\ \mathrm{GeV}.
	\end{split}
	\label{CMScut}
	\end{equation}
	In the simulation, we set the anomalous coupling coefficient $ a_1=a_2=a_3=1 $. Differential cross sections with varying these coupling parameters can be obtained easily by multiplying a rescale factor. Interference between SM Higgs mediate/box process and the anomalous $HZZ$ process is expected to be dominent according to coupling order analysis($a_1\approx 1$, $a_2$ and $a_3$ should be small).Since the distribution of $ \theta_1 $ is similar to that of $ \theta_2 $, we only list the distribution of $ \cos\theta_1 $ here.
		
	\subsection{Simulation results and analysis of angular distribution}
	
	In this section, we draw various differential cross sections in the four lepton final states. As aforementioned, four angular variables $\phi$, $\phi_1$, $\theta_1$ and $\theta^*$  are studied here. There are large numbers of differential cross sections with considering three ingredients as follows. 1. The energy region can be divided into Higgs on-shell~($m_{2e2\mu}<130~\mbox{GeV}$) and off-shell ~($m_{2e2\mu}>220~\mbox{GeV}$) regions. 2. At the amplitude level, the four lepton final states can be generated through SM-Higgs-Mediated~(abbreviated to SM), BSM CP-even(abbreviated to CP-even), BSM CP-odd(abbreviated to CP-odd), and SM box(abbreviated to box) processes. 3. Cut criteria may change the differential cross sections apparently. We compared the differential cross section with and without CMS cuts mentioned in Eq.~(\ref{CMScut}).
	
	In Appendix~\ref{Appendix-figure}, Fig.~\ref{self-Higgs-0n} to \ref{signal-inter-SM-CPodd-on} show a complete collection of angular differential cross sections with different combinations of the above three situations. Specific situations are explained in the caption of each diagram. Each row of these Figures contain four differential cross sections, which are 1. Angle $\phi$ between the two $Z$ decay planes in Higgs rest frame, 2. Angle $\phi_1$ between $Z_1$ production and decay plane in the Higgs rest frame, 3. Decay polar angle $\theta_1$ in $Z_1$ rest frame, 4. Angle $\theta^*$ between the momentum of $Z_1$ and the Z-axis in Higgs rest frame. It can be seen that parts of these distributions are not so distinguishable (Eg. the plat pattern of $d\sigma/d \cos \theta^*$ in most cases). Nevertheless, we retain them to form a complete collection of the distributions.

We analysis these figures according to the three aspects as mentioned in the Introduction section.
	
\subsection*{The advantage of differential cross section compare to the integrated total cross section}
Although analyzing the experimental angular cross section needs more statistics, the advantage of doing so is obvious. For cases where distributions are odd functions, such as contributions from interference between CP-odd and box in off-shell region(the third row in Fig.~\ref{inter-box-off}) \& CP-even and CP-odd(the second row in Fig.~\ref{inter-Higgs-on} and \ref{inter-Higgs-off}), SM and CP-odd(the third row in Fig.~\ref{inter-Higgs-on} and \ref{inter-Higgs-off}) in both on-shell and off-shell Higgs energy region, the integrated cross section is zero. 
Only differential distributions can exhibit these anti-symmetric pattern.
	
\subsection*{Constraints from off-shell region}
The success of the constraints of upper limit of the Higgs width~\cite{CMS:2014quz, CMS:2022ley} shows an practical method of measuring small physics parameters with limited experimental data sets. For the BSM couplings in the Lagrangian in the off shell Higgs region, although there is a disadvantage coming from the suppress in the Higgs propagator, the cross section can be enhanced by the vertices which are proportional to $p^\mu p^\nu$~\cite{Feng:2021izk}. The overall effects make the cross section of the interference~(Fig.~\ref{inter-box-off} and Fig.~\ref{inter-Higgs-off} ) comparable to that of the self-conjugate~(Fig.~\ref{self-Higgs-off}) when we take the anomalous coefficients $a_1=a_2=a_3=1$.
	
\subsection*{Interference effects} 
\begin{figure}[htb]
	\includegraphics[width=5.5cm]{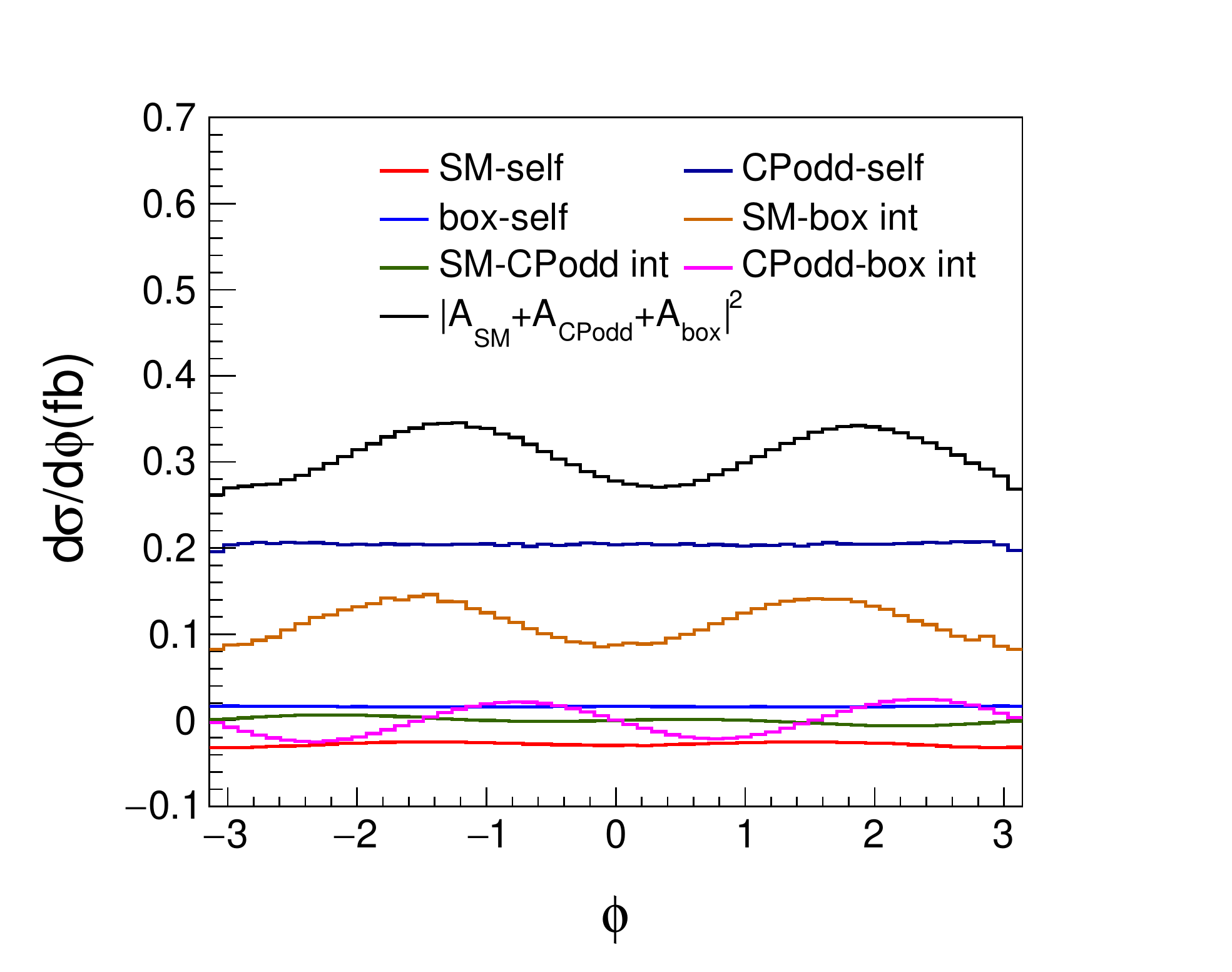}
	\includegraphics[width=5.5cm]{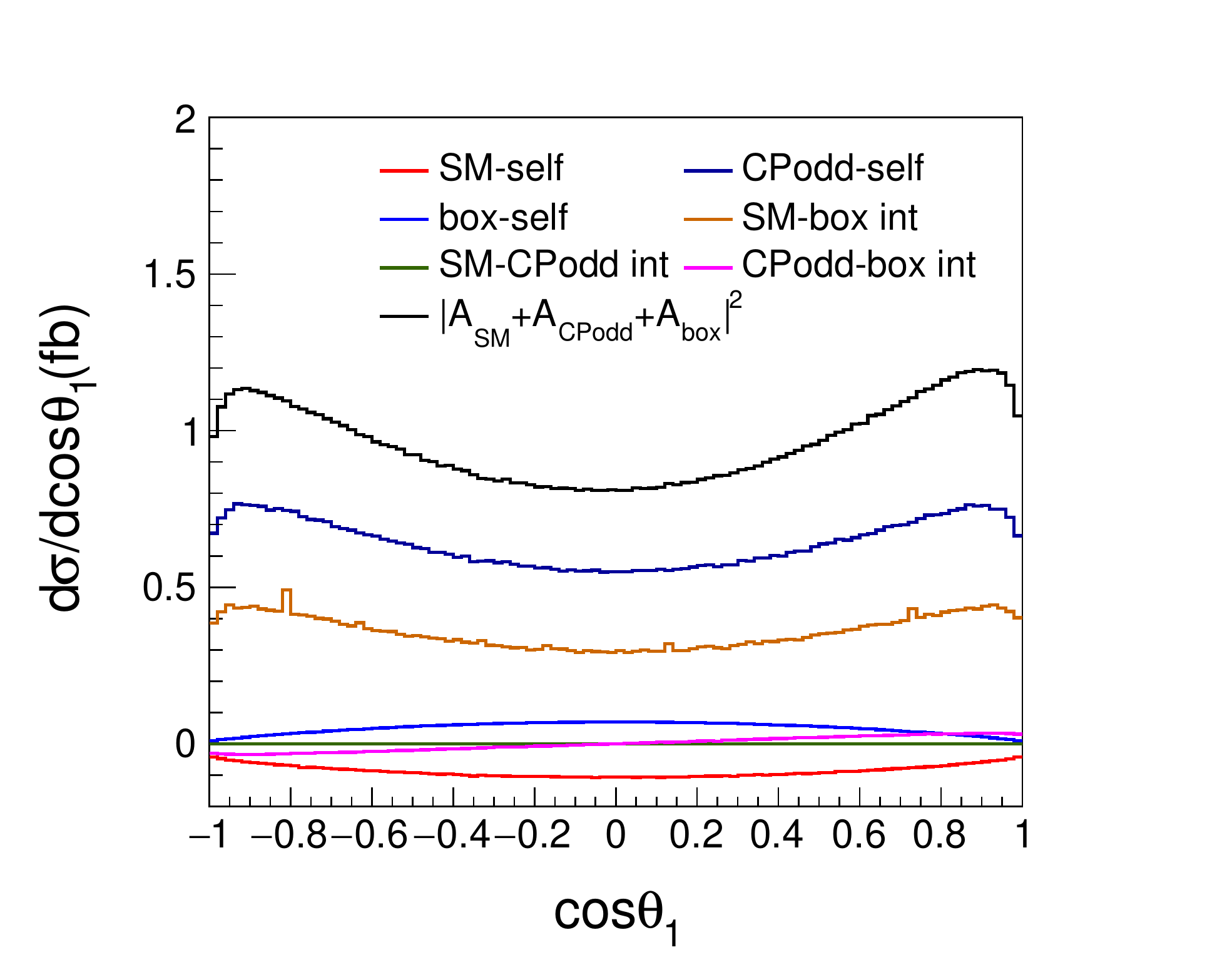}
	\\
	\includegraphics[width=5.5cm]{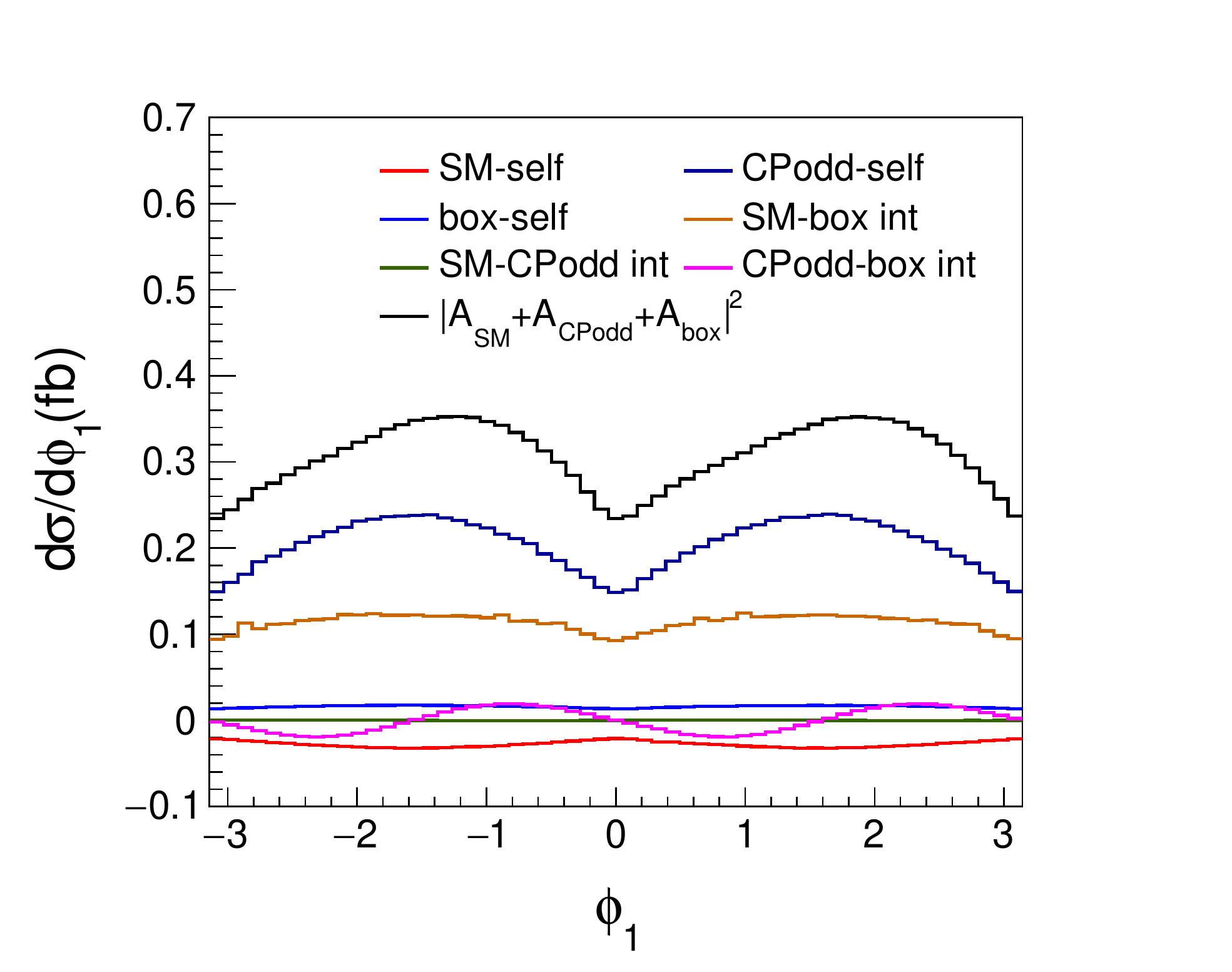}
	\includegraphics[width=5.5cm]{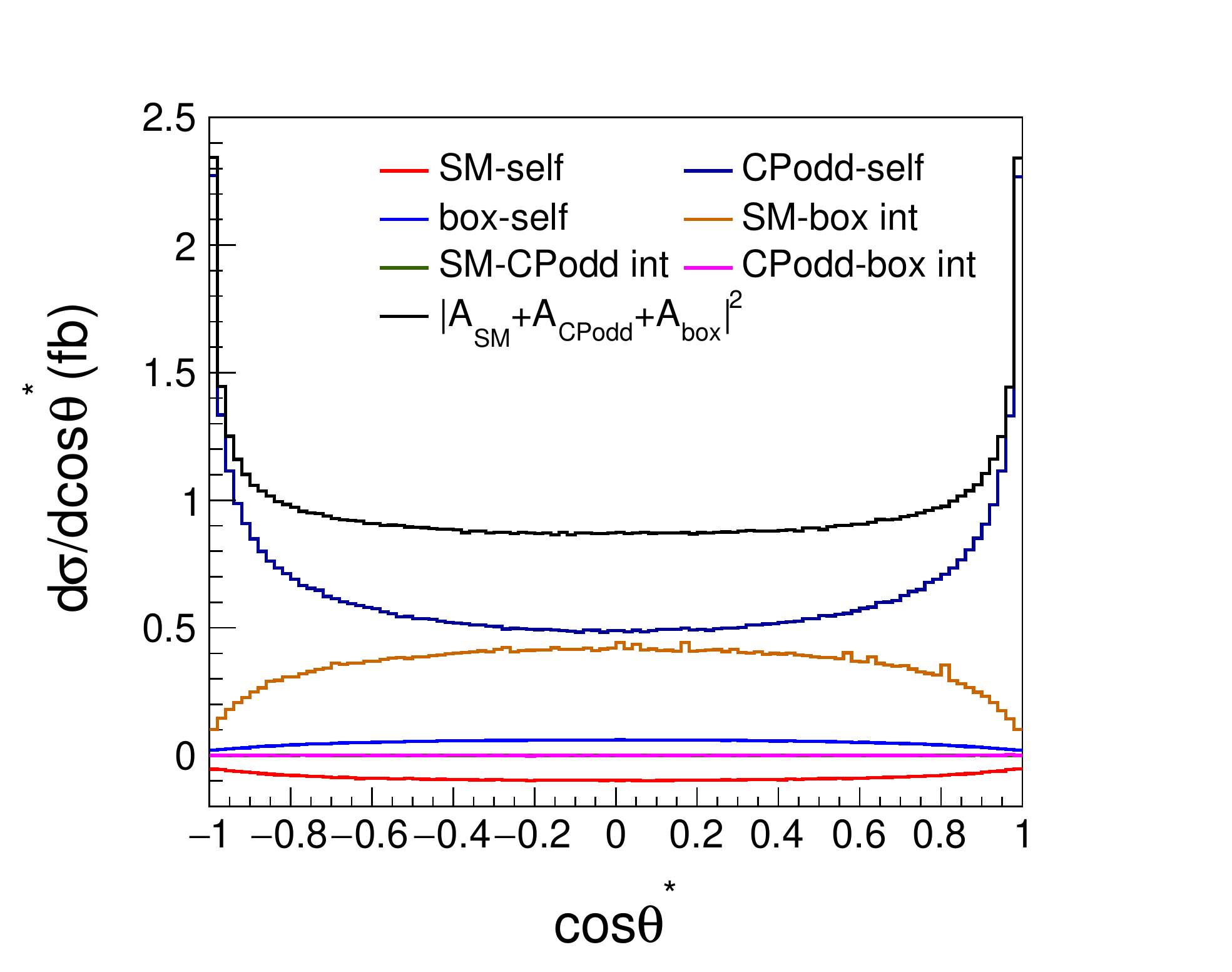}
	\caption{Off-shell summed differential cross section of all signal process and its interference between each other. }\label{totcross}
\end{figure}

For the interference diagrams in Fig.~\ref{inter-box-on},\ref{inter-box-off},\ref{inter-Higgs-on} and \ref{inter-Higgs-off} in both Higgs on-shell and off-shell regions, we notice that the similarities of the anti-symmetric patterns in these diagrams are that all of them  are interference contributions from CP-odd process and another process. So we consider that these kind of odd functions can be constructed into a CP-sensitive variable measurement to indicate the CP-odd term in the effective Lagrangian. This again shows one advantage of our analytical calculation. Although one can not separate contributions from each source in the real experiment, which are a summation of all sources in the overall differential cross sections as shown in Fig.~$\ref{totcross}$, only analytical theoretical calculations can exhibit separately the contribution  from these interference. Experimentally, one can divide the angular distribution cross section into two parts: symmetric and antisymmetric functions. The symmetric differential cross sections are contributed from the self or conjugate interference between CP-even amplitudes, which include SM, CP-even and  box diagrams, additional with the square of the CP-odd amplitude. Therefore, it can be used to study the BSM CP-even term in $ {\cal L}_{eff}$. The antisymmetric functions are contributed from the interference between CP-odd process and above three CP-even process(SM, CP-even and  box). So we can construct a CP-odd sensitive observable by using this antisymmetric cross section, similar as the well known Forward and Backward Asymmetry $A_{FB}$. 
\clearpage
	
Furthermore, the form of  Eq.~(\ref{dsigdphi-even-odd})/Eq.~(\ref{dsigdphi-SM-odd})(or the explicit form of Eq.(\ref{even-odd})/Eq.~(\ref{SM-odd}) in Appendix~\ref{Appendix-formula}) shows that the odd function of the  differential interference cross section to angle $\phi$ between CP-even/SM and CP-odd process(the second/third row, the first column in Fig.\ref{inter-box-on} and Fig.~\ref{inter-box-off}) are constructed by a linear combination of $\sin 2\phi$(with period $\pi$) and $\sin \phi$ (with period $2\pi$). For comparison, the form of Eq.~(\ref{dsigdphi-SM-even})(or the explicit form of Eq.~(\ref{SM-even}) in Appendix~\ref{Appendix-formula}) shows that the even function of the differential interference cross section between SM and CP-even process~(the first row, the first column in Fig.\ref{inter-box-on} and Fig.~\ref{inter-box-off}) are constructed by a linear combination of $\cos 2\phi$(with period $\pi$) and $\cos \phi$ (with period $2\pi$). The above analyses show that conveniently a linear combination of $\sin n\phi (n=1,2)$ or $\cos n\phi$ series can be adopted in the fitting of the experimental differential cross section $d\sigma /d \phi$.

In real experimental measurements, the observed angular distributions are a summation of
all contributions. At the amplitude level, we have discussed four sources~(SM, CP-even, CP-odd
and box). Our analytical calculations can separate contribution clearly from each self-conjugate or interference sources. Here we show a method to extract the CP-odd coefficient in the effective dimension-6 Lagrangian Eq.~\ref{Effective Lagrangian} in case of $a_1= 1$ and $a_2 = 0$.
	
\begin{figure}[ht]

		\includegraphics[width=5.5cm]{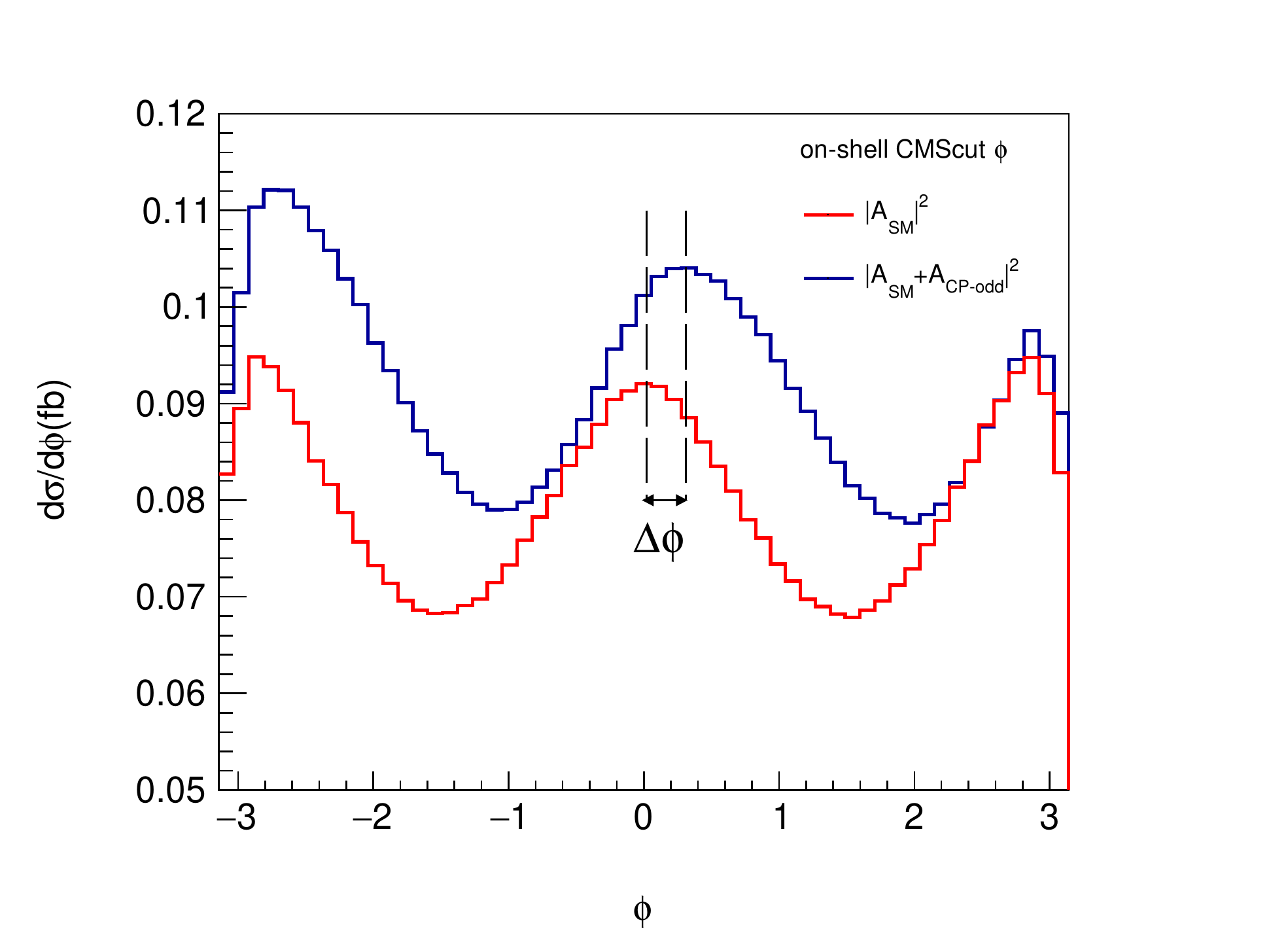}
		\includegraphics[width=5.5cm]{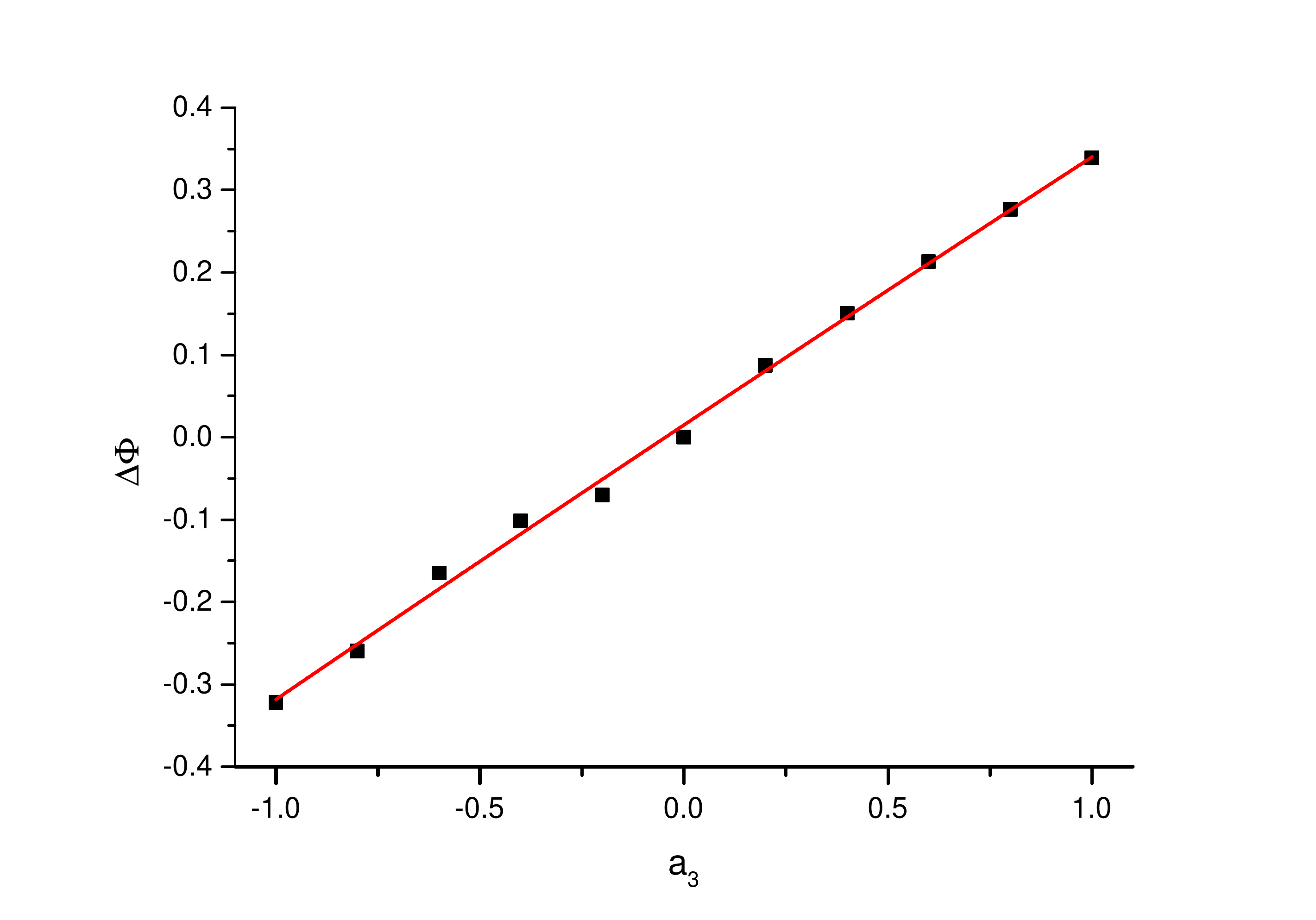}
		\caption{ A linear relationship between the CP-odd coefficient $a_3$ and the shift angle $\Delta\phi$ in $d\sigma/d\phi$ angular distribution with $a_1= 1$ and $a_2 = 0$. }
		\label{Delta-phi-linear}
	\end{figure}

\begin{figure}[htb]
		
		\centering
		
		\includegraphics[width=5.5cm]{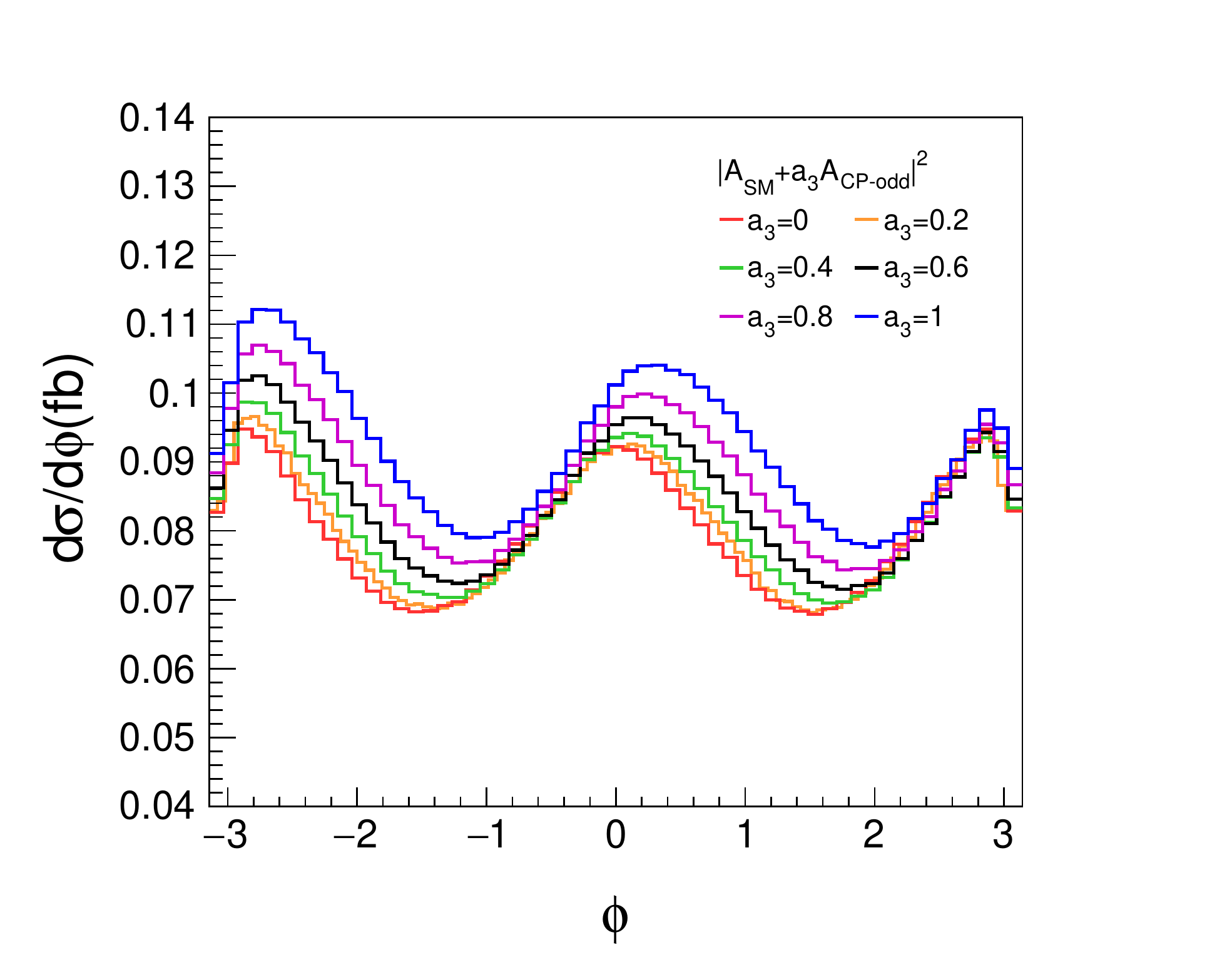}
		\includegraphics[width=5.5cm]{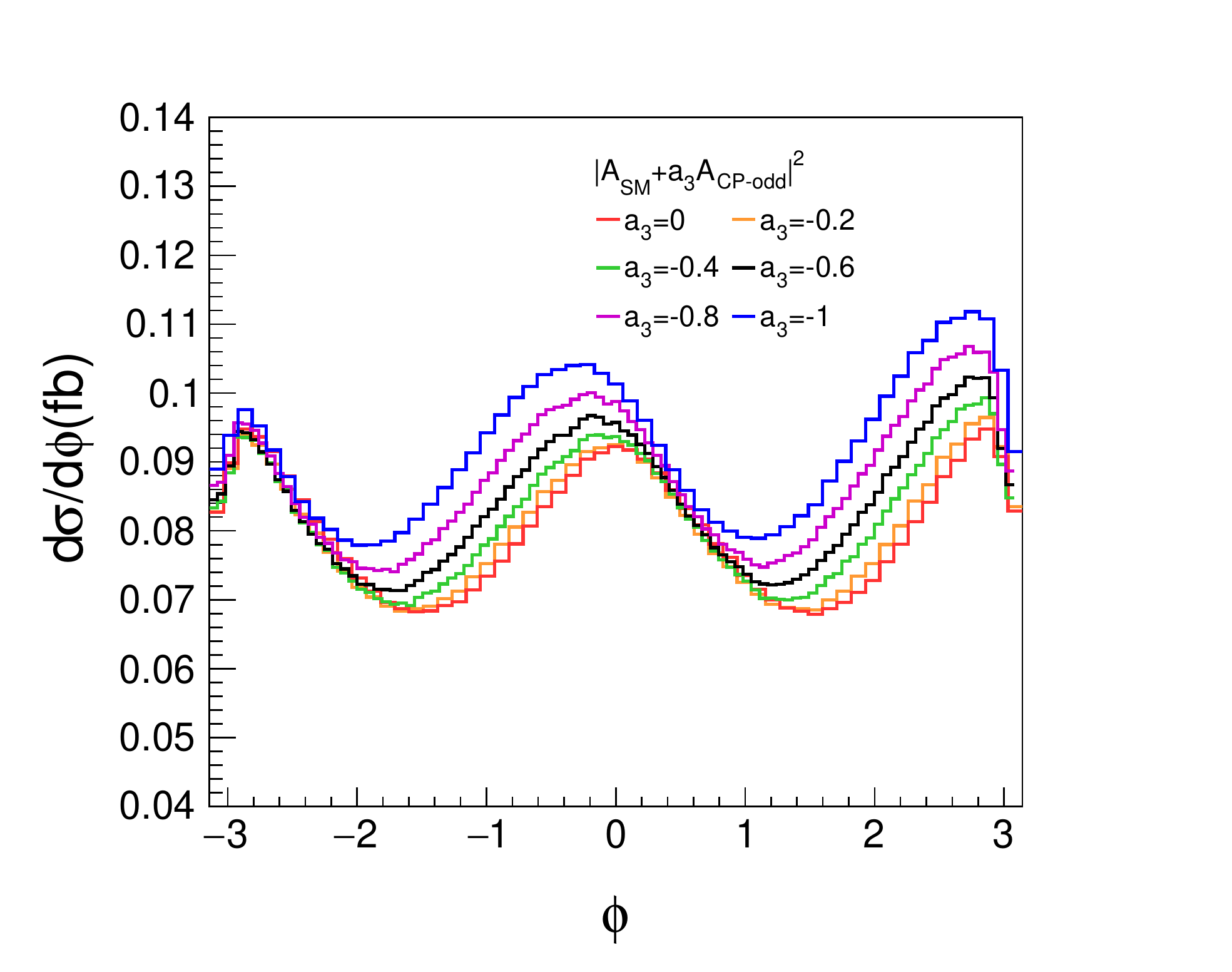}
		
\caption{The shifts of the angular distributions $d\sigma/d\phi$ by varying CP-odd coefficient $a_3$ in Higgs on-shell region with CMS basic cuts.}
\label{SM-CPodd on-shell}
	\end{figure}

As discussed above, the angle $\phi$ defined between the $Z_1$ and $Z_2$ decay planes in the Higgs rest frame can be a CP sensitive angle.  Fig.~\ref{Delta-phi-linear} and \ref{SM-CPodd on-shell} show that by changing the value of the CP-odd coupling term $a_3$, the peak of the differential cross section with SM plus CP odd coupling $d\sigma/d\phi$ distribution in the on-shell region can have a linear shift correspondingly.
	
Therefore, $\delta\phi$ can be a CP-odd sensitive experimental variable to constrain the coefficient $a_3$.	

\clearpage

\section{summary}\label{summary}
CP properties of the Higgs boson attract lots of attentions in searching of New Physics Beyond the Standard Model. In this paper, by adopting the dimension-6 effective operators, we study the angular distribution cross section in $gg\to H\to ZZ\to e^+e^-\mu^+\mu^-$ process. We give analytical formulas of the angular dependence at the cross section level. Experimentally, the angle $\phi$, which is defined as the angle between two $Z$ decay planes in the Higgs rest frame, the Forward and Backward asymmetry constructed from the antisymmetric pattern of the angular distribution, and the linear relationship between $\Delta\phi$ and the CP-odd coefficient $a_3$, are emphasized as the CP sensitive observables to extract BSM $HZZ$ CP couplings.  
    
\begin{acknowledgements}
The work is supported by the National Natural Science Foundation of China under Grant No.~11847168, the Fundamental Research Funds for the Central Universities of China under Grant No.~GK202003018, and the Natural Science Foundation of Shannxi Province, China (2019JM-431, 2019JQ-739).
\end{acknowledgements}
\begin{appendix}

\section{Amplitudes Formulas} 
\label{Appendix-formula}
\setcounter{equation}{0}
The SM, BSM CP even and BSM CP odd Higgs decay amplitudes can be written as
	\begin{equation}
		\begin{split}
			\mathcal{A}_{\rm SM}^{H\to Z^*Z^*\to2e2\mu}
			=&i\dfrac{M_Z^4}{2v^3}g^{\mu\nu}\dfrac{1}{k_1^2-M_Z^2}\dfrac{1}{k_2^2-M_Z^2}[\overline{u(3)}\gamma_{\mu}(c_v+\gamma^5)v(4)]\\&\times[\overline{u(5)}\gamma_{\nu}(c_v+\gamma^5)v(6)],
		\end{split}
	\end{equation}
	
	\begin{equation}
		\begin{split}
			\mathcal{A}_{\rm CP\raisebox{0mm}{-}even}^{H\to Z^*Z^*\to2e2\mu}
			=&-i\dfrac{M_Z^2}{v^3}(k_1^\nu k_2^\mu-k_1\cdot k_2g^{\mu\nu})\dfrac{1}{k_1^2-M_Z^2}\dfrac{1}{k_2^2-M_Z^2}\\
			&\times[\overline{u(3)}\gamma_{\mu}(c_v+\gamma^5)v(4)][\overline{u(5)}\gamma_{\nu}(c_v+\gamma^5)v(6)], \\
		\end{split}
	\end{equation}
	
	\begin{equation}
		\begin{split}
			\mathcal{A}_{\rm CP\raisebox{0mm}{-}odd}^{H\to Z^*Z^*\to2e2\mu}
			=&-i\dfrac{M_Z^2}{v^3}\varepsilon^{\mu\nu\rho\sigma}k_{1\rho}k_{2\sigma}\dfrac{1}{k_1^2-M_Z^2}\dfrac{1}{k_2^2-M_Z^2}\\
			&\times[\overline{u(3)}\gamma_{\mu}(c_v+\gamma^5)v(4)][\overline{u(5)}\gamma_{\nu}(c_v+\gamma^5)v(6)].
		\end{split}
	\end{equation}
	
	Therefore, the interference terms of the amplitude squared BSM CP-even process and BSM CP-odd process can be expressed as 
	\begin{equation}
		\begin{split}
			2{\mbox{Re}}(\mathcal{A}_{\rm CP-even}\mathcal{A}_{\rm CP-odd}^*)
			=&|\mathcal{A}^{gg\to H}P_H|^2\dfrac{M_Z^4}{v^6}\dfrac{128}{(k_1^2-M_Z^2)^2(k_2^2-M_Z^2)^2}\\
			&\times\{(c_v^2+1)^2[(p_3\cdot p_5)(p_4\cdot p_6)-(p_4\cdot p_5)(p_3\cdot p_6)]\\
			&+4c_v^2(p_3\cdot p_4)(p_5\cdot p_6)\}\varepsilon^{\mu\nu\rho\sigma}p_{3\mu}p_{4\nu}p_{5\rho}p_{6\sigma}.
		\end{split}
	\end{equation}
	The interference terms of the amplitude squared SM process and BSM CP-odd process can be expressed as 
	\begin{equation}
		\begin{split}
			2{\mbox{Re}}(\mathcal{A}_{\rm SM}\mathcal{A}_{\rm CP-odd}^*)
			=&|\mathcal{A}^{gg\to H}P_H|^2\dfrac{M_Z^6}{v^6}\dfrac{16}{(k_1^2-M_Z^2)^2(k_2^2-M_Z^2)^2}\\
			&\times[(c_v^2+1)^2(p_3\cdot p_5+p_4\cdot p_6-p_4\cdot p_5-p_3\cdot p_6)+4c_v^2k_1\cdot k_2]\\
			&\times\varepsilon^{\mu\nu\rho\sigma}p_{3\mu}p_{4\nu}p_{5\rho}p_{6\sigma}.
		\end{split}
	\end{equation}
	The interference terms of the amplitude squared BSM CP-even process and SM process can be expressed as 
	\begin{equation}
		\begin{split}
			2{\mbox{Re}}(\mathcal{A}_{\rm SM}\mathcal{A}_{\rm CP-even}^*)
			=&|\mathcal{A}^{gg\to H}P_H|^2\dfrac{M_Z^6}{v^6}\dfrac{16}{(k_1^2-M_Z^2)^2(k_2^2-M_Z^2)^2}\\
			&\times\{(c_v^2+1)^2[(p_3\cdot p_5+p_4\cdot p_6-p_3\cdot p_6-p_4\cdot p_5)\\
			&\times((p_3\cdot p_5)(p_4\cdot p_6)-(p_3\cdot p_6)(p_4\cdot p_5))+(p_3\cdot p_4)(p_5\cdot p_6)(k_1\cdot k_2)]\\
			&+4c_v^2[(p_3\cdot p_4)(p_5\cdot p_6)(p_3\cdot p_5+p_4\cdot p_6-p_3\cdot p_6-p_4\cdot p_5)\\
			&+k_1\cdot k_2((p_3\cdot p_5)(p_4\cdot p_6)-(p_3\cdot p_6)(p_4\cdot p_5))]\}\label{eq:zfpf12. 3}.
		\end{split}
	\end{equation}

\begin{equation}
	\begin{split}
			2{\mbox{Re}}(\mathcal{A}_{\rm CP-even}\mathcal{A}_{\rm CP-odd}^*)
		=&-|\mathcal{A}^{gg\to H}P_H|^2\dfrac{4M_Z^4}{v^6}\dfrac{\lambda\sqrt{st}}{\left(s-\dfrac{M_Z^2}{k^2}\right)^2\left(t-\dfrac{M_Z^2}{k^2}\right)^2}\\
		&\times\{[(c_v^2+1)^2st c_{\theta_2} c_{\theta_1}+4c_v^2]s_{\theta_2} s_{\theta_1} s_{\phi}-2(c_v^2+1)^2g_H\sqrt{st}s^2_{\theta_2} s^2_{\theta_1} c_{\phi} s_{\phi}\}.
	\end{split}
\end{equation}

\begin{equation}~\label{even-odd}
	\dfrac{d\sigma_{\rm even\, odd}}{dc_{\theta_2} dc_{\theta_1} d\phi}=-(A_{\rm SM}c_{\theta_2} c_{\theta_1} s_{\theta_2} s_{\theta_1}+A_{\rm CP\raisebox{0mm}{-}even}s_{\theta_2} s_{\theta_1})s_\phi+A_{\rm CP\raisebox{0mm}{-}odd}s^2_{\theta_2} s^2_{\theta_1} c_\phi s_\phi,
\end{equation}
where $ c_{\theta} $ and $ s_{\theta} $ represent $ \cos \theta $ and $ \sin \theta $. $ A_{\rm SM} $, $ A_{\rm CP\raisebox{0mm}{-}even} $ and $ A_{\rm CP\raisebox{0mm}{-}odd} $ are unknown coefficients.
Following the same steps, we can get
\begin{equation}
	\begin{split}
		&2{\mbox{Re}}(\mathcal{A}_{\rm SM}\mathcal{A}_{\rm CP-odd}^*)\\
		=&|\mathcal{A}^{gg\to H}P_H|^2\dfrac{M_Z^6}{v^6k^2}\dfrac{\lambda\sqrt{st}}{\left(s-\dfrac{M_Z^2}{k^2}\right)^2\left(t-\dfrac{M_Z^2}{k^2}\right)^2}\\
		&\times\{-[(c_v^2+1)^2g_H c_{\theta_2} c_{\theta_1}+4c_v^2g_H]s_{\theta_2} s_{\theta_1} s_\phi+2(c_v^2+1)^2g_H\sqrt{st}s^2_{\theta_2} s^2_{\theta_1} c_\phi s_\phi\}.
	\end{split}
\end{equation}

\begin{equation}\label{SM-odd}
	\dfrac{d\sigma_{\rm SM\, odd}}{dc_{\theta_2} dc_{\theta_1} d\phi}=-(B_1c_{\theta_2} c_{\theta_1} s_{\theta_2} s_{\theta_1}+B_2s_{\theta_2} s_{\theta_1})s_\phi+B_3s^2_{\theta_2} s^2_{\theta_1} c_\phi s_\phi,
\end{equation}
in which $ B_1 $, $ B_2 $ and $ B_3 $ are unknown coefficients.

\begin{equation}
	\begin{split}
		&2{\mbox{Re}}(\mathcal{A}_{\rm SM}\mathcal{A}_{\rm CP-even}^*)\\
		=&|\mathcal{A}^{gg\to H}P_H|^2\dfrac{M_Z^6}{v^6k^2}\dfrac{1}{\left(s-\dfrac{M_Z^2}{k^2}\right)^2\left(t-\dfrac{M_Z^2}{k^2}\right)^2}\\
		&\times\{(c_v^2+1)^2[2g_H st+2g_H stc_{\theta_2}^2 c_{\theta_1}^2-(g_H^2\sqrt{st}+4(st)^{\frac{3}{2}})c_{\theta_2} s_{\theta_1} c_{\theta_1} s_{\theta_1} c_\phi\\
		&+2g_H sts_{\theta_2}^2 s_{\theta_1}^2 c_\phi^2]+4c_v^2[4g_H st c_{\theta_2} c_{\theta_1}-(g_H^2\sqrt{st}+4(st)^{\frac{3}{2}})s_{\theta_2} s_{\theta_1} c_\phi]\}.
	\end{split}
\end{equation}

\begin{equation}~\label{SM-even}
	\dfrac{d\sigma_{\rm SM\, even}}{dc_{\theta_2} dc_{\theta_1} d\phi}=C_1+C_2 c^2_{\theta_2} c^2_{\theta_1}-C_3 c_{\theta_2} s_{\theta_2} c_{\theta_1} s_{\theta_1} c_{\phi} + C_4 s^2_{\theta_2} s^2_{\theta_1} c^2_{\phi}+ C_5 c_{\theta_2} c_{\theta_1}-C_6 s_{\theta_2} s_{\theta_1} c_{\phi},
\end{equation}
in which $ C_1 $, $ C_2 $, $ C_3 $, $ C_4 $, $ C_5 $ and $ C_6 $ are unknown coefficients.
	
\section{Differential cross sections} 
\label{Appendix-figure}
The simulation results under different conditions and cuts are shown here.
\clearpage	
	\begin{figure}[htbp]
		\centering
		\includegraphics[width=3.6cm]{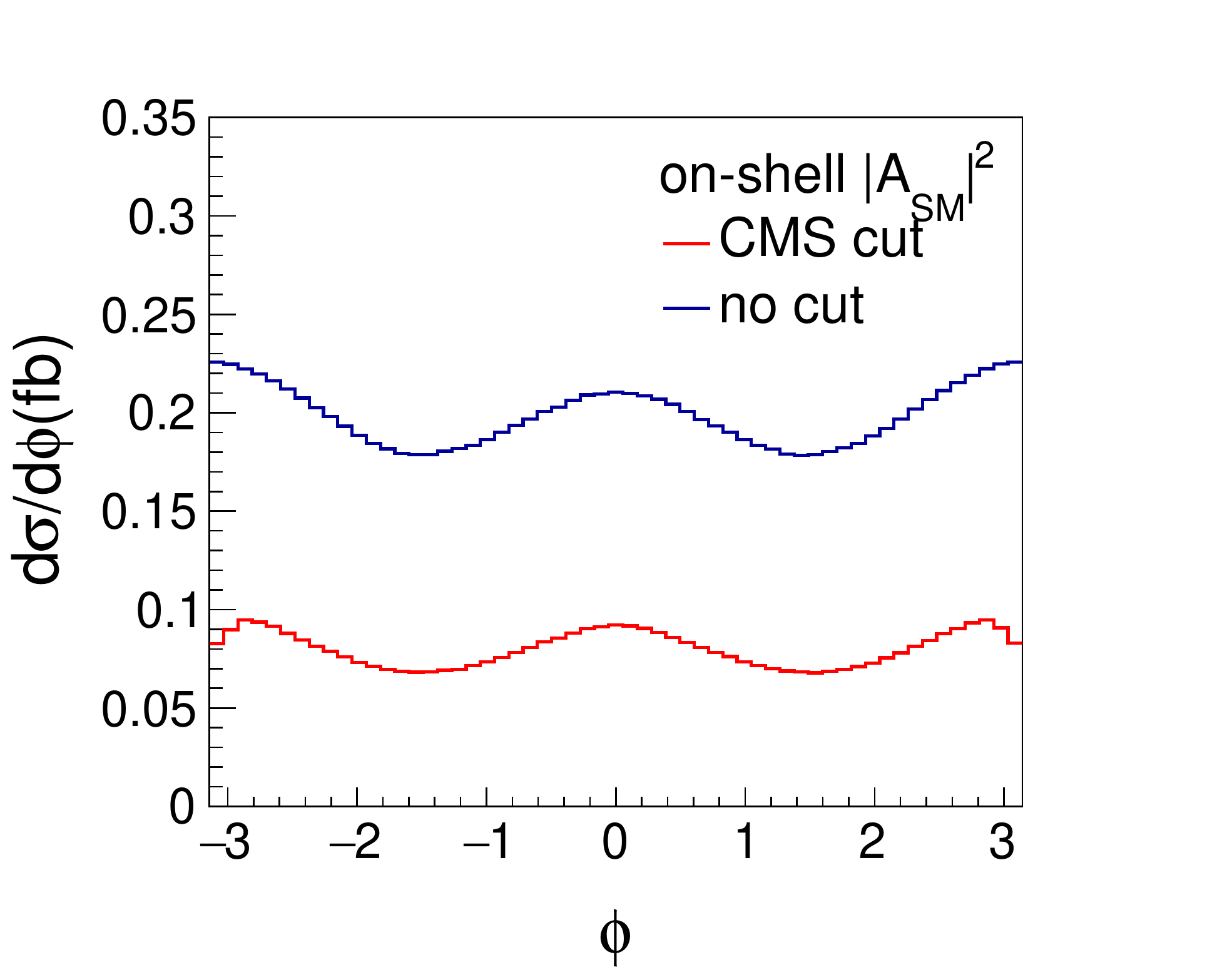}
		\includegraphics[width=3.6cm]{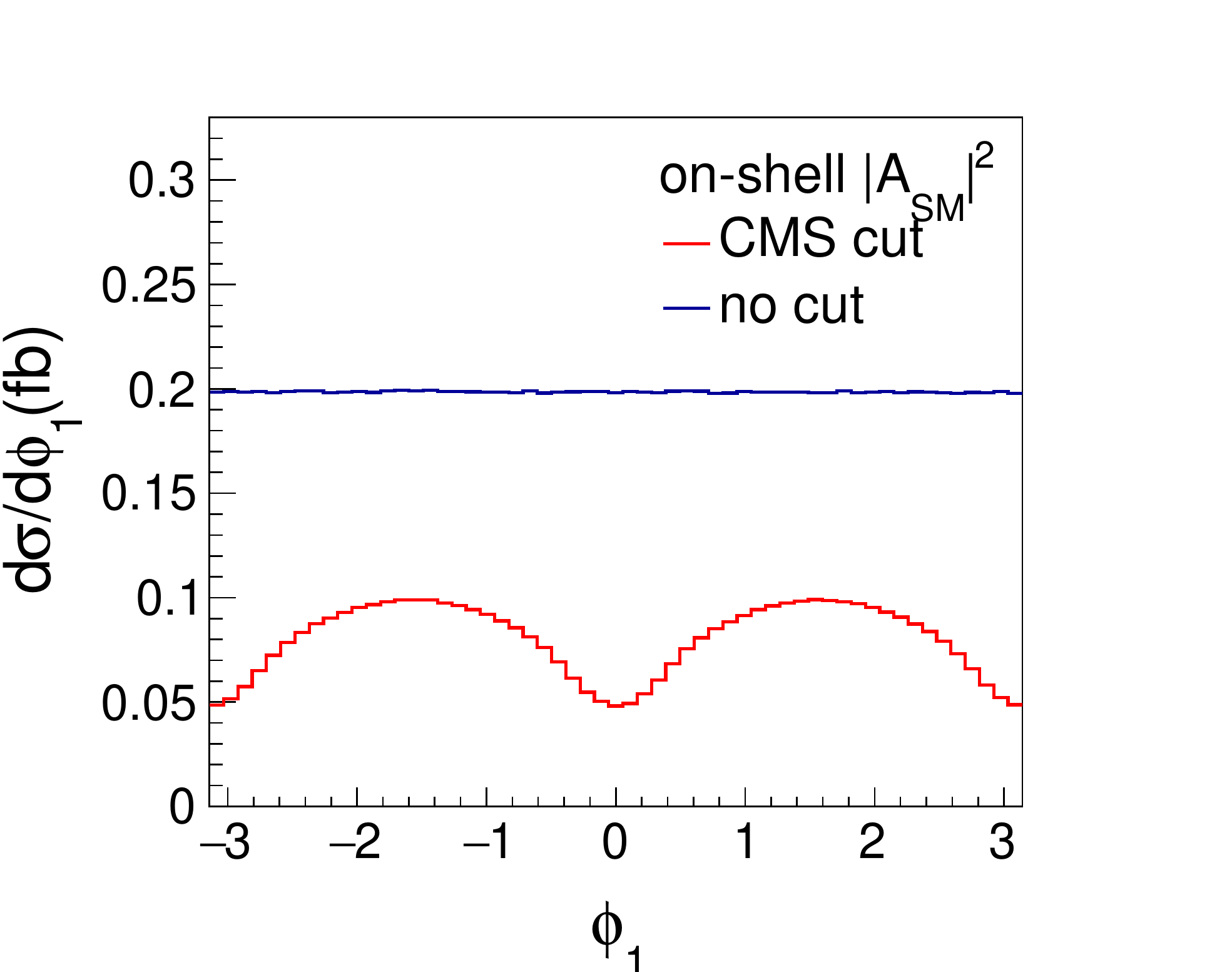}
		\includegraphics[width=3.6cm]{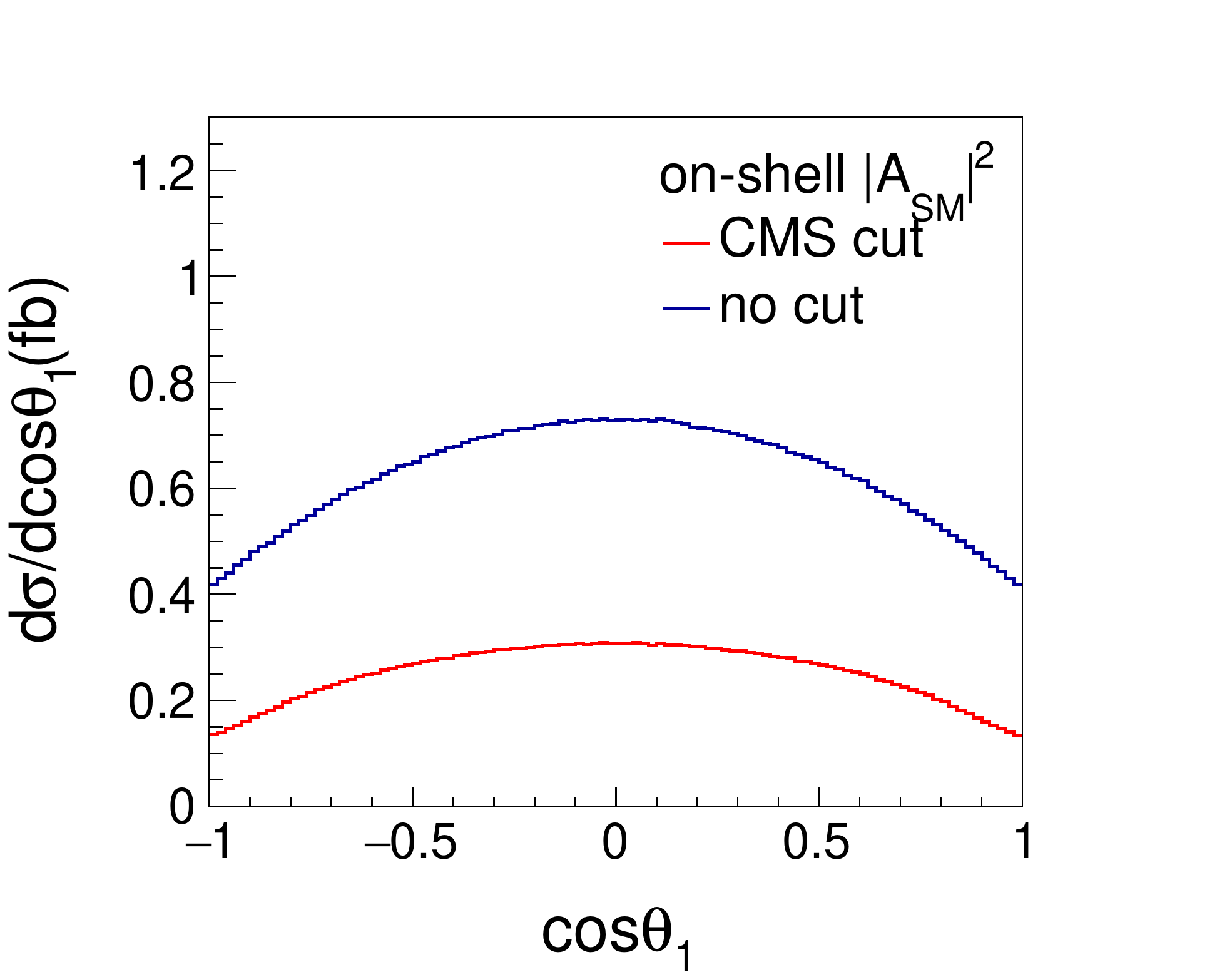}
		\includegraphics[width=3.6cm]{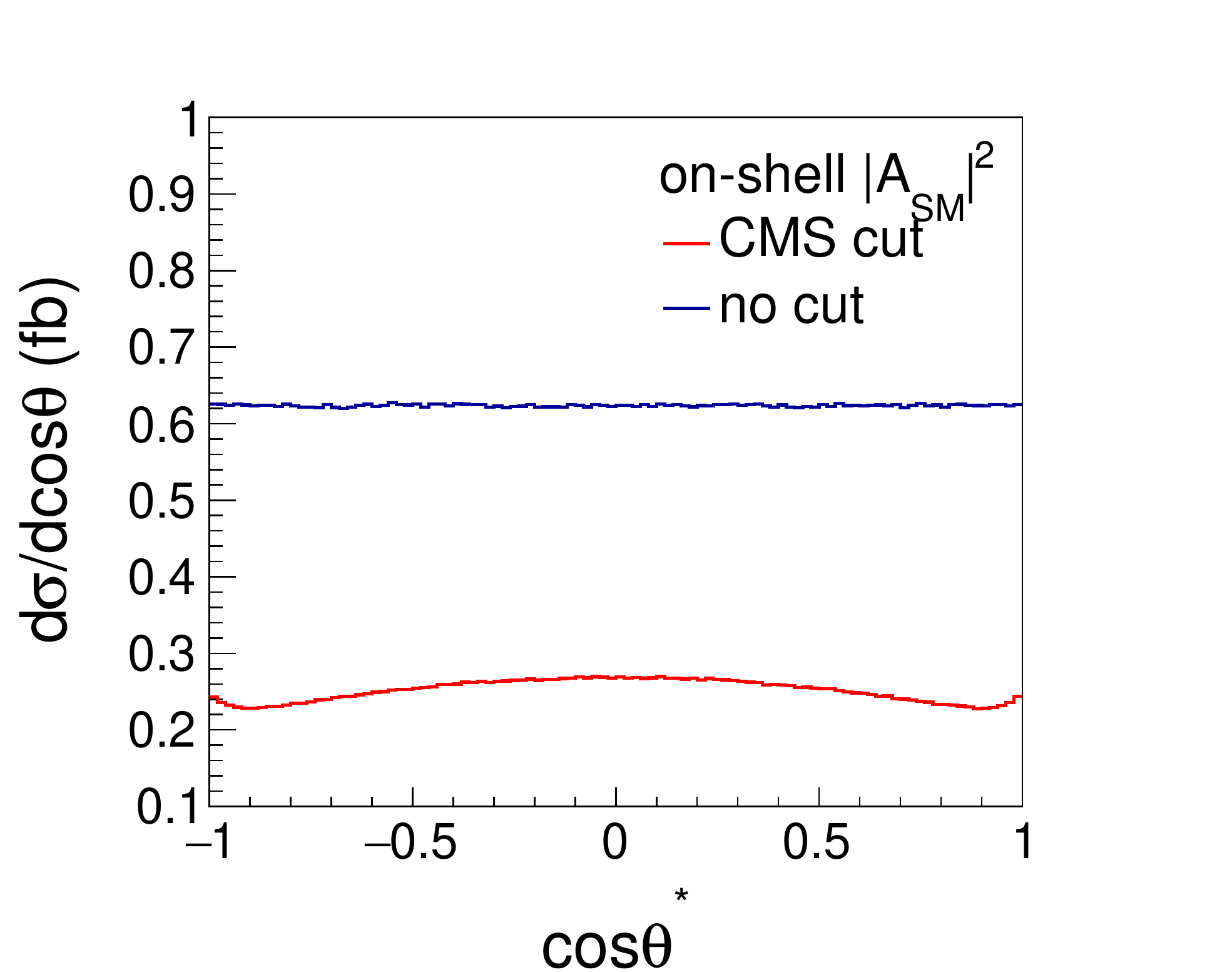}
		\\
		\includegraphics[width=3.6cm]{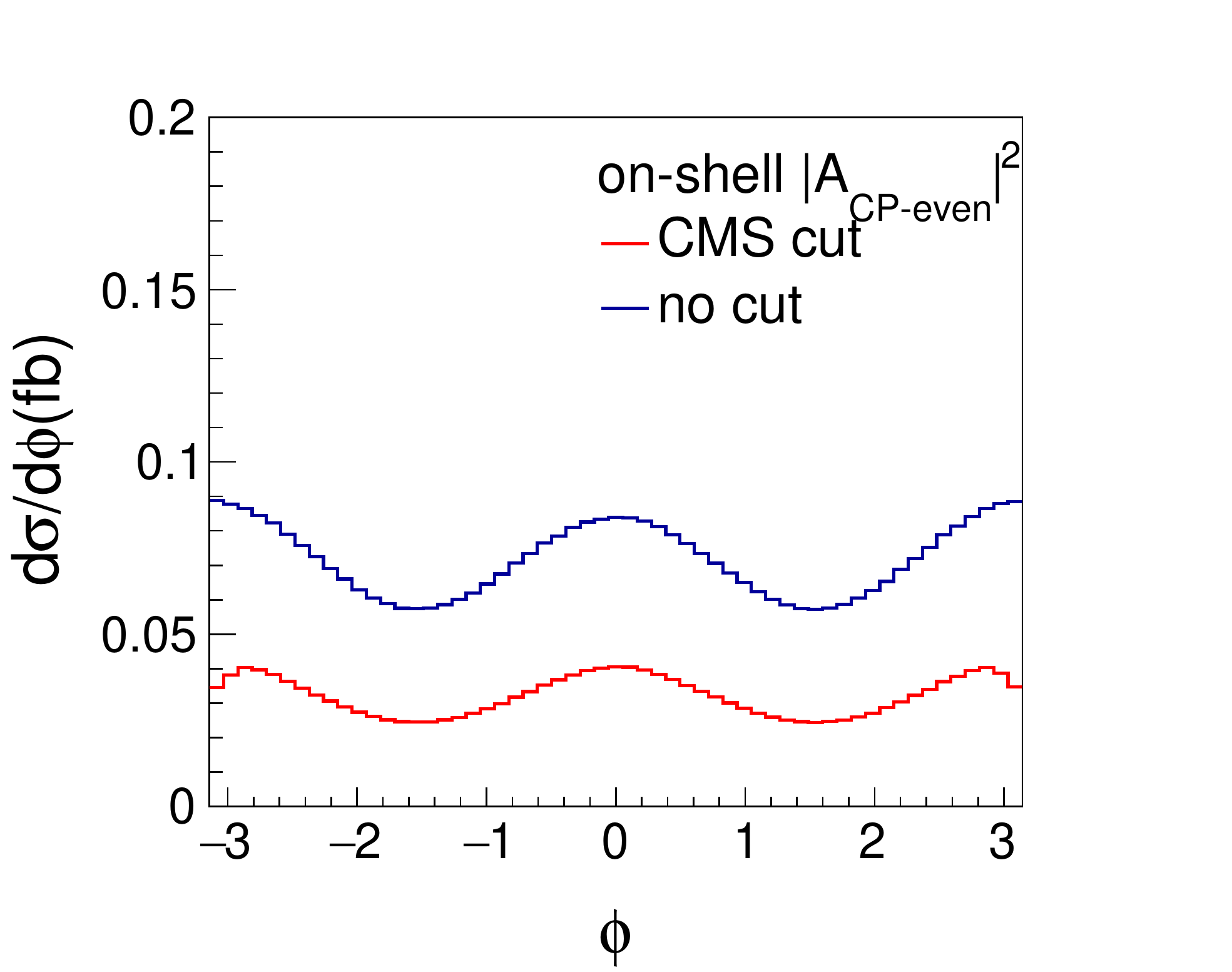}
		\includegraphics[width=3.6cm]{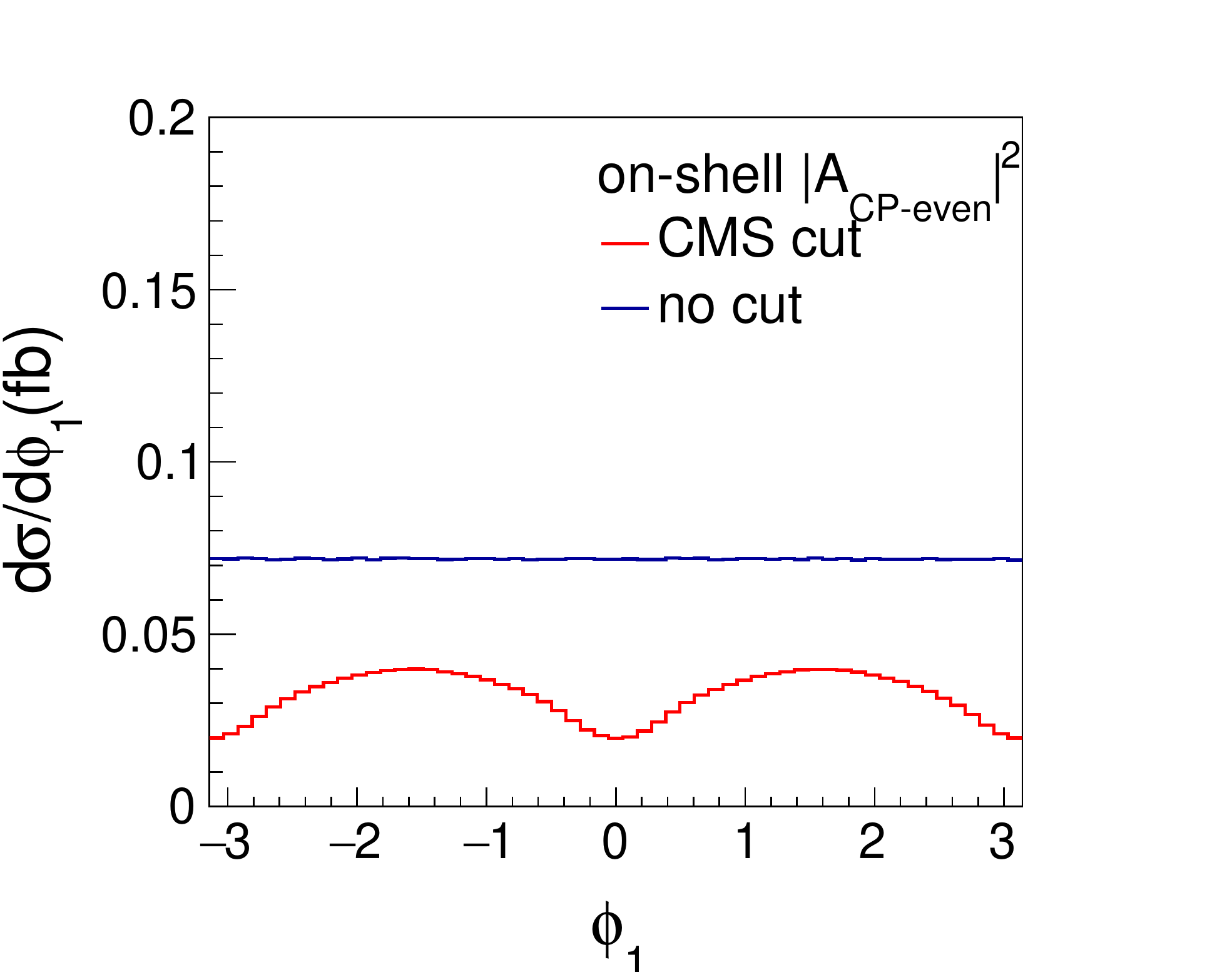}
		\includegraphics[width=3.6cm]{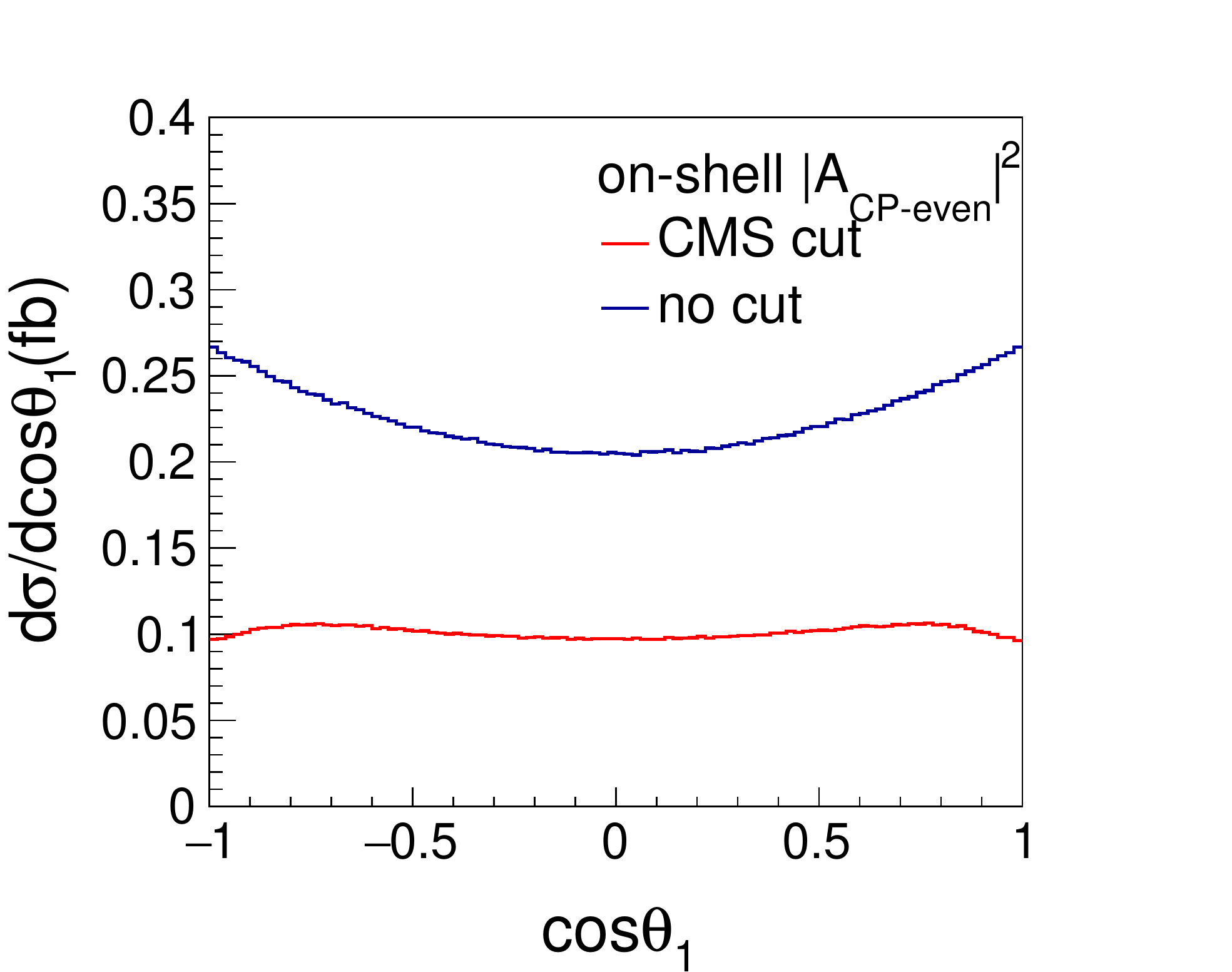}
		\includegraphics[width=3.6cm]{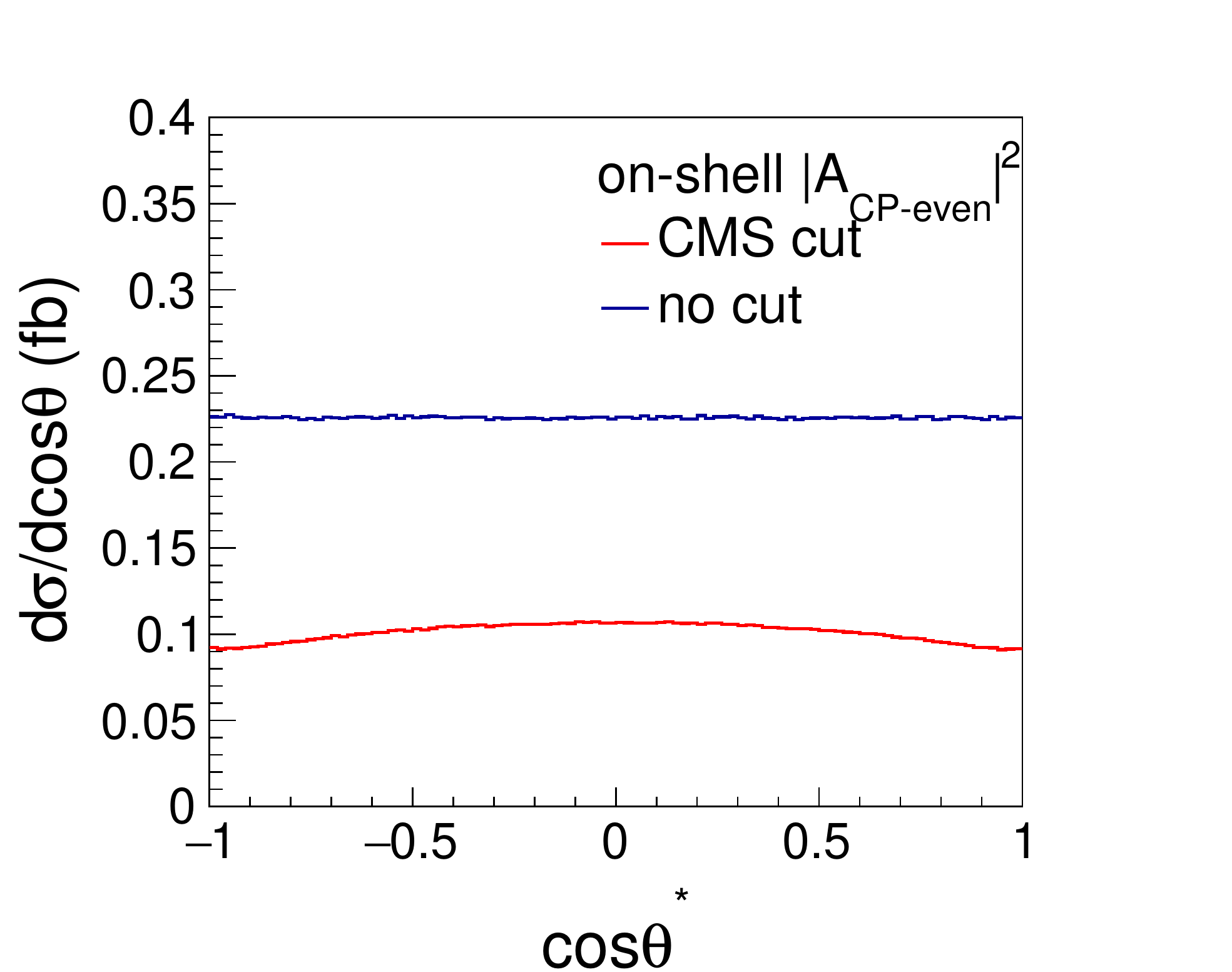}
		\\
		\includegraphics[width=3.6cm]{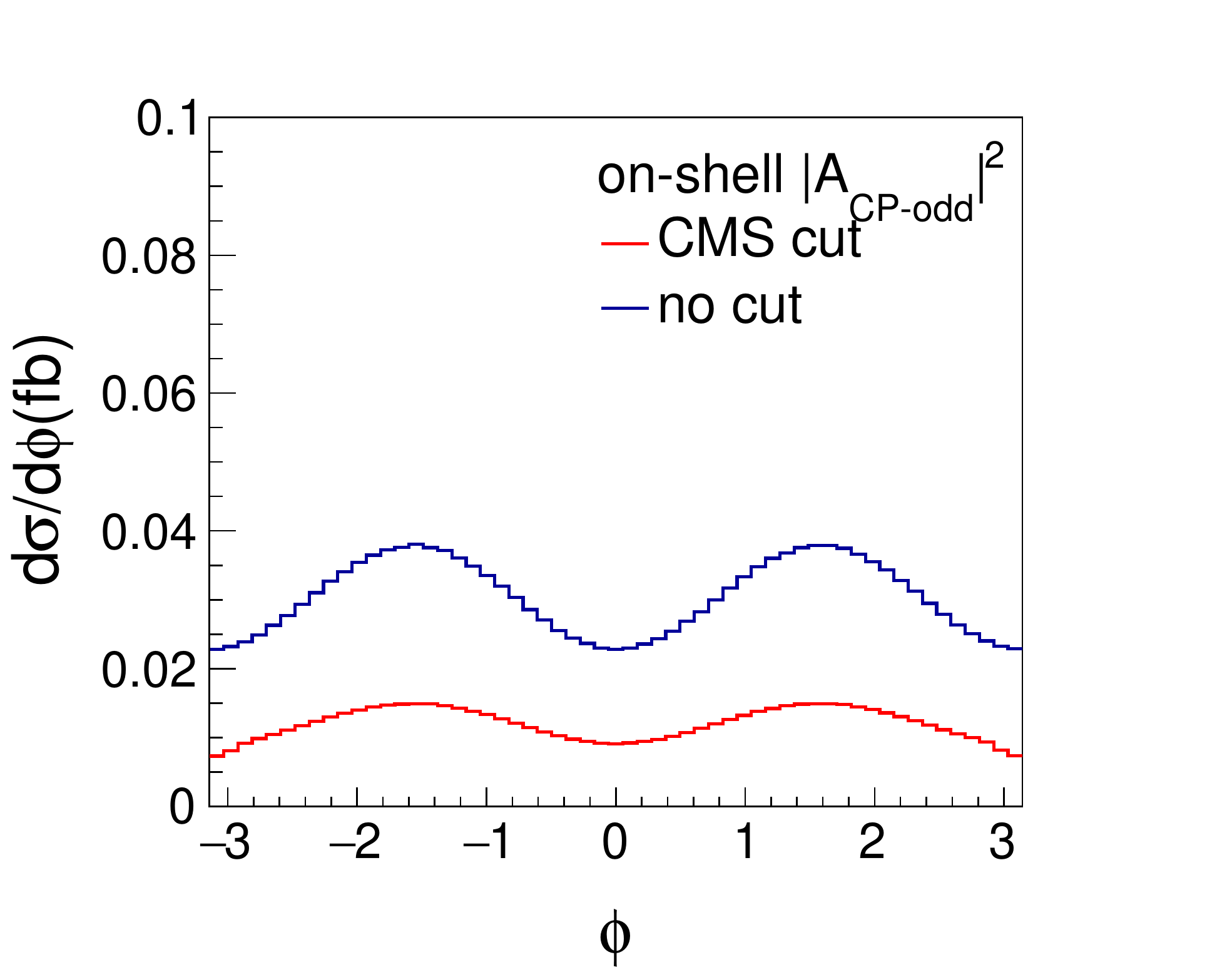}
		\includegraphics[width=3.6cm]{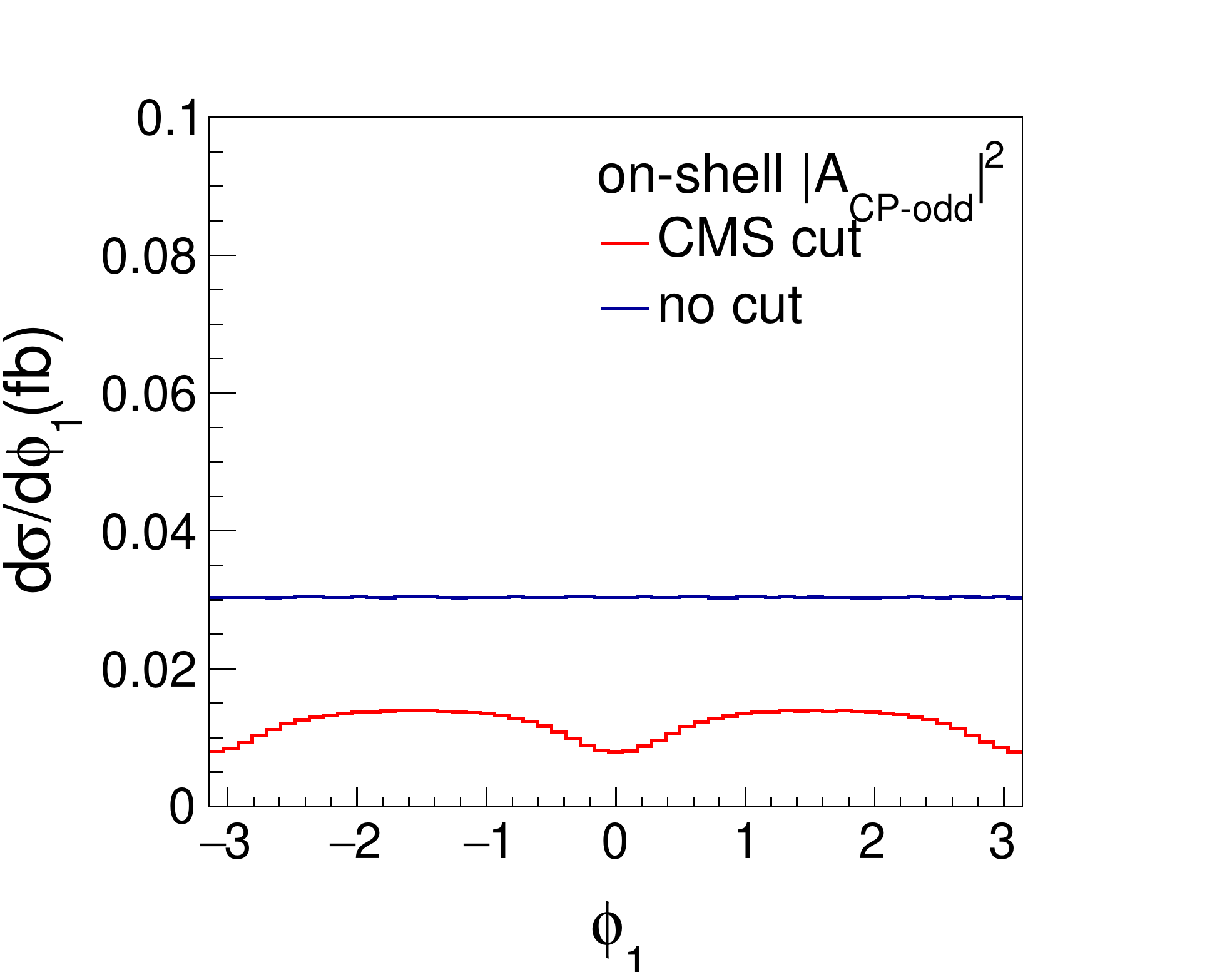}
		\includegraphics[width=3.6cm]{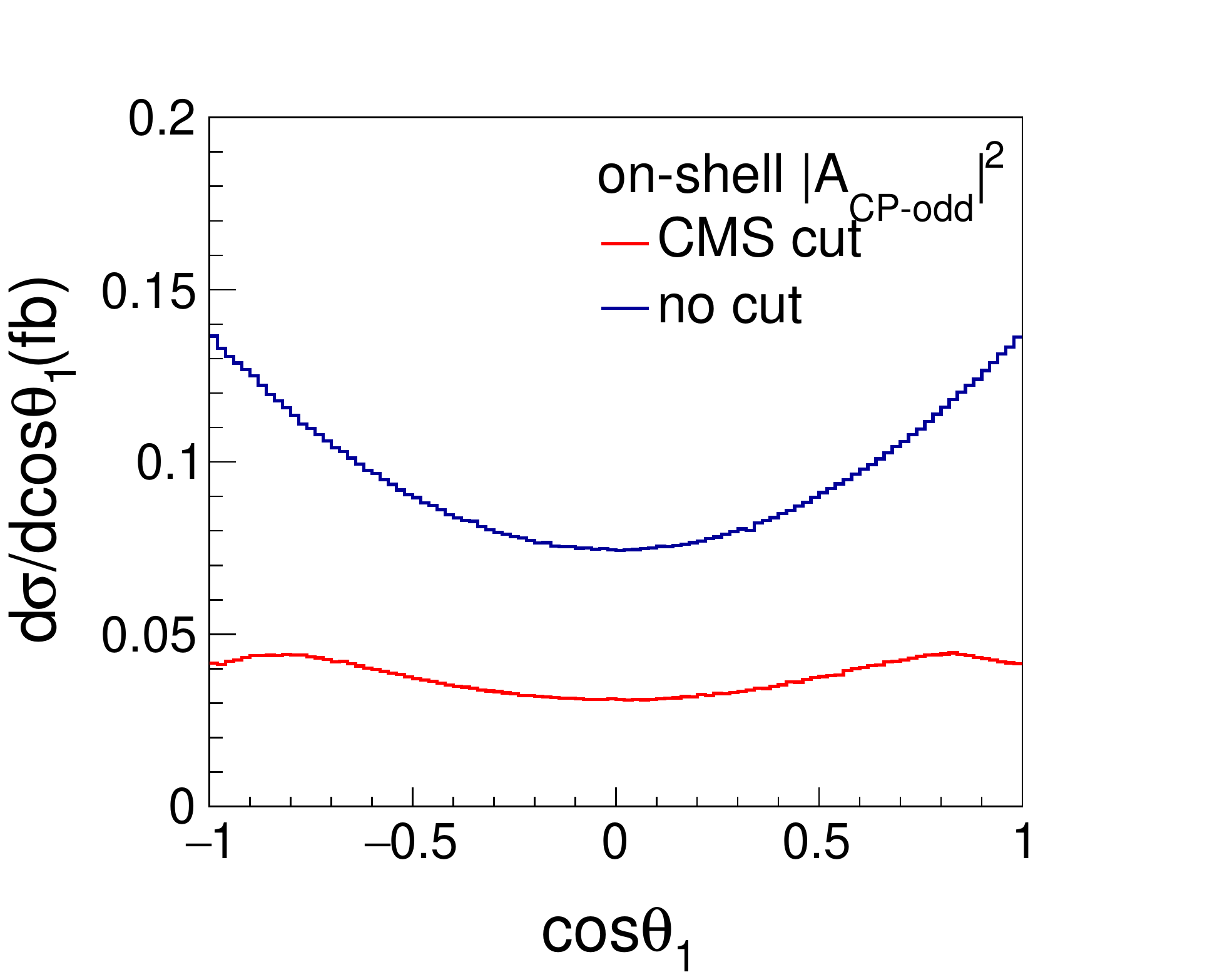}
		\includegraphics[width=3.6cm]{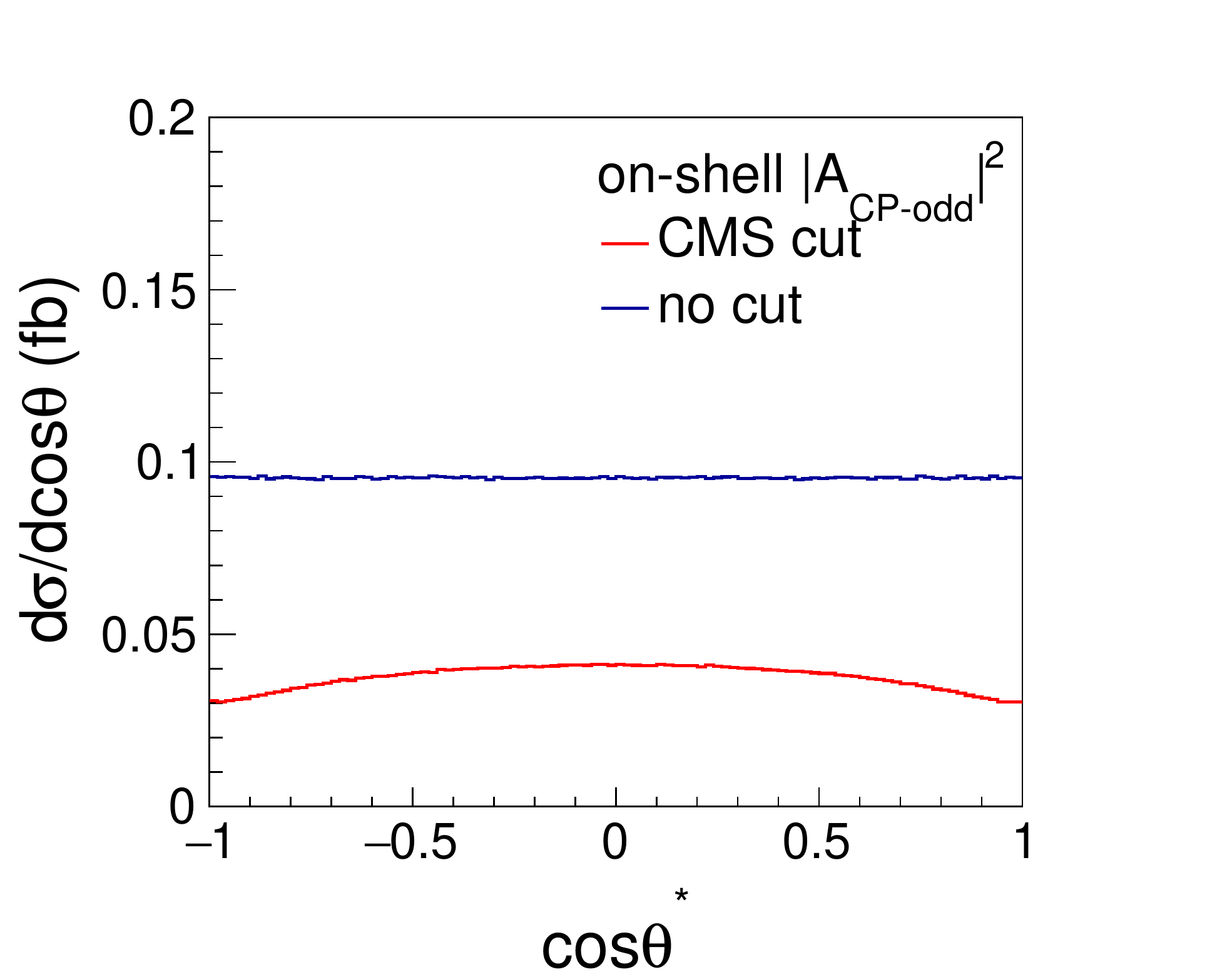}
		\caption{The angular differential cross sections from self-conjugate of Higgs mediated processes in Higgs on-shell region. }
		\label{self-Higgs-0n}
	\end{figure}

	\begin{figure}[htbp]
		\centering
		\includegraphics[width=3.6cm]{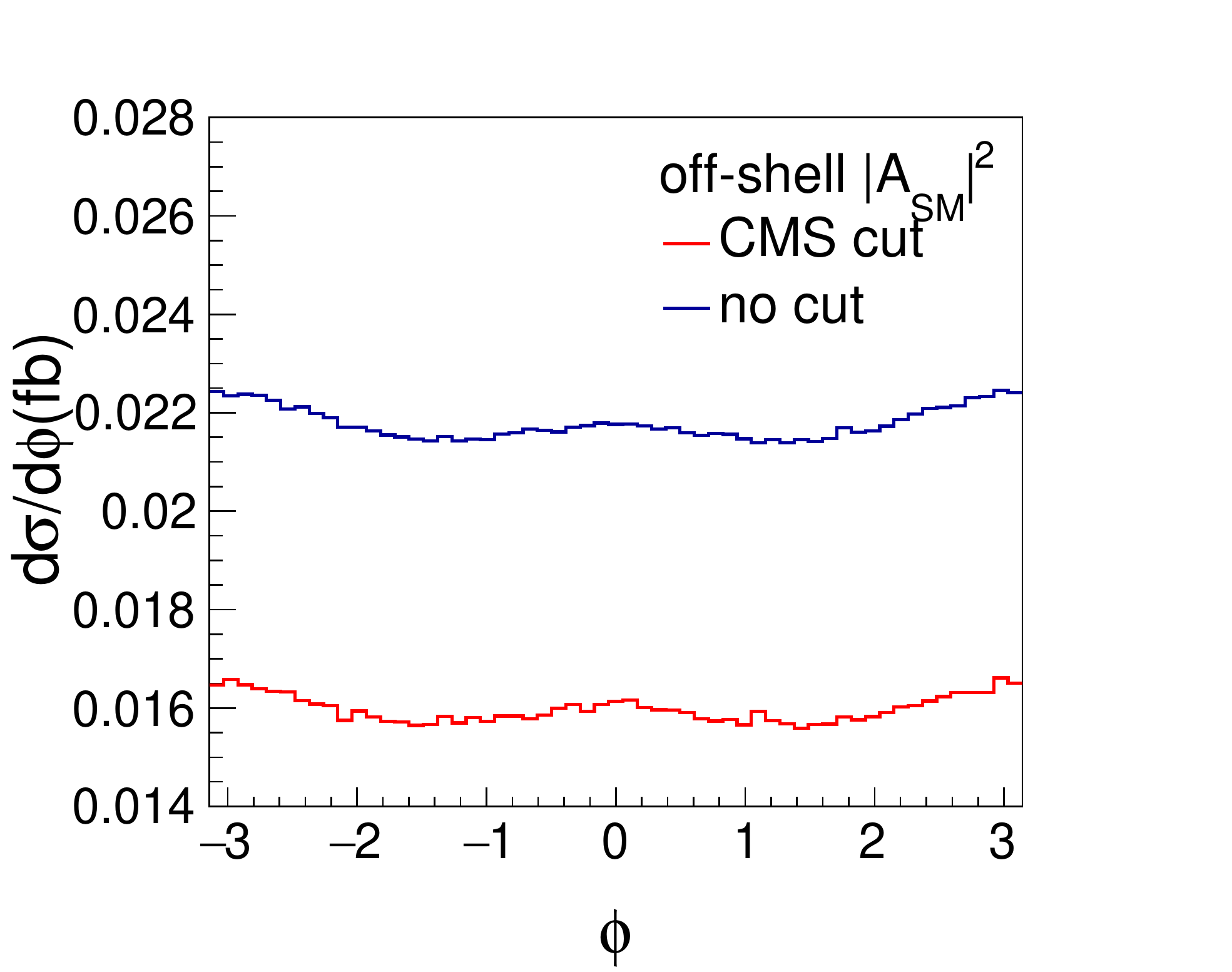}
		\includegraphics[width=3.6cm]{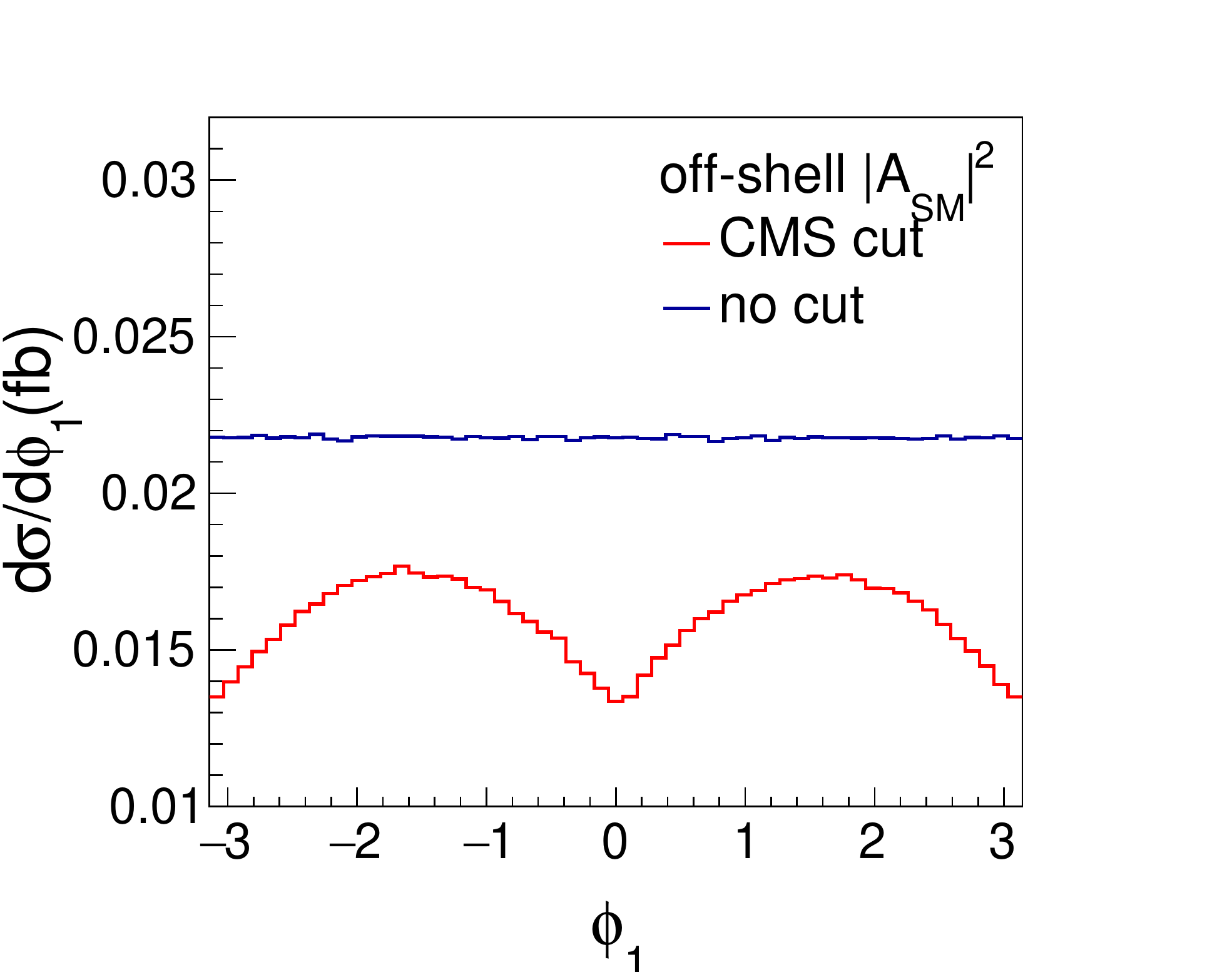}
		\includegraphics[width=3.6cm]{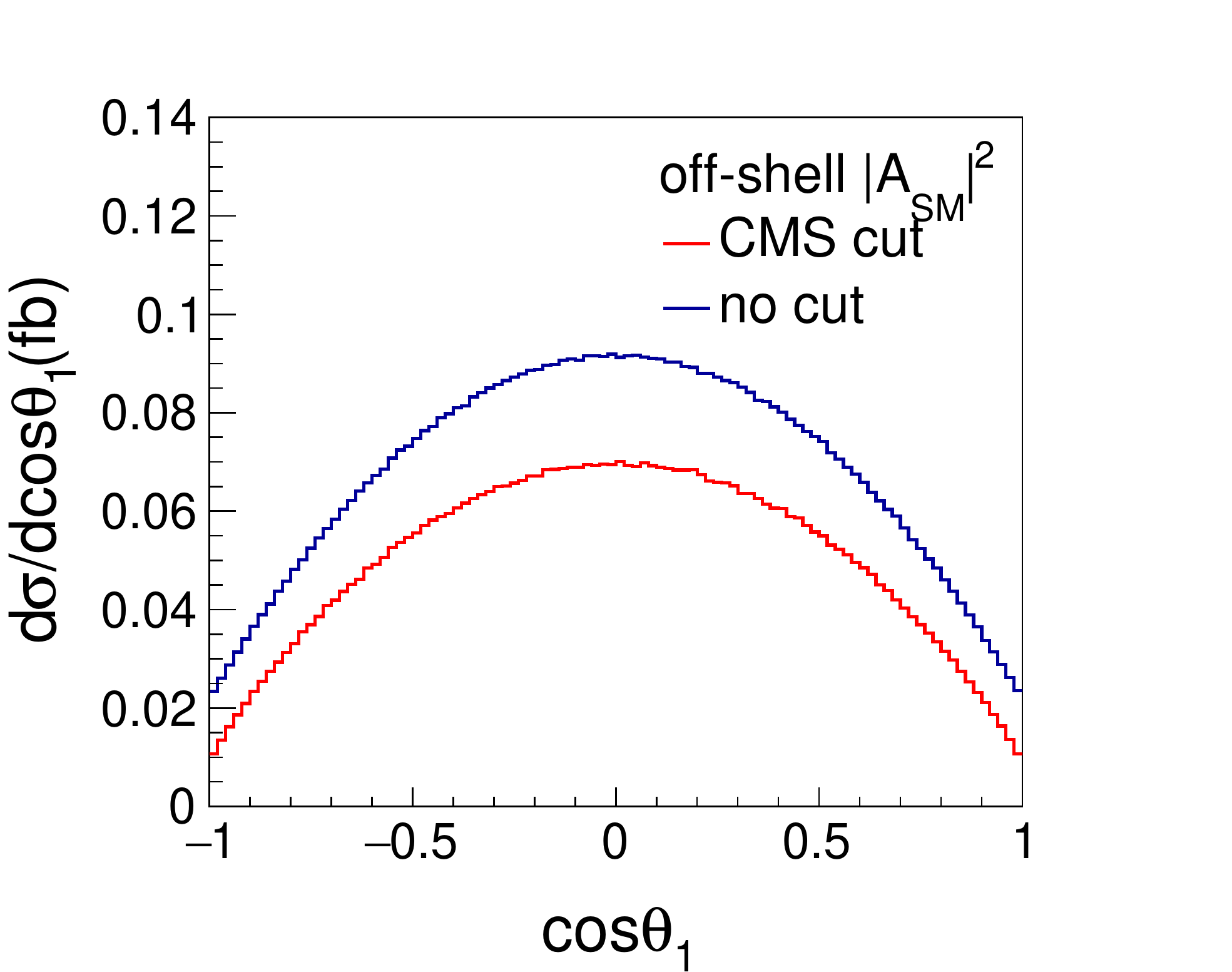}
		\includegraphics[width=3.6cm]{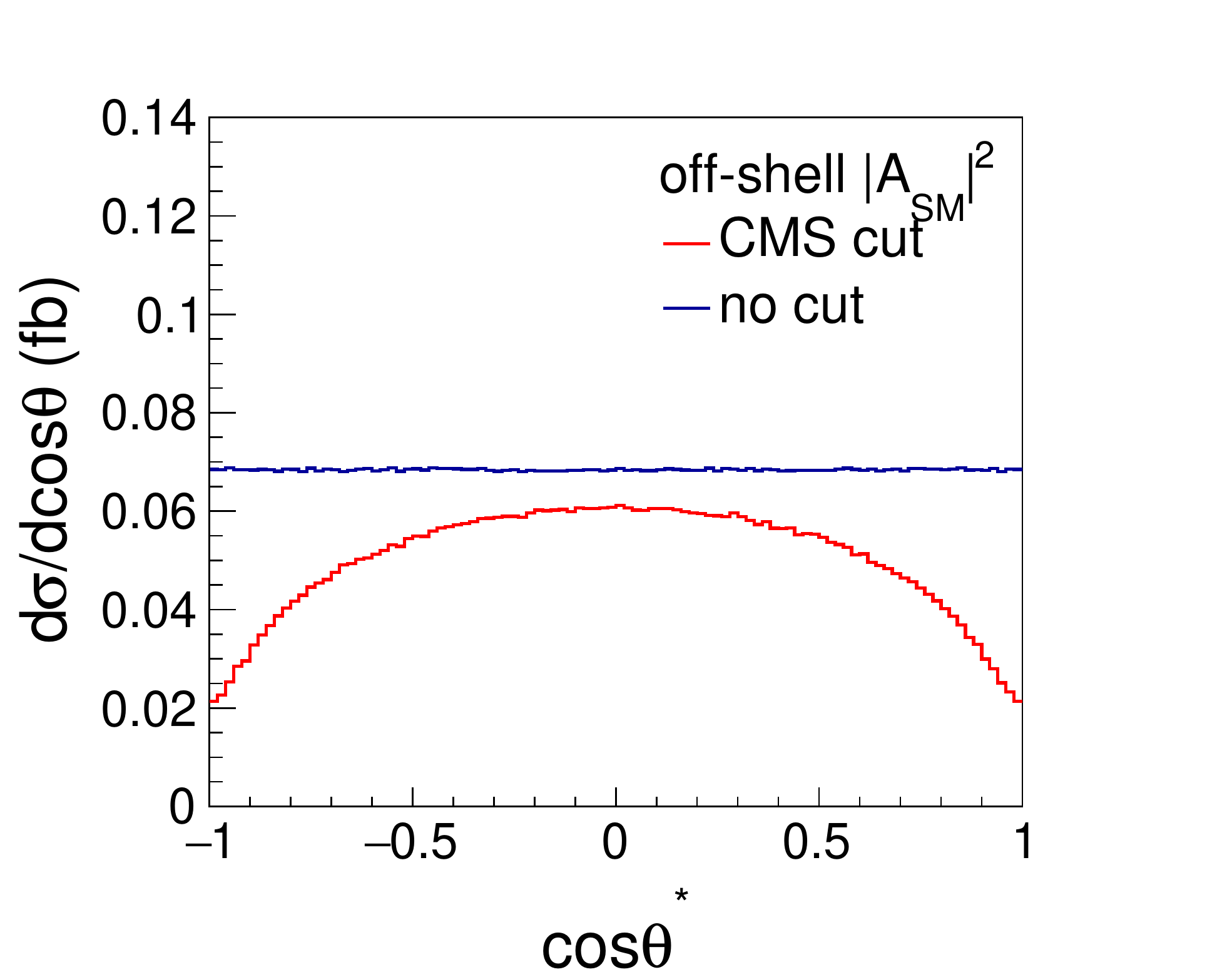}
		\\
		\includegraphics[width=3.6cm]{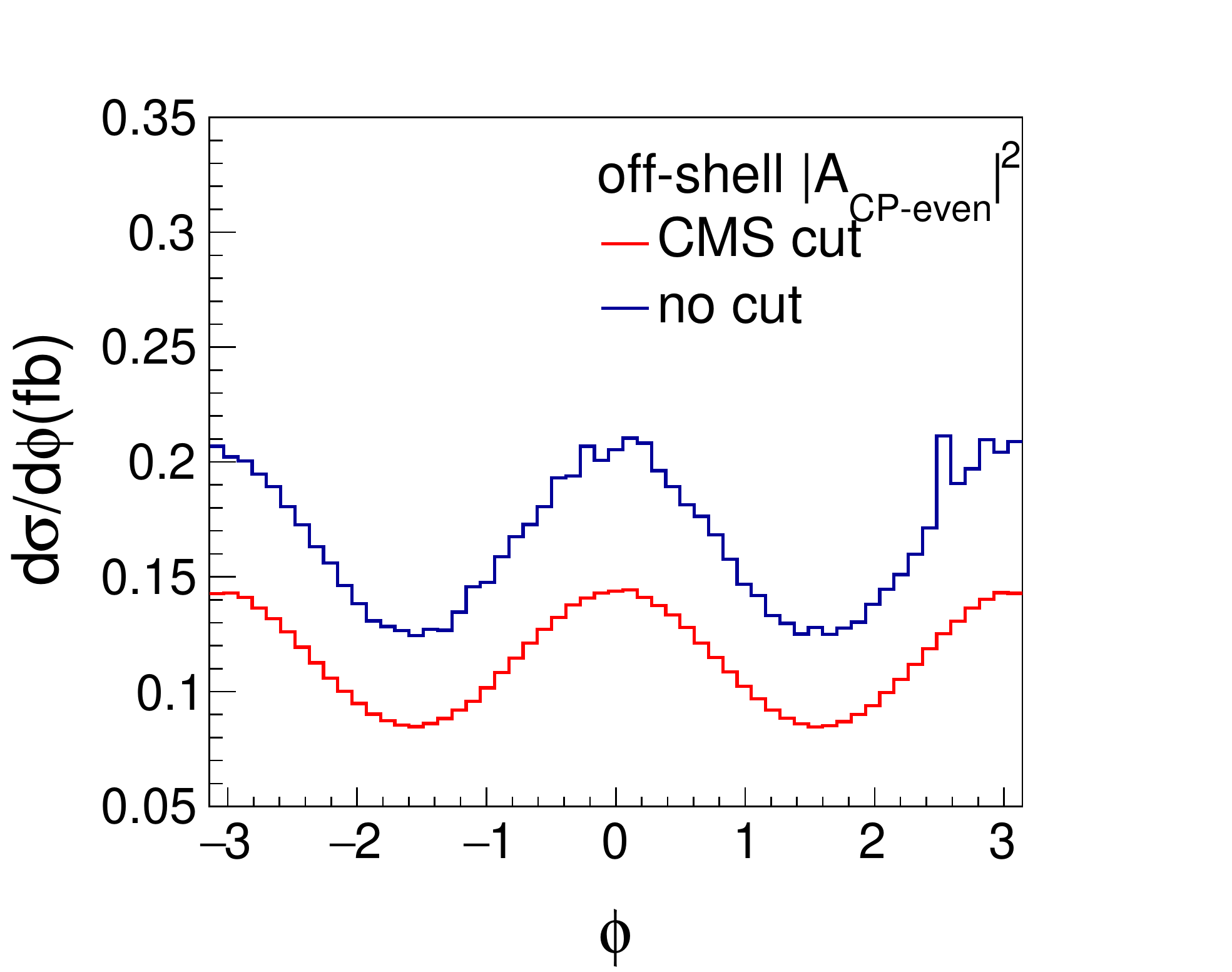}
		\includegraphics[width=3.6cm]{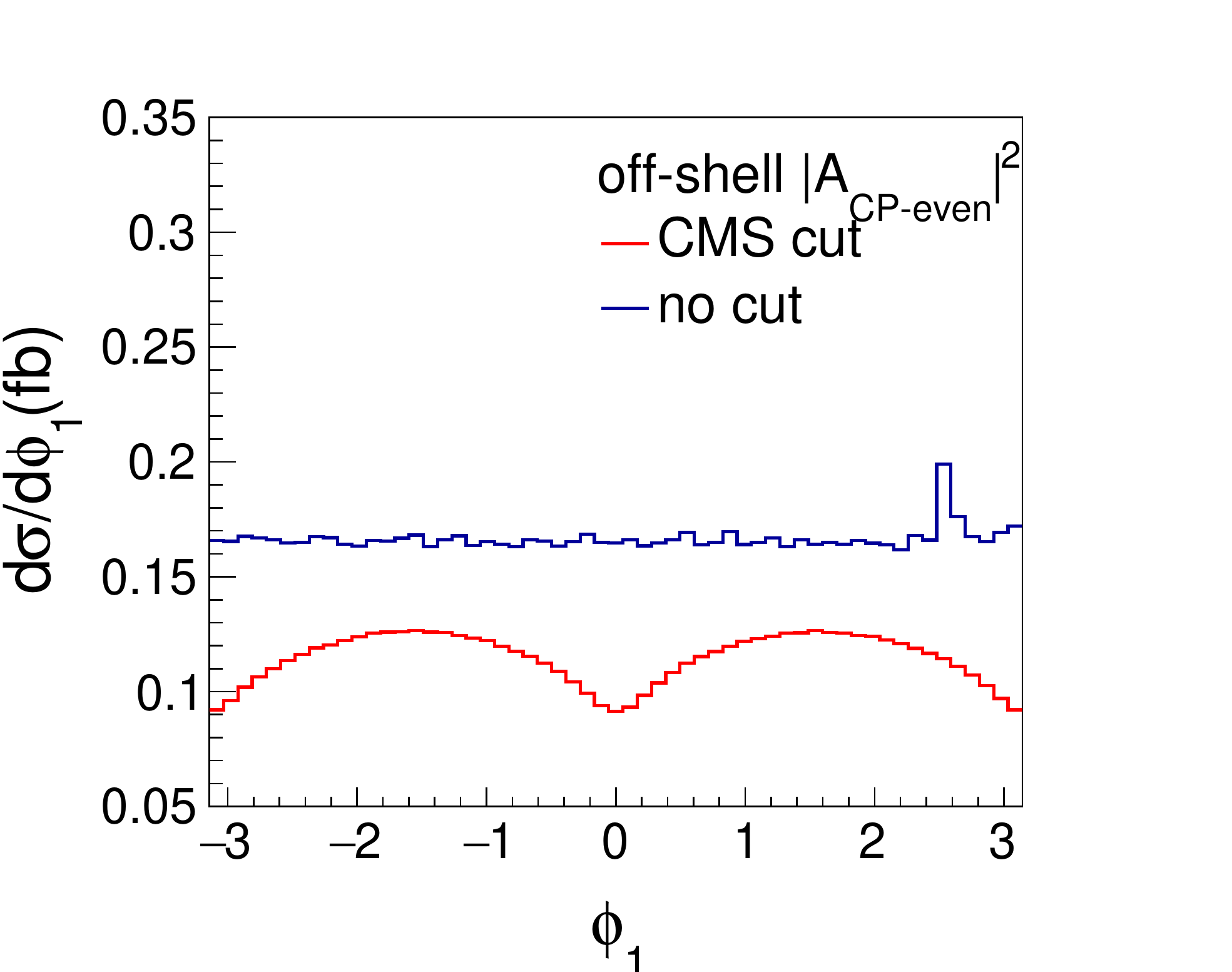}
		\includegraphics[width=3.6cm]{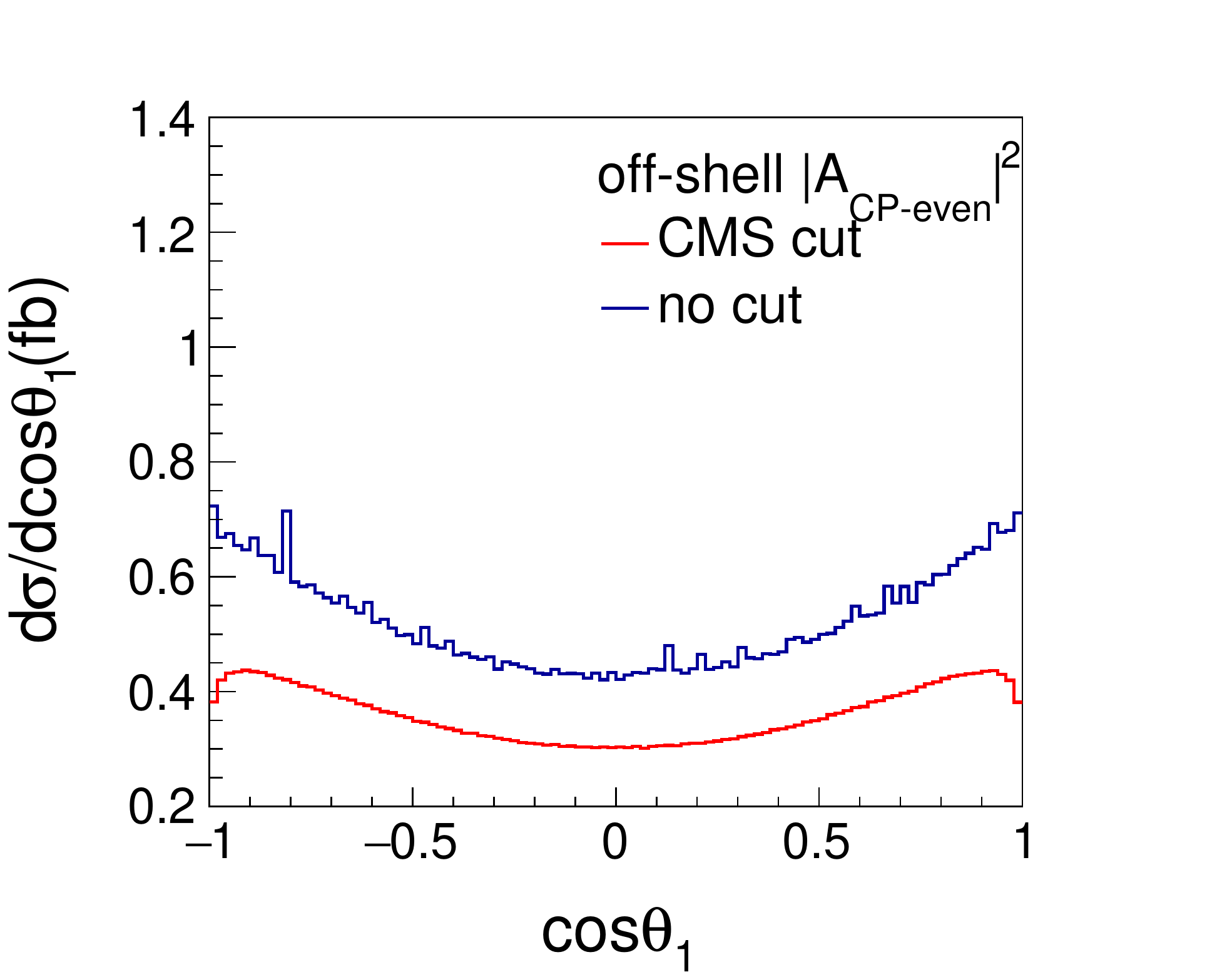}
		\includegraphics[width=3.6cm]{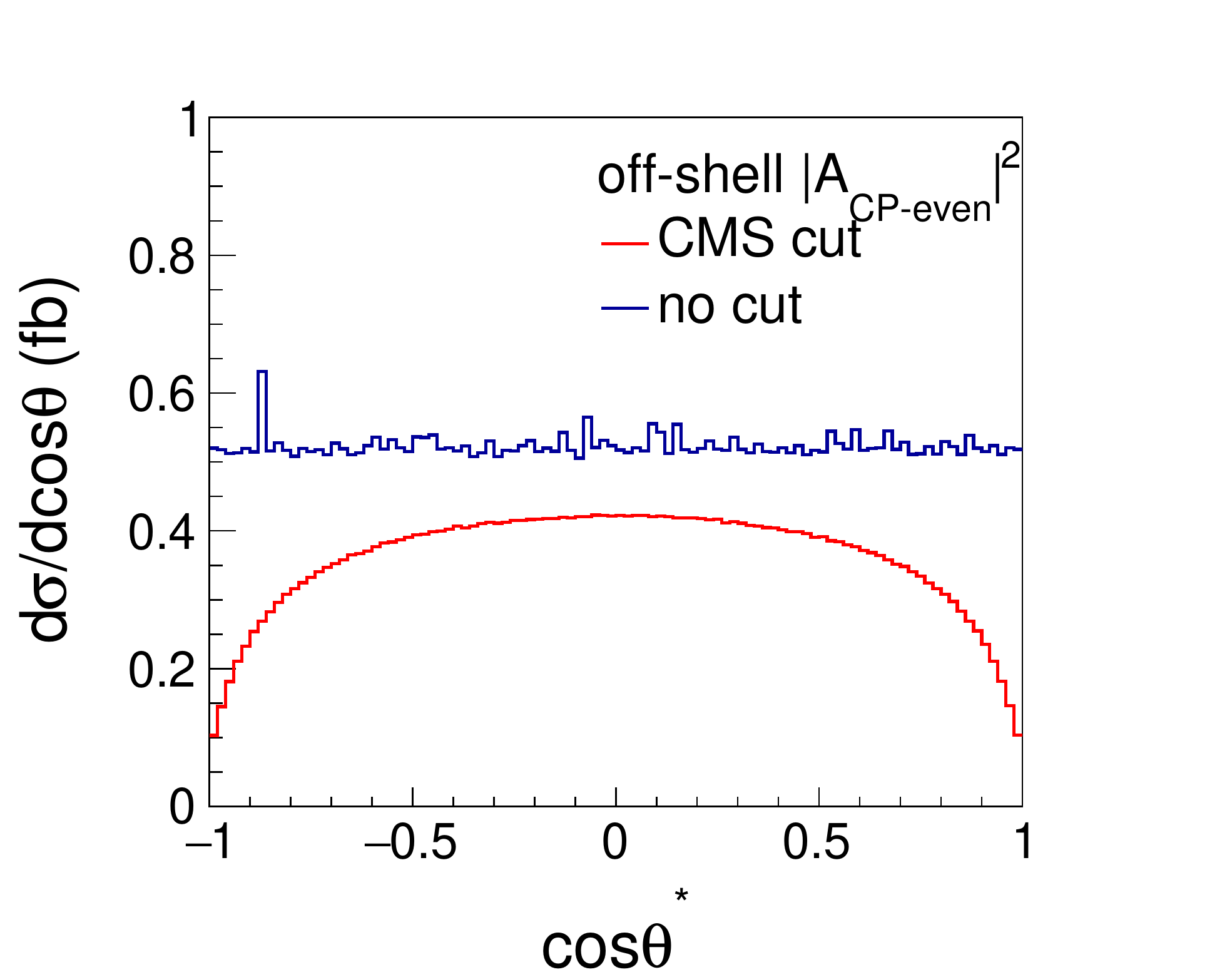}
		\\
		\includegraphics[width=3.6cm]{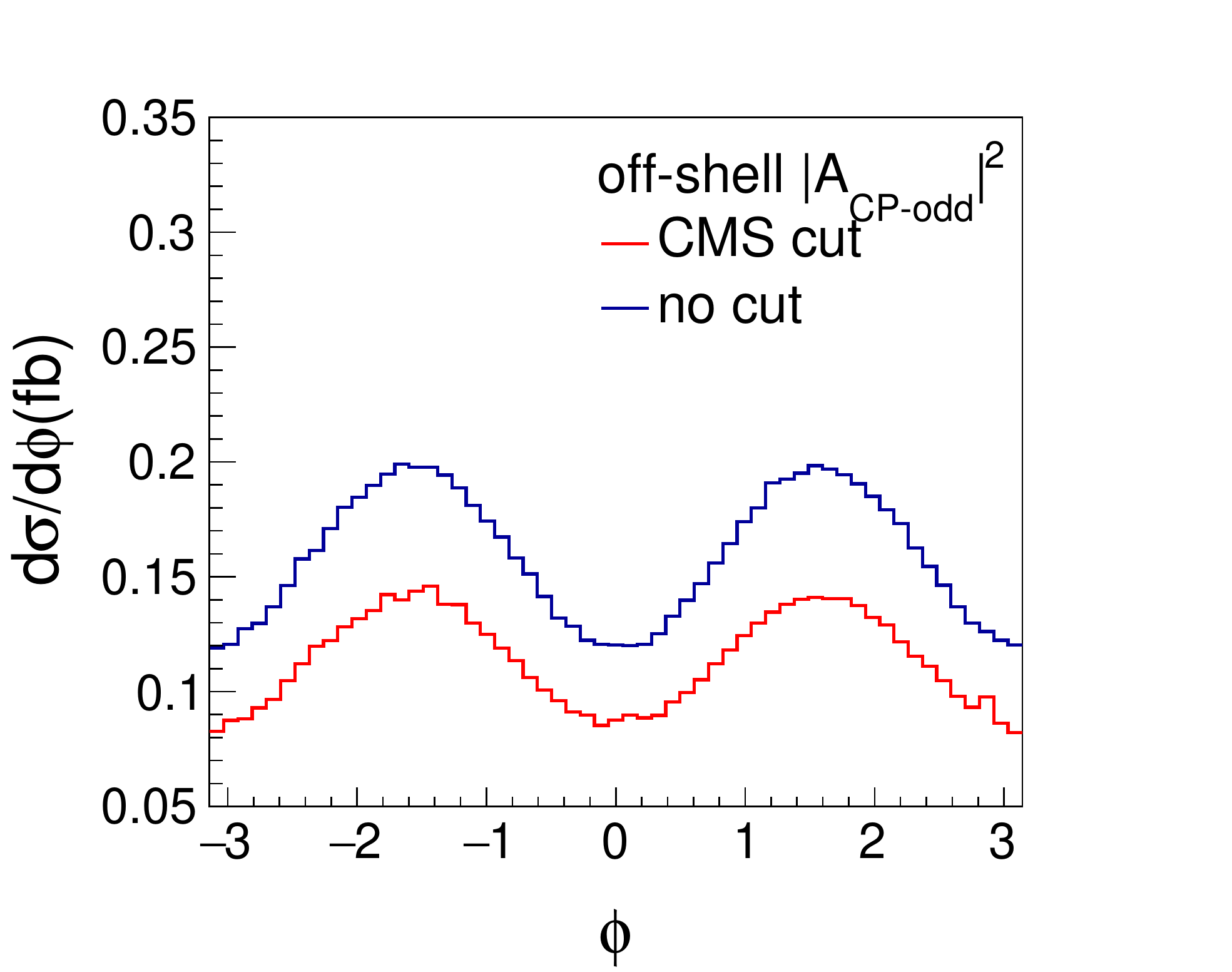}
		\includegraphics[width=3.6cm]{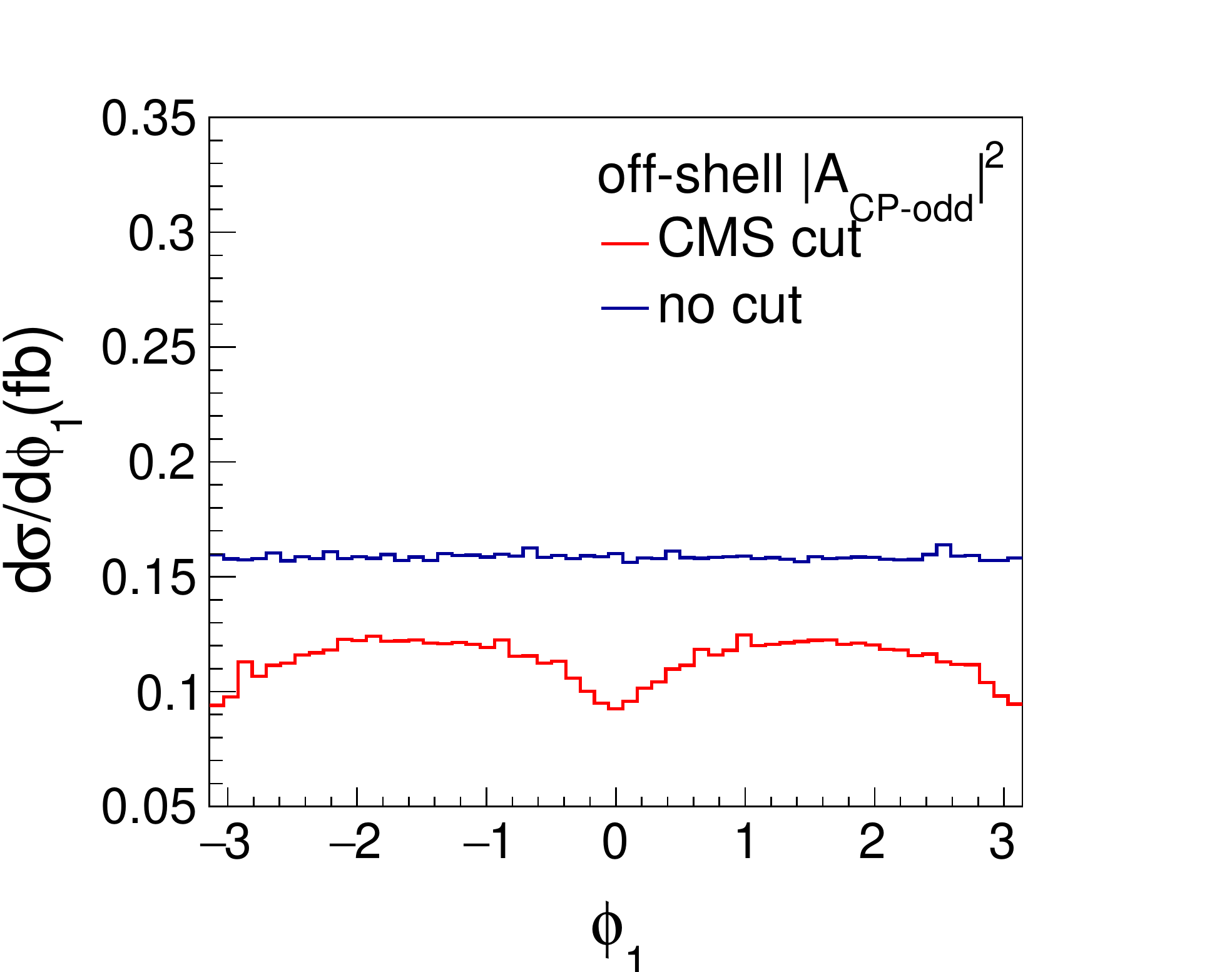}
		\includegraphics[width=3.6cm]{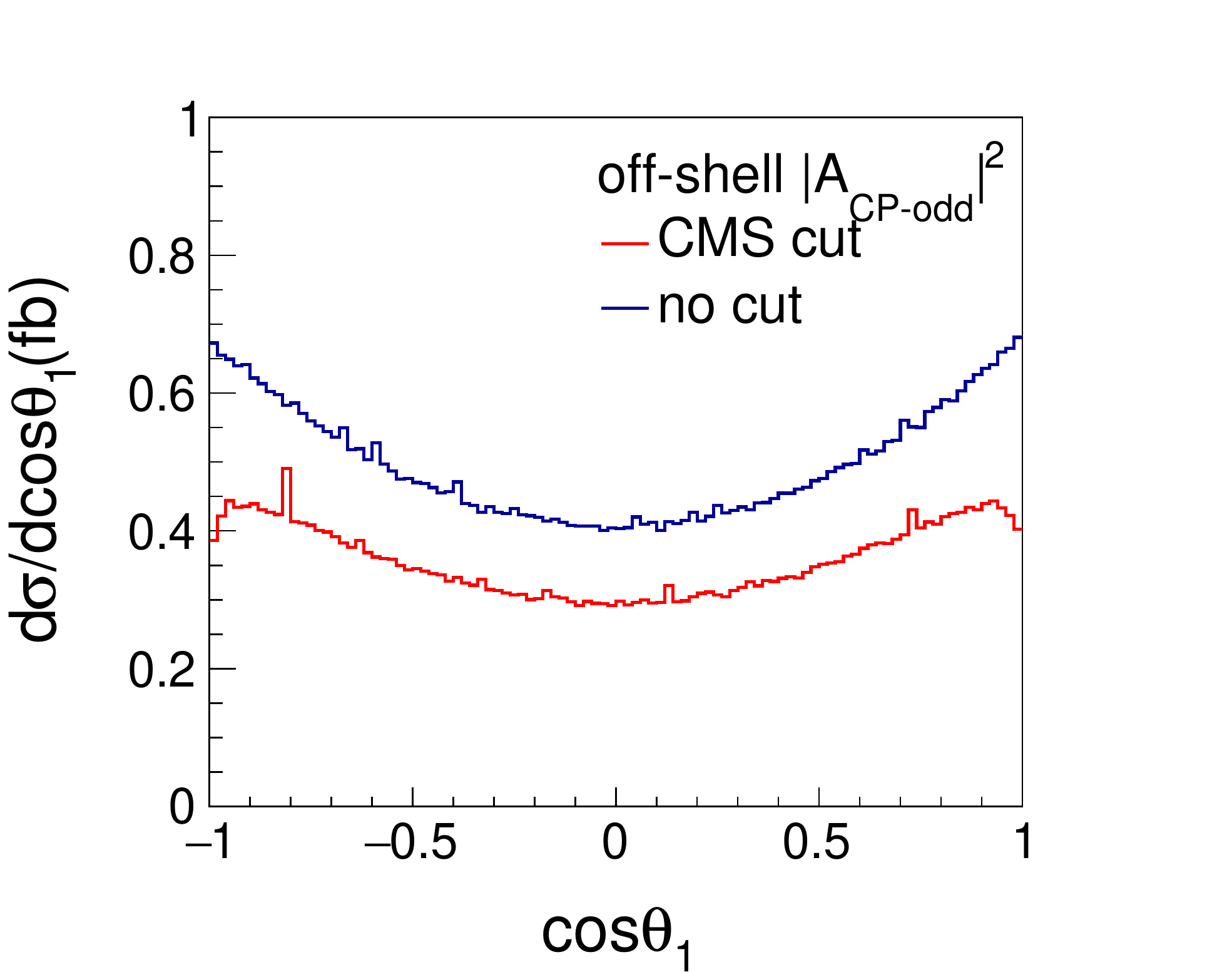}
		\includegraphics[width=3.6cm]{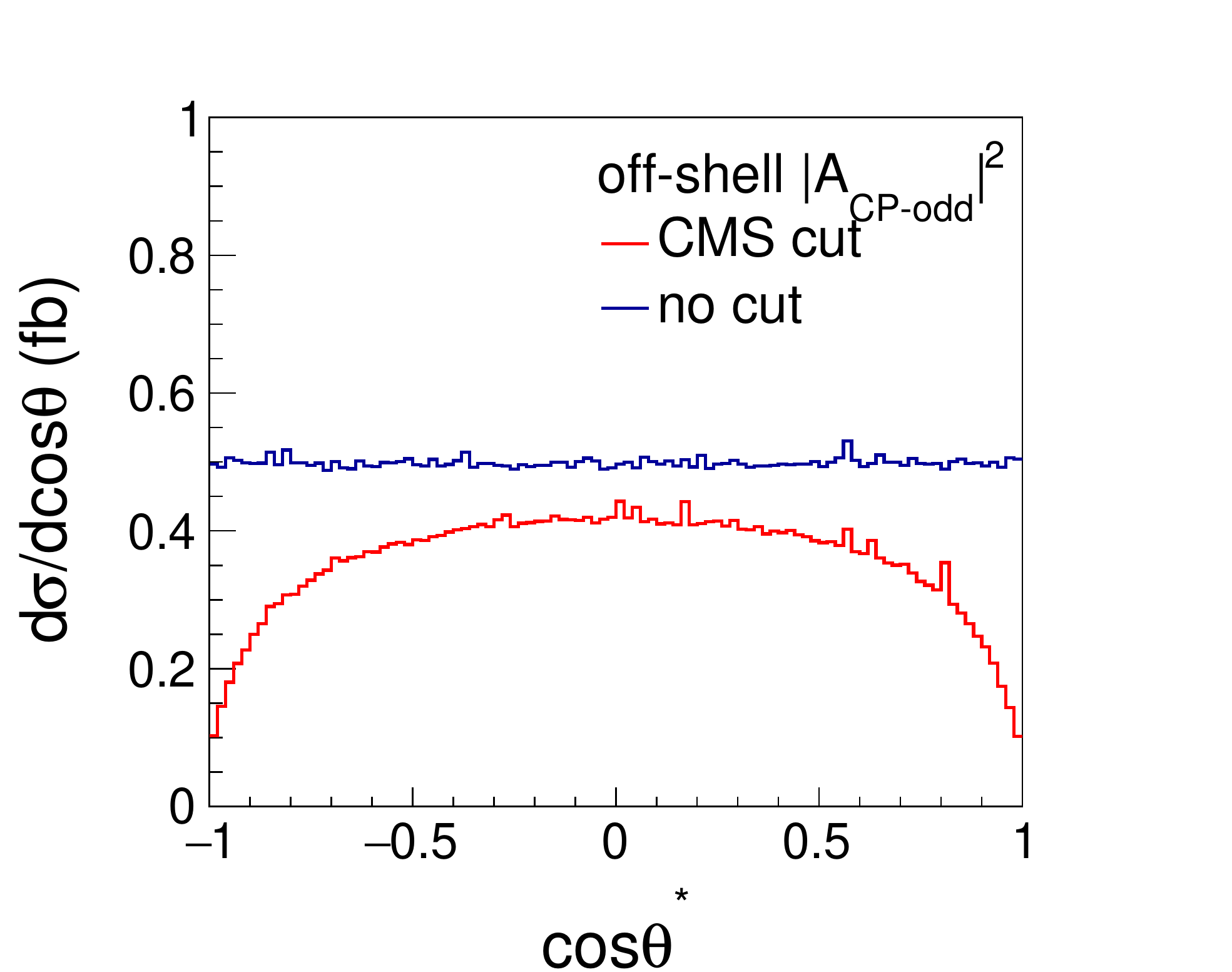}
		
		\caption{The angular differential cross sections from self-conjugate of Higgs mediated processes in Higgs off-shell region. }
		\label{self-Higgs-off}
	\end{figure}
	
	\begin{figure}[htbp]
		\centering
		\includegraphics[width=3.6cm]{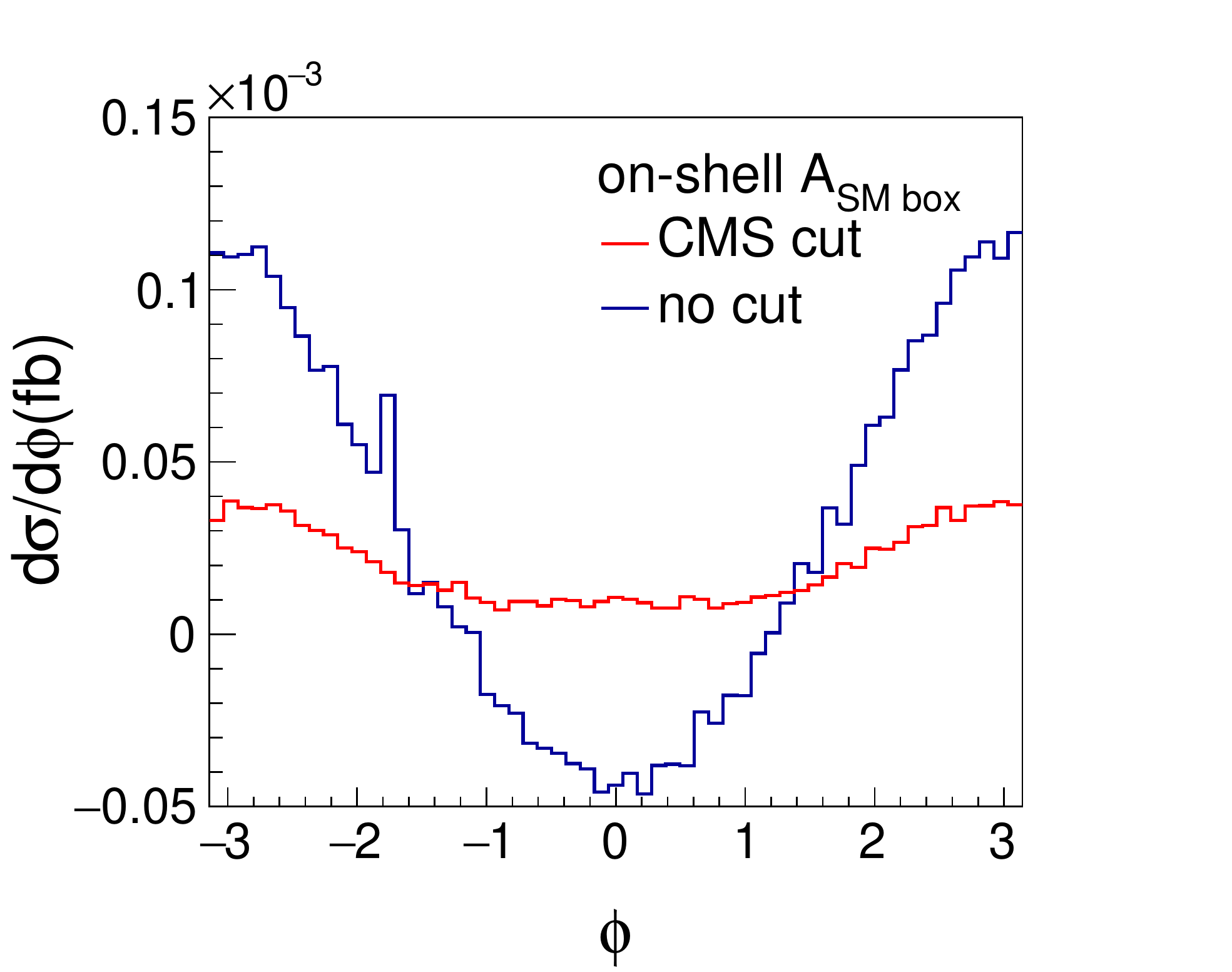}
		\includegraphics[width=3.6cm]{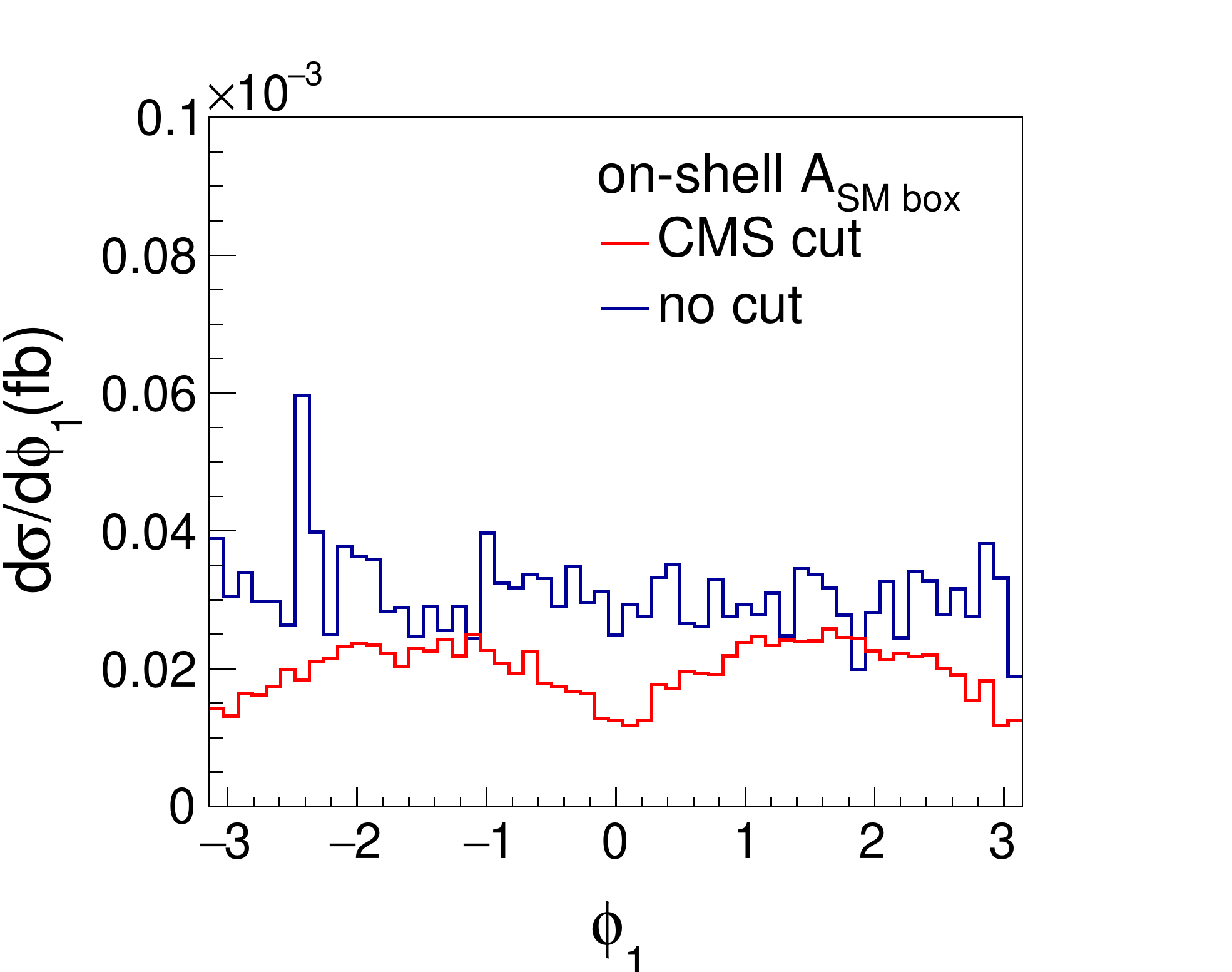}
		\includegraphics[width=3.6cm]{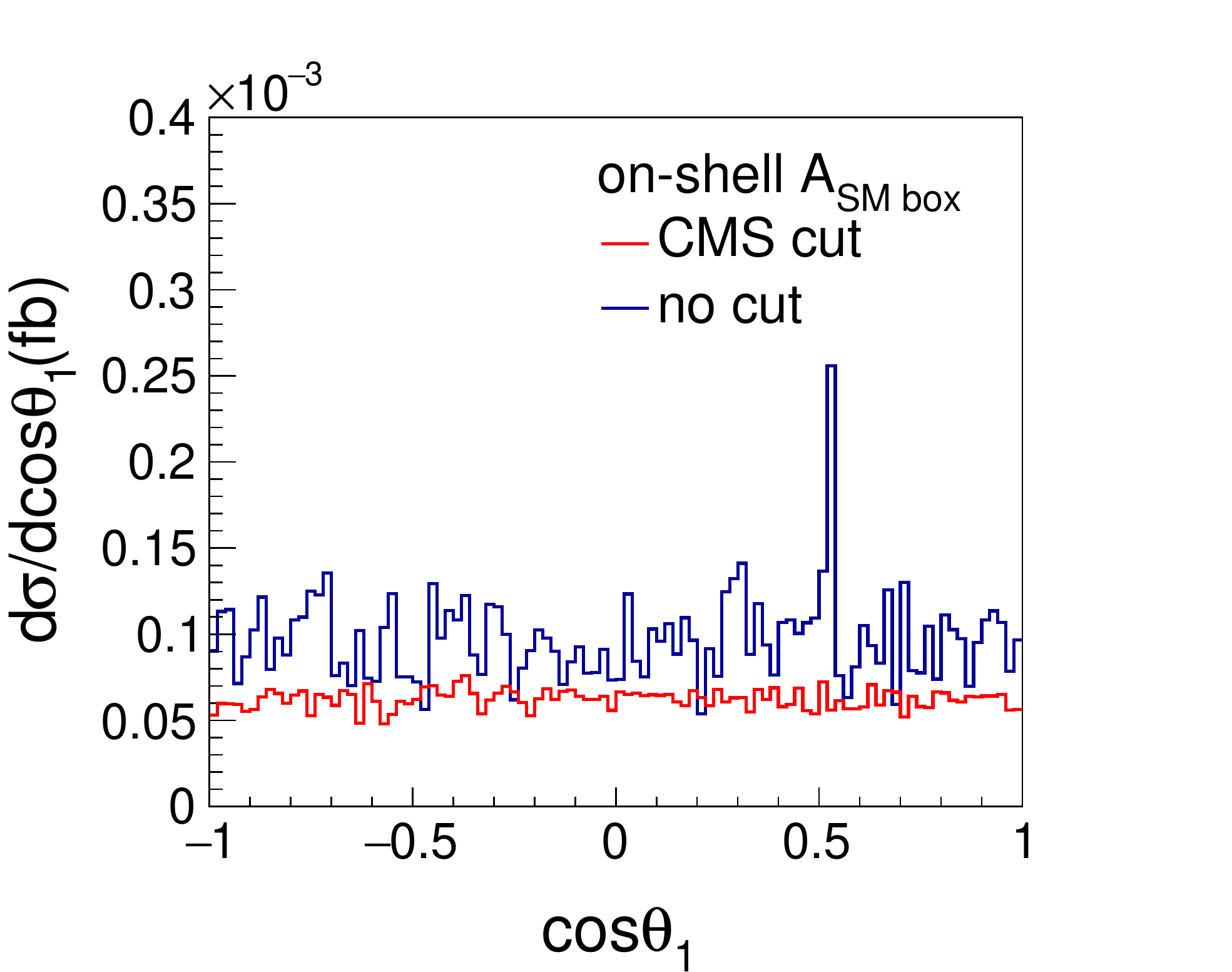}
		\includegraphics[width=3.6cm]{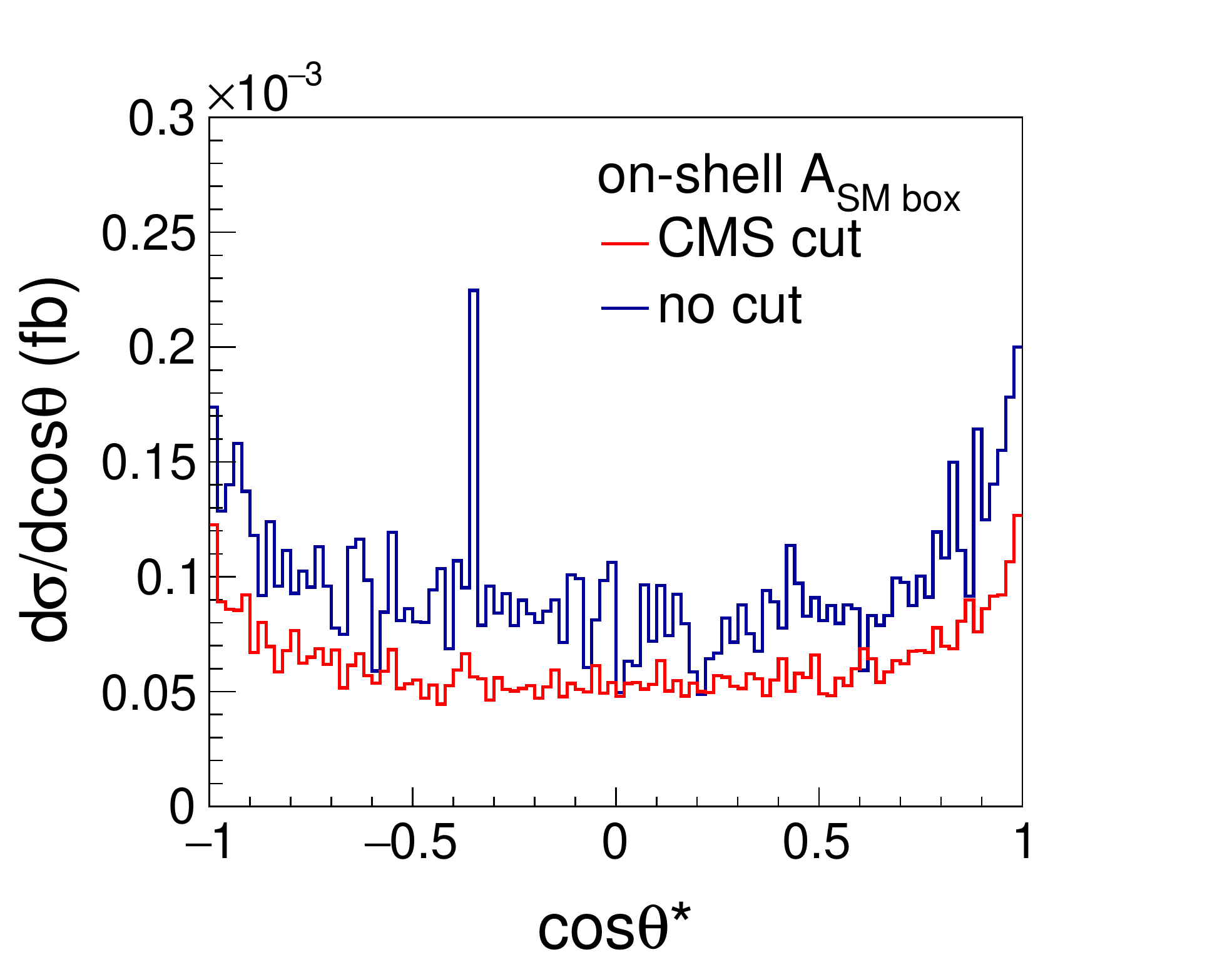}
		\\
		\includegraphics[width=3.6cm]{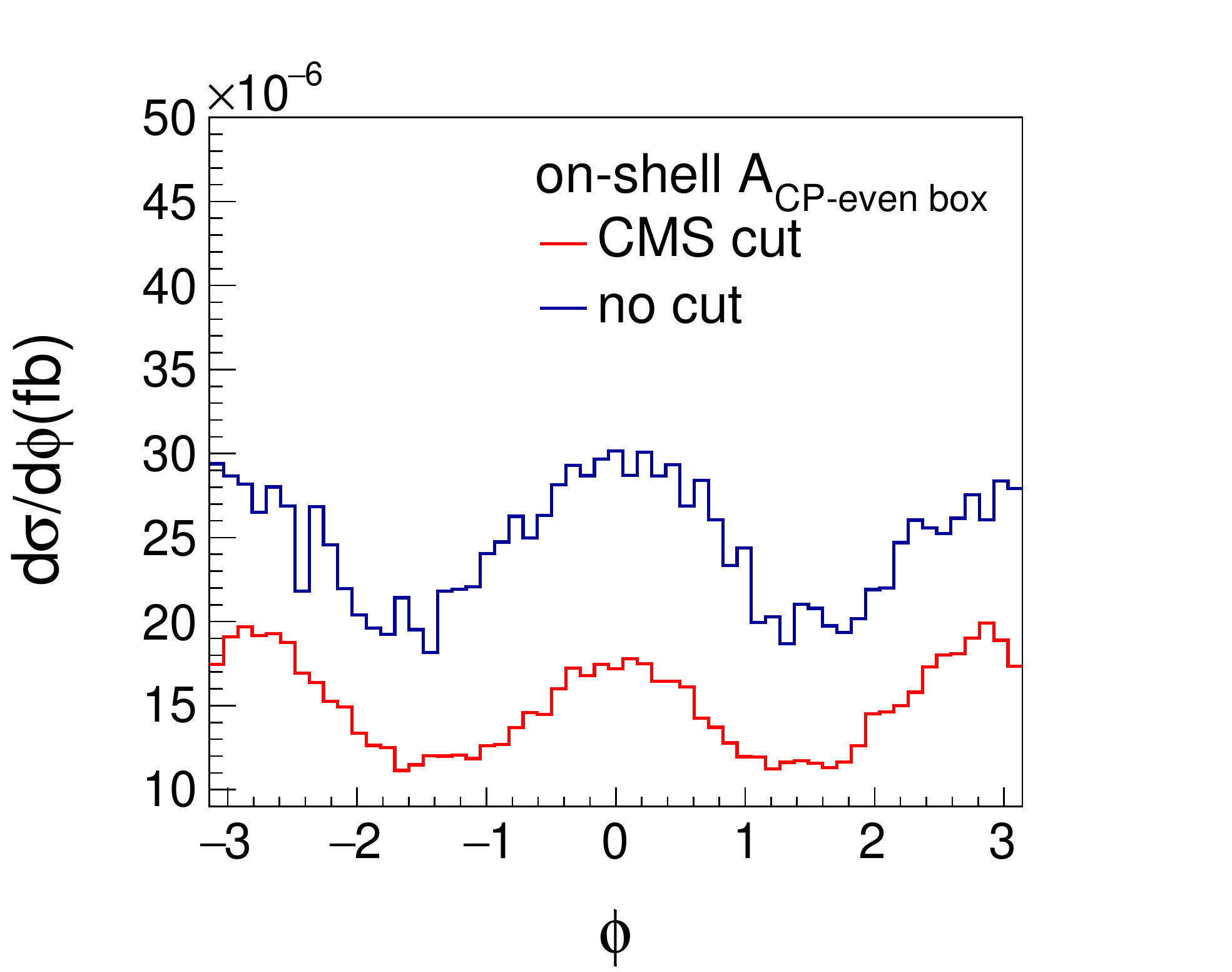}
		\includegraphics[width=3.6cm]{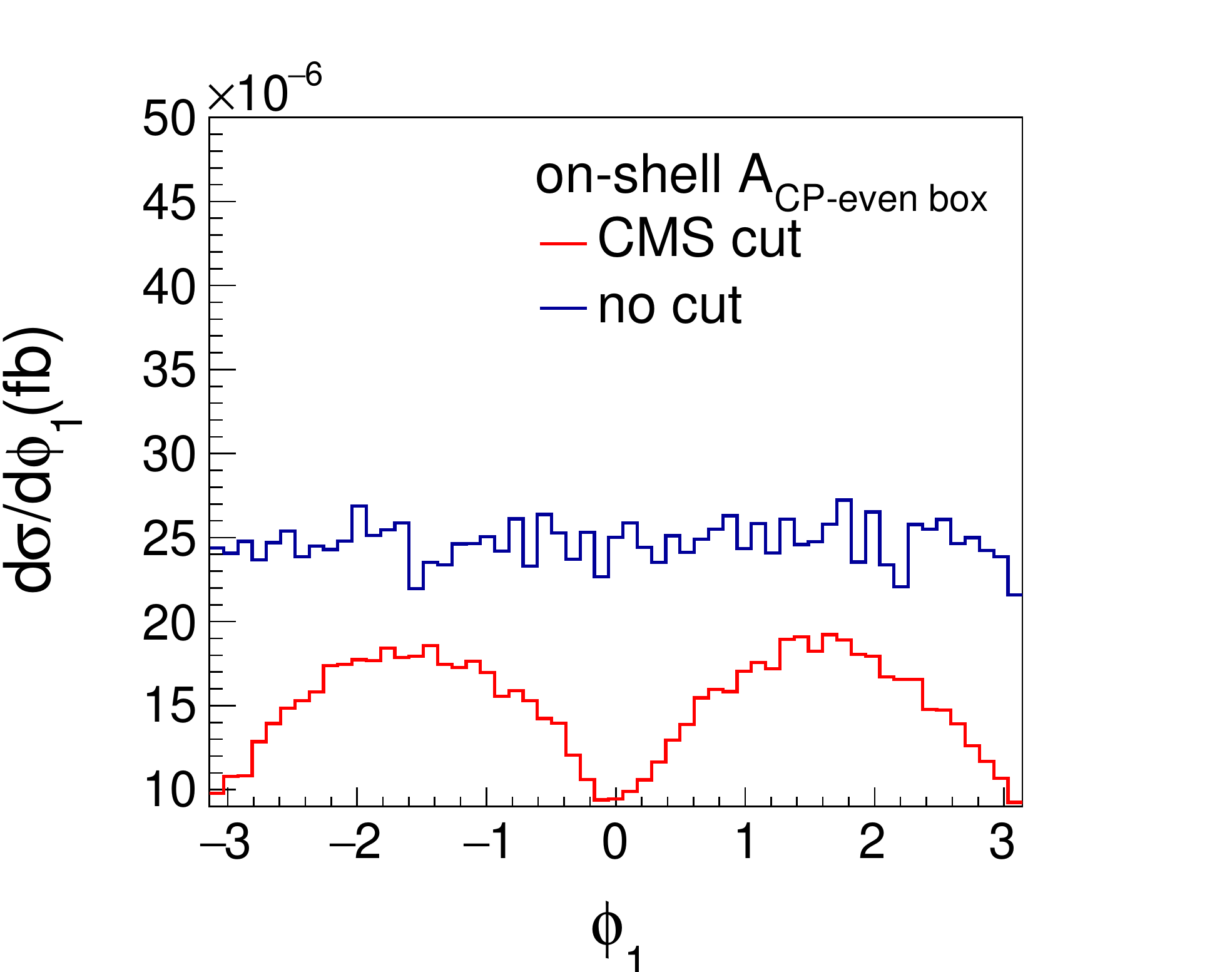}
		\includegraphics[width=3.6cm]{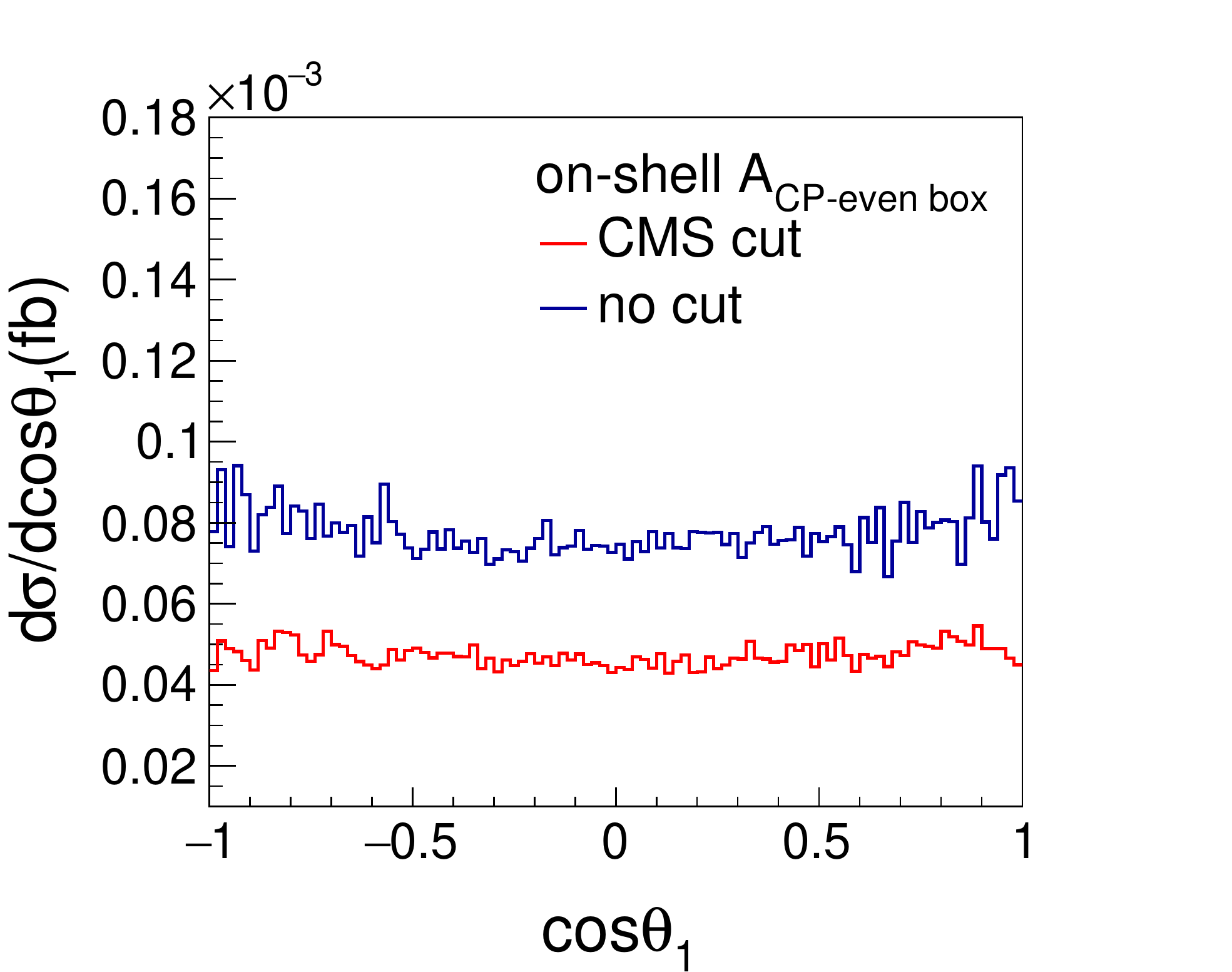}
		\includegraphics[width=3.6cm]{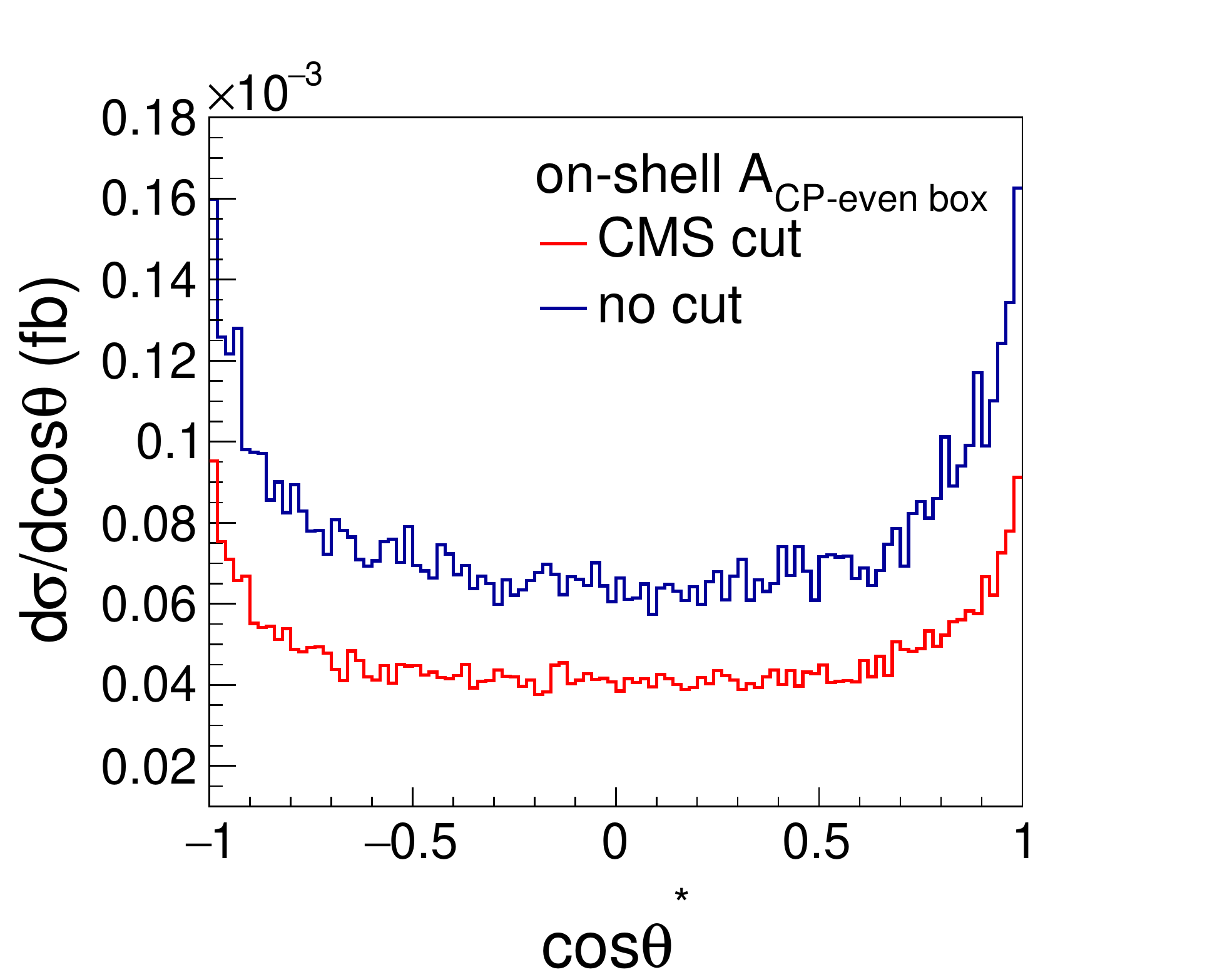}
		\\
		\includegraphics[width=3.6cm]{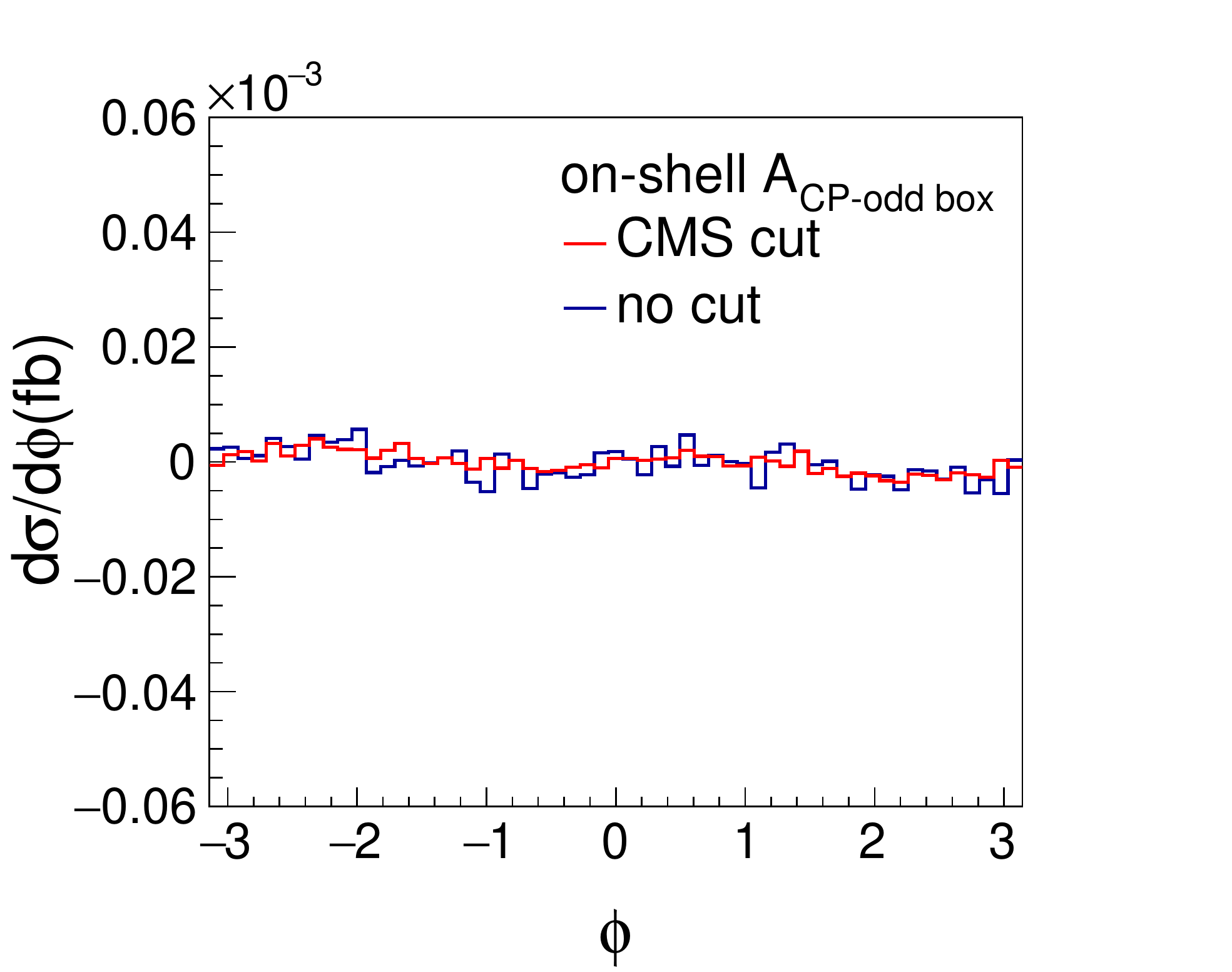}
		\includegraphics[width=3.6cm]{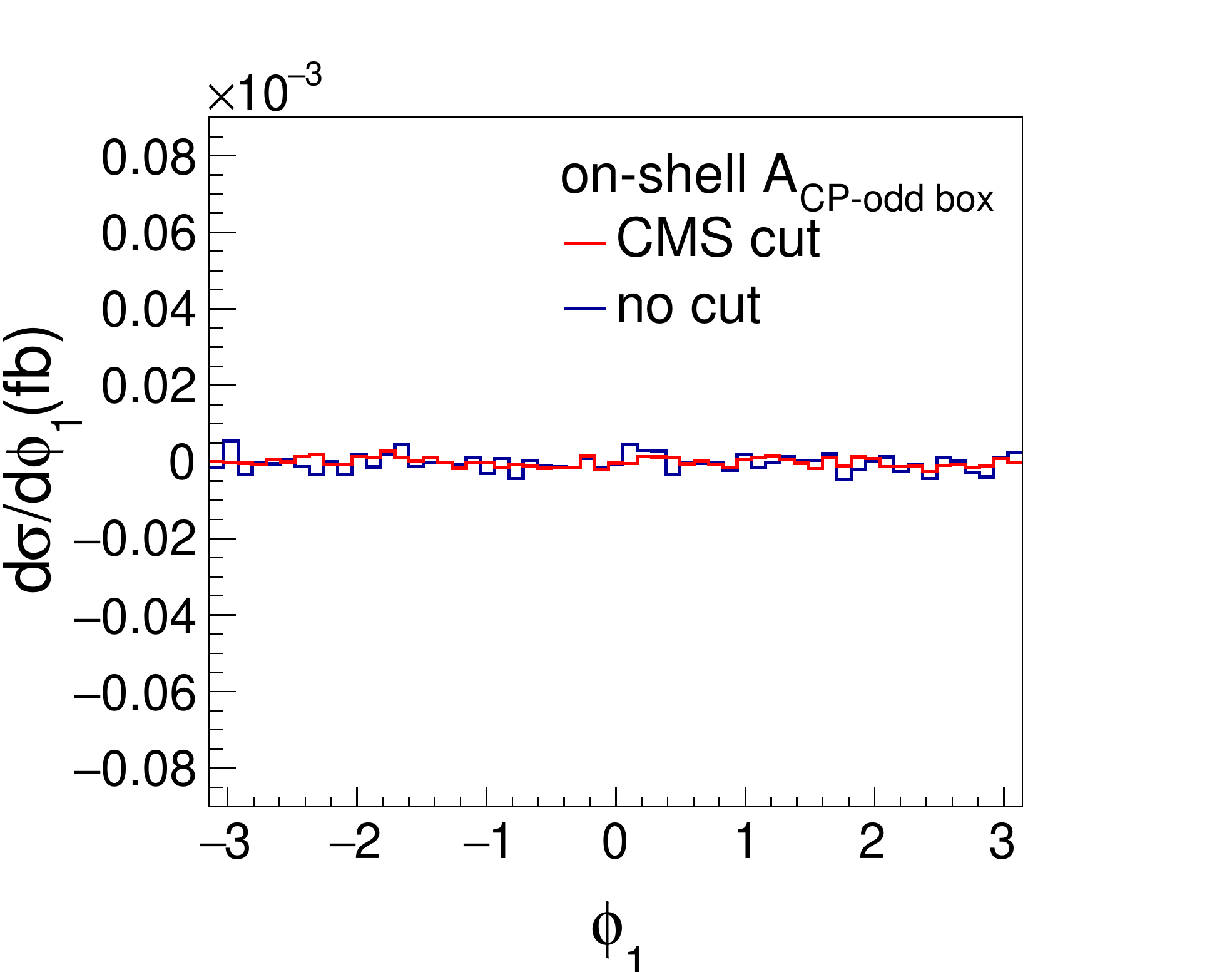}	
		\includegraphics[width=3.6cm]{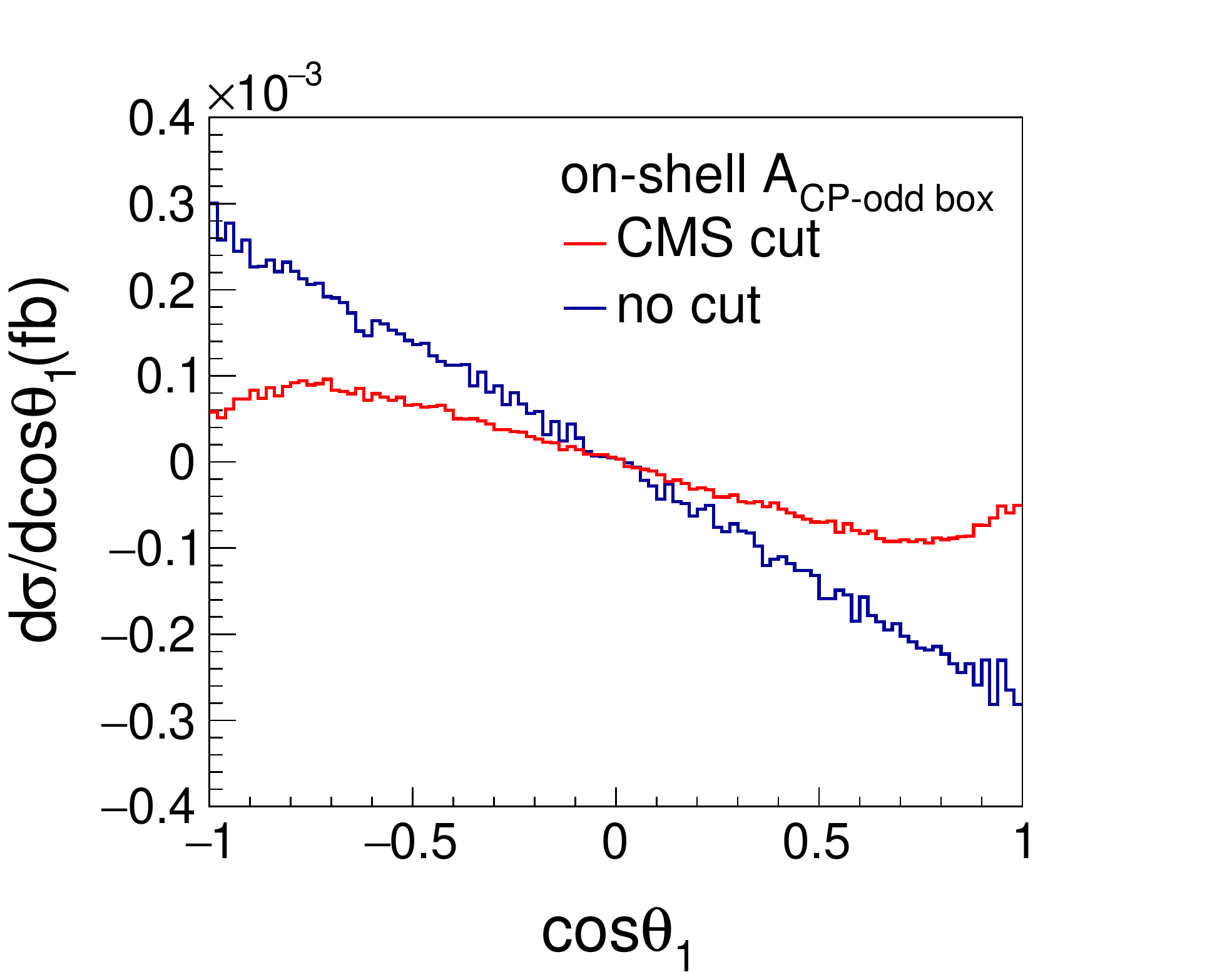}
		\includegraphics[width=3.6cm]{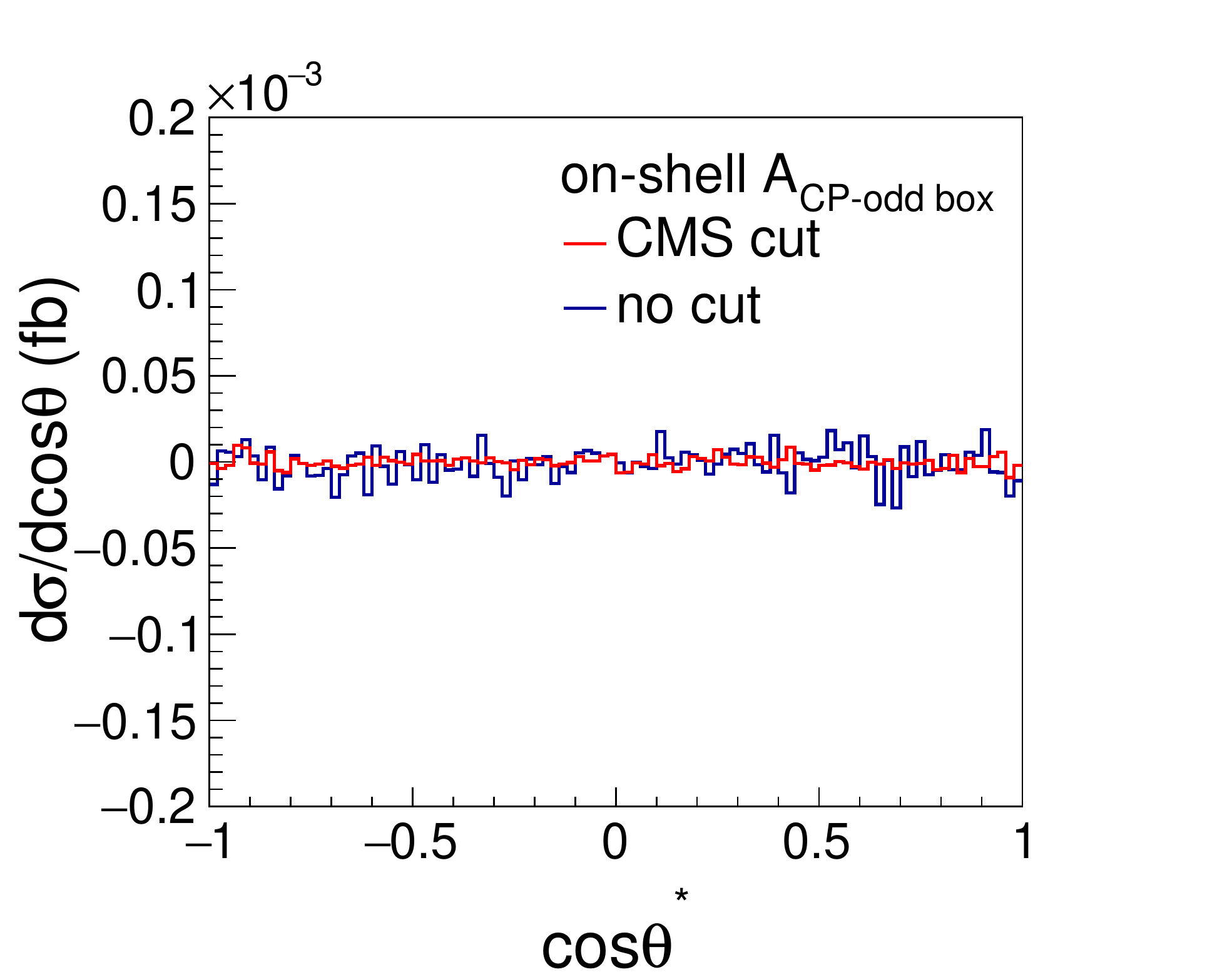}

		\caption{The angular differential cross sections from Higgs mediated processes interference with box in Higgs on-shell region. }
	\label{inter-box-on}	
	\end{figure}
	
	\begin{figure}[htbp]
		\centering
		\includegraphics[width=3.6cm]{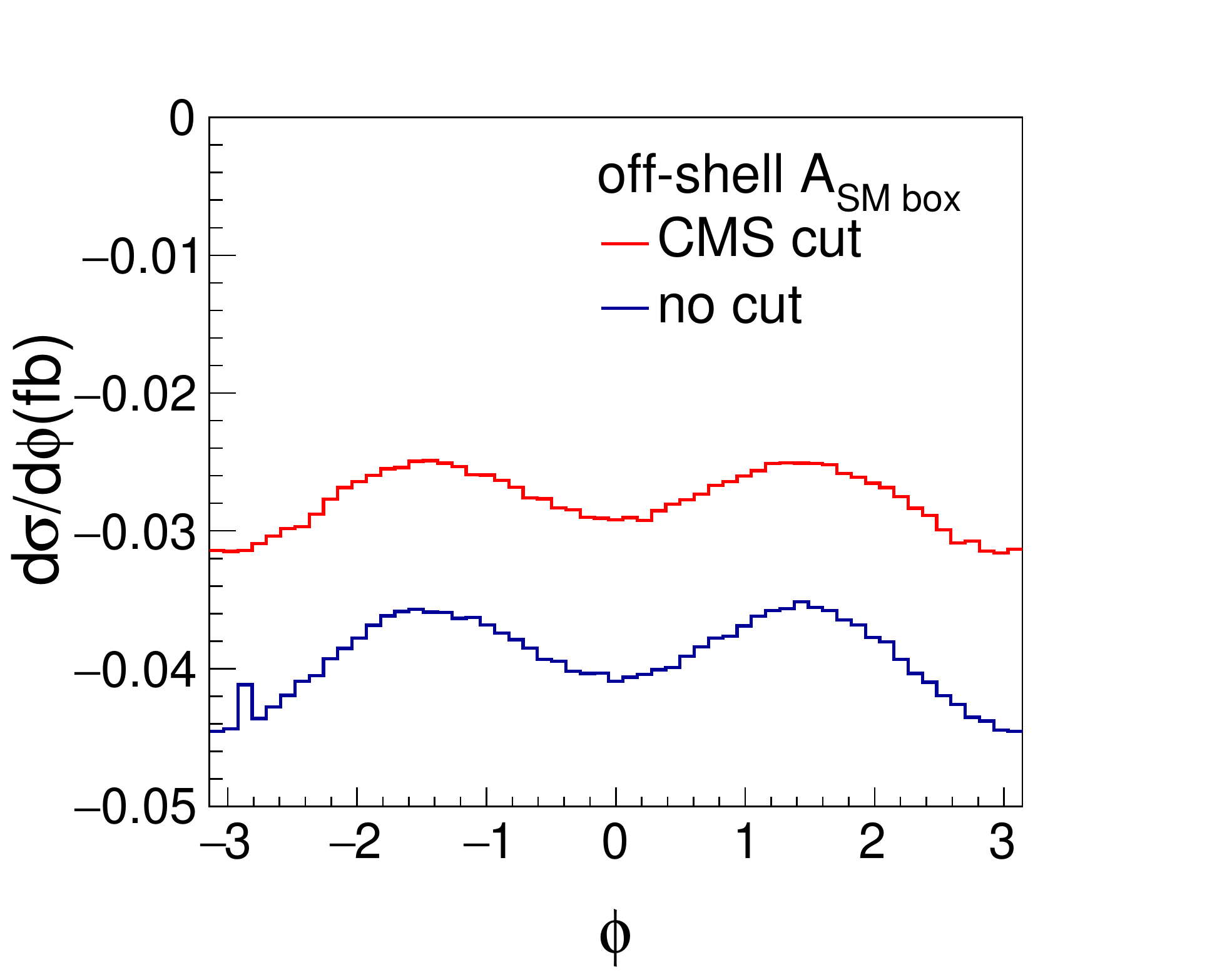}
		\includegraphics[width=3.6cm]{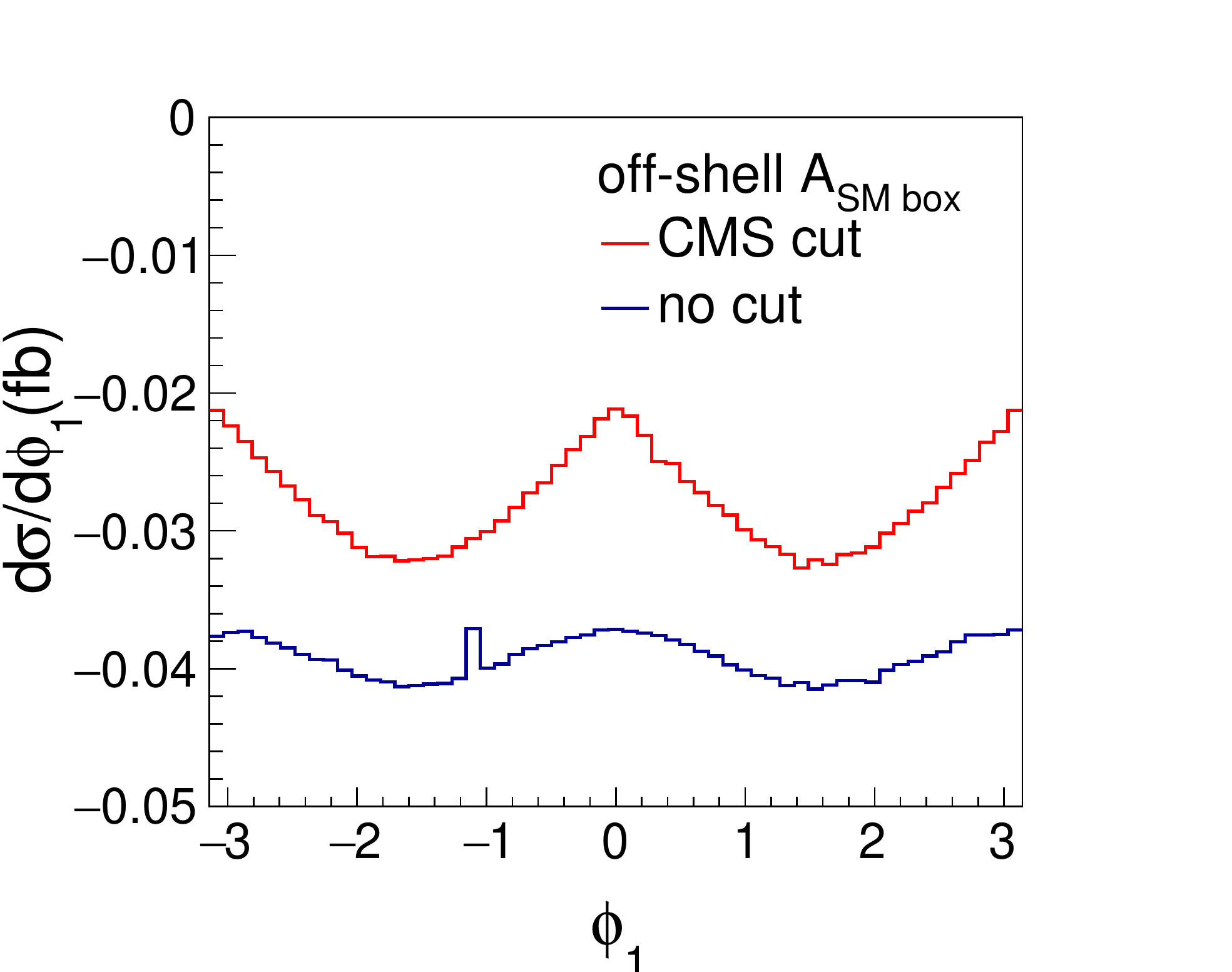}
		\includegraphics[width=3.6cm]{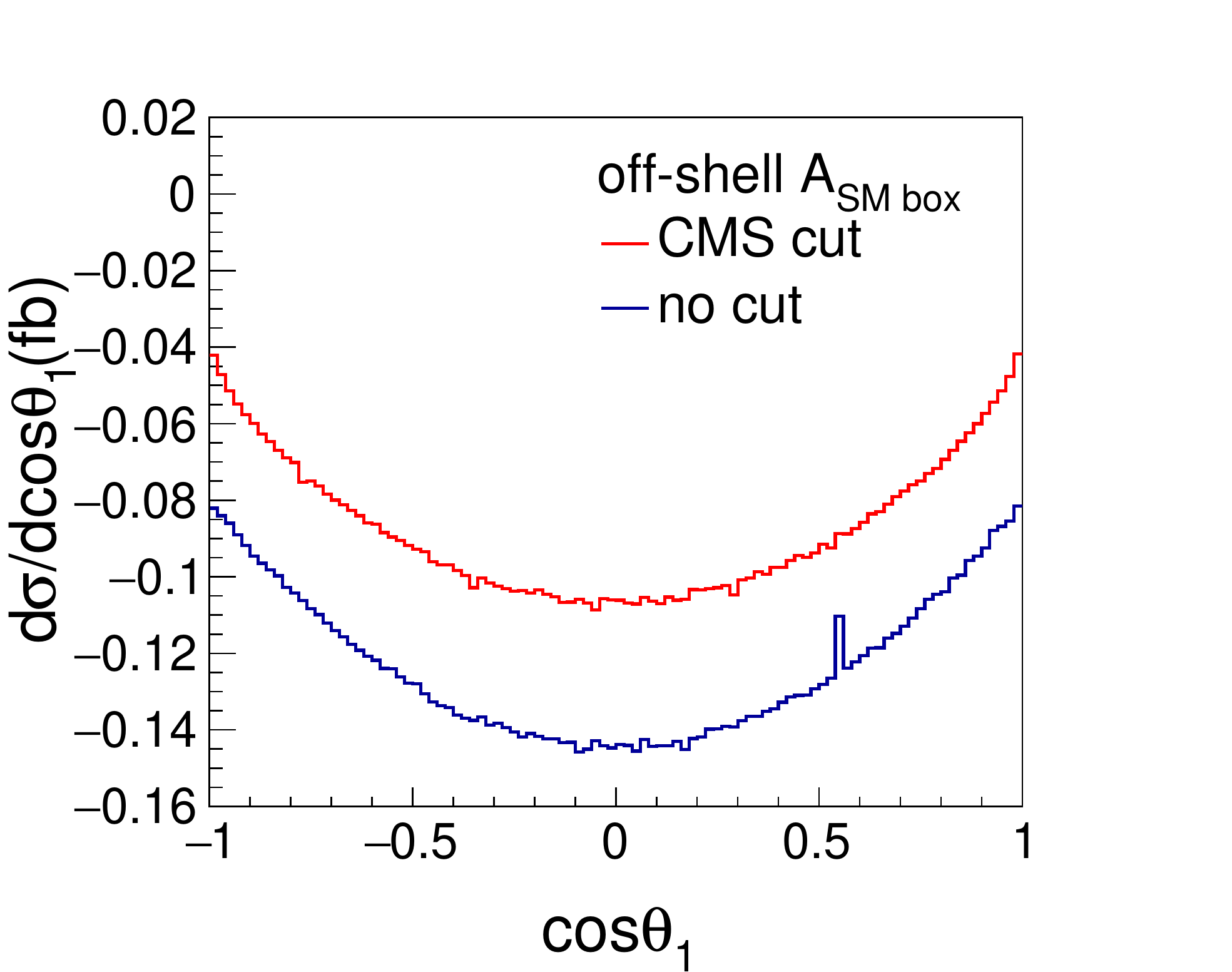}
		\includegraphics[width=3.6cm]{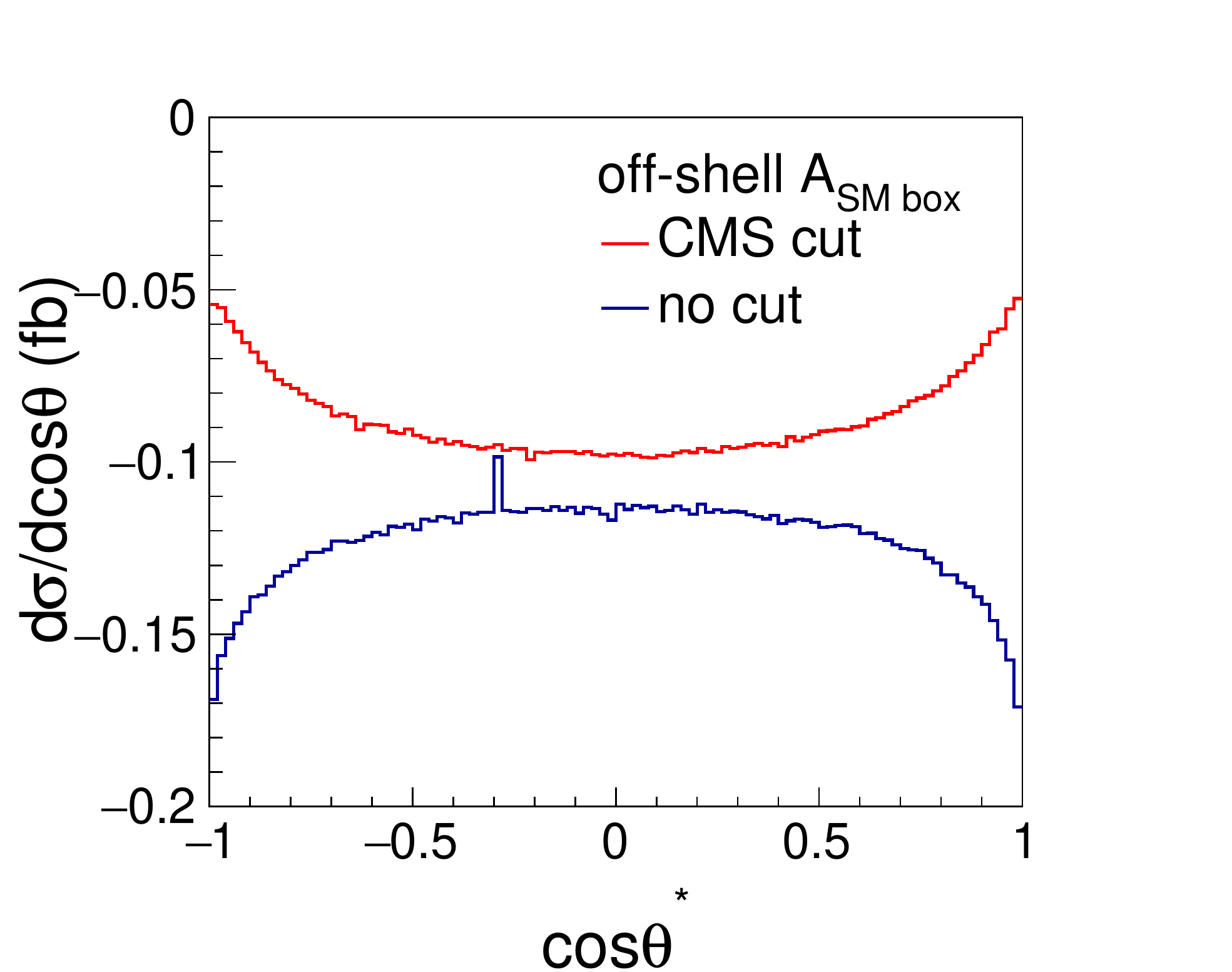}
		\\
		\includegraphics[width=3.6cm]{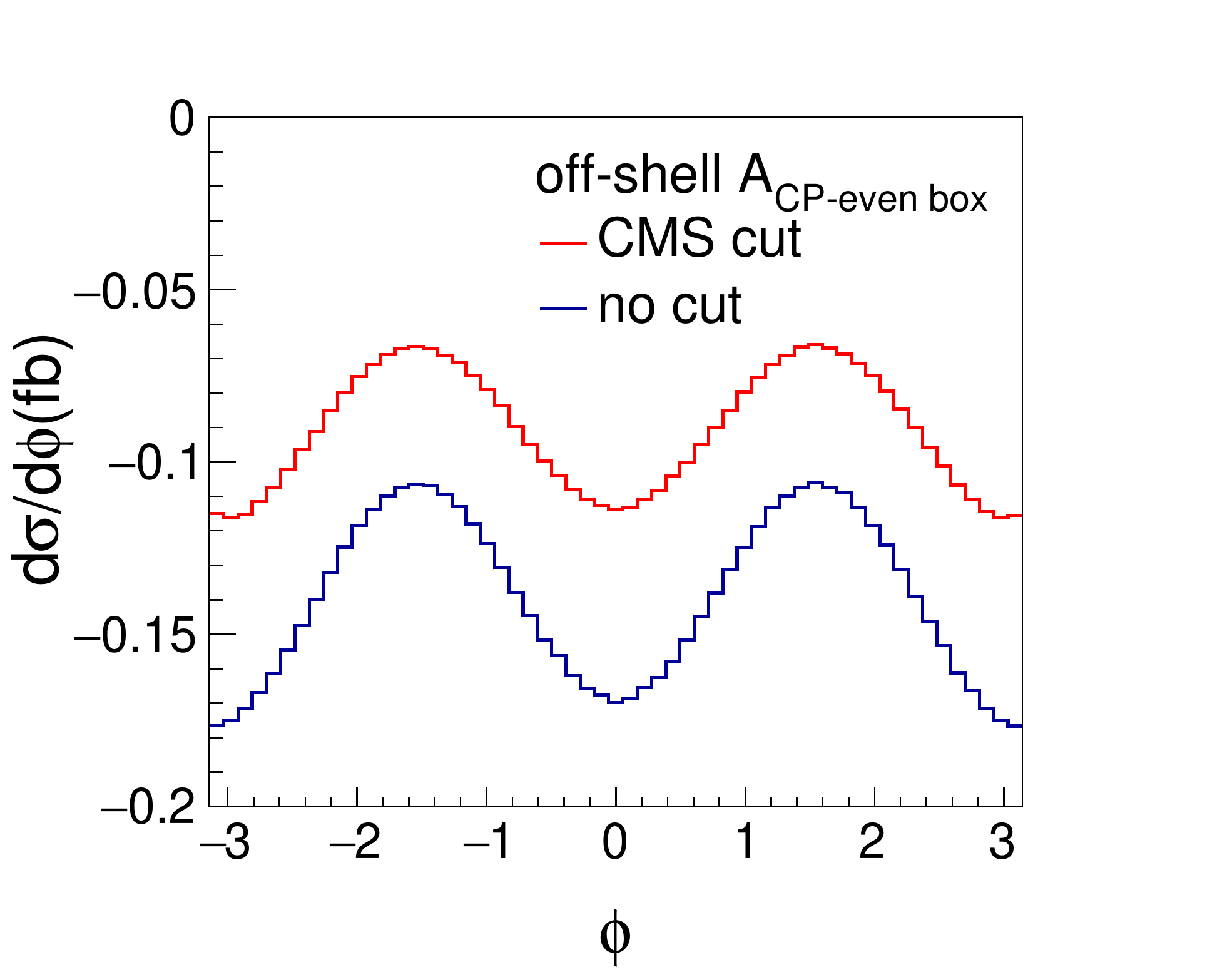}
		\includegraphics[width=3.6cm]{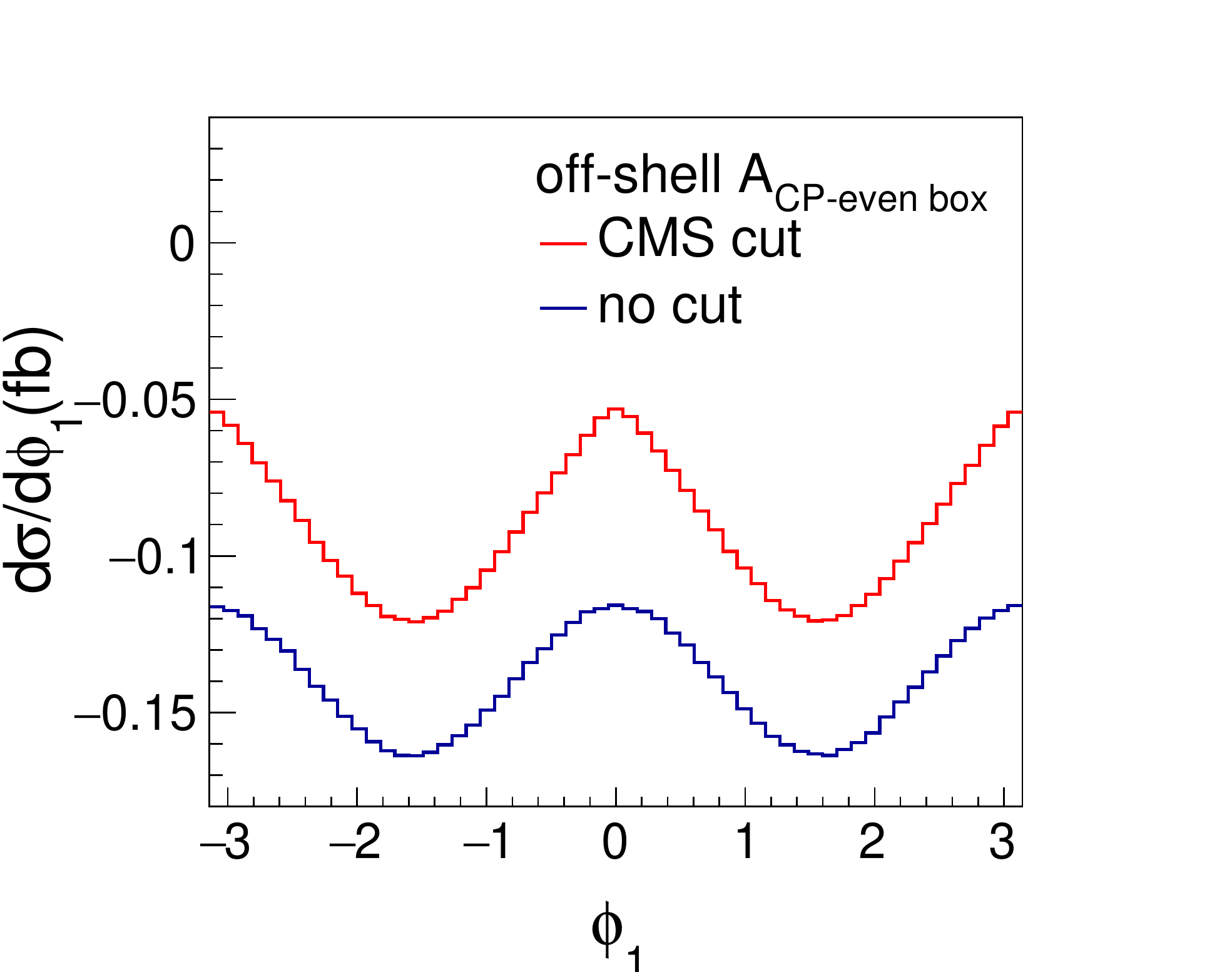}
		\includegraphics[width=3.6cm]{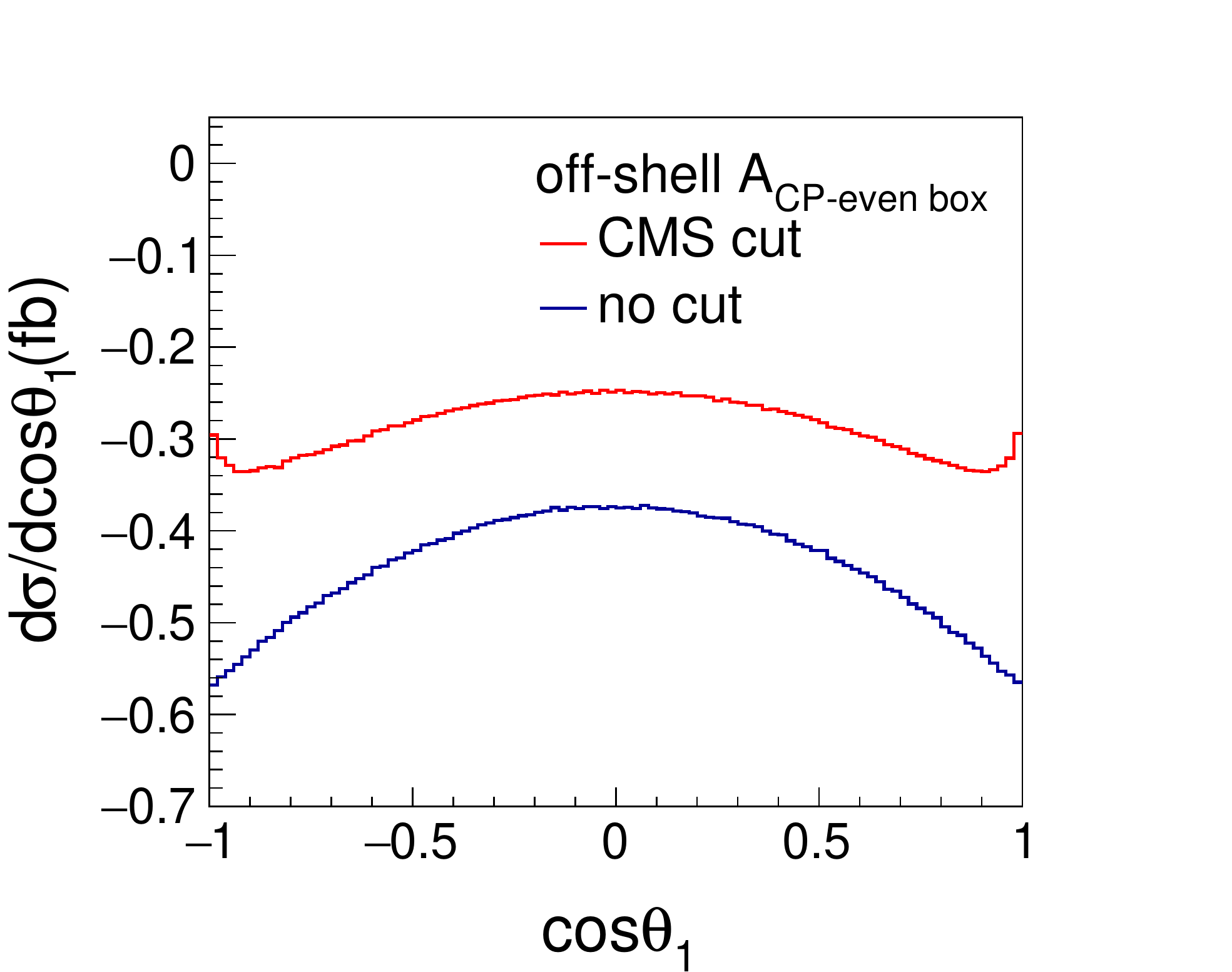}
		\includegraphics[width=3.6cm]{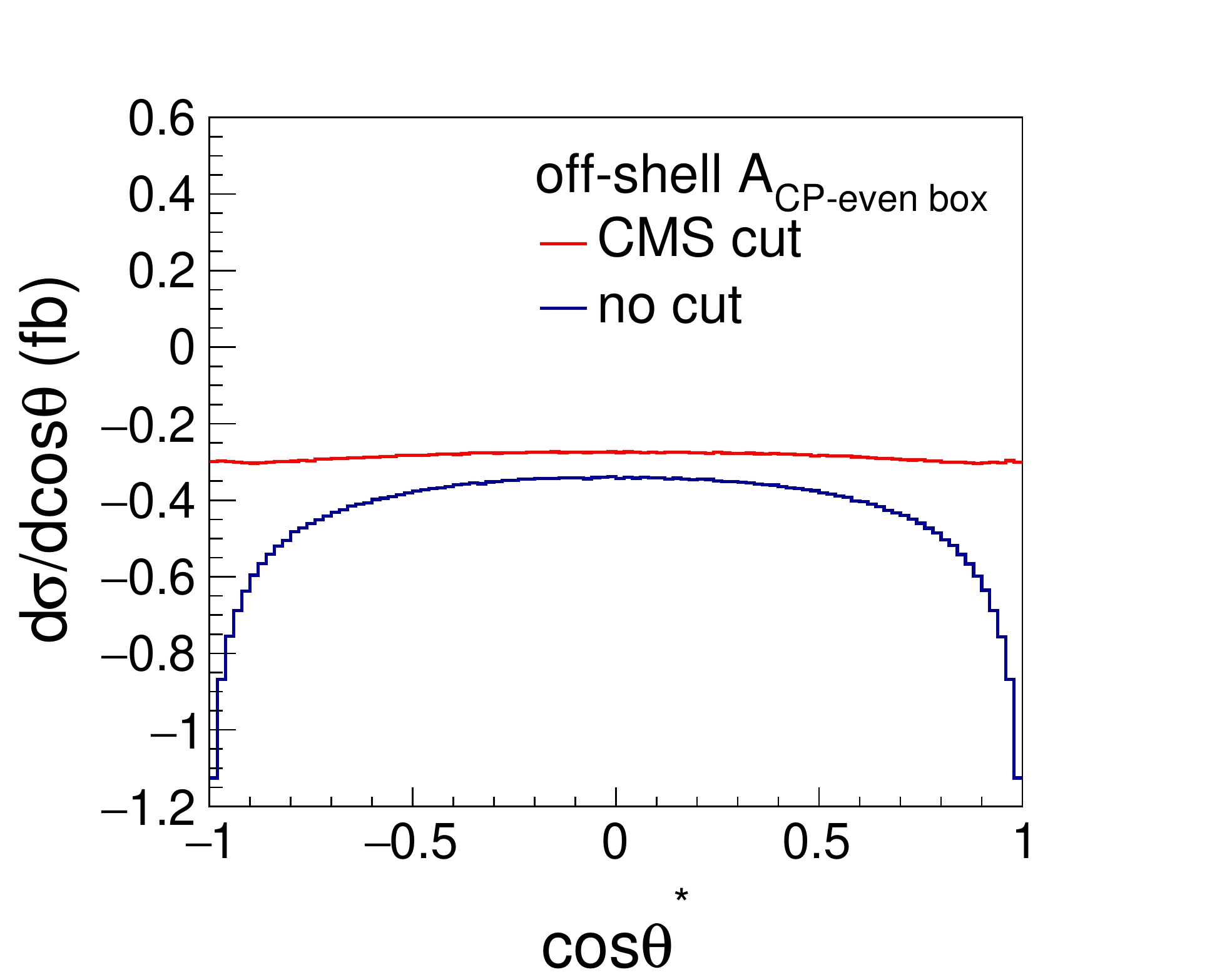}
		\\
		\includegraphics[width=3.6cm]{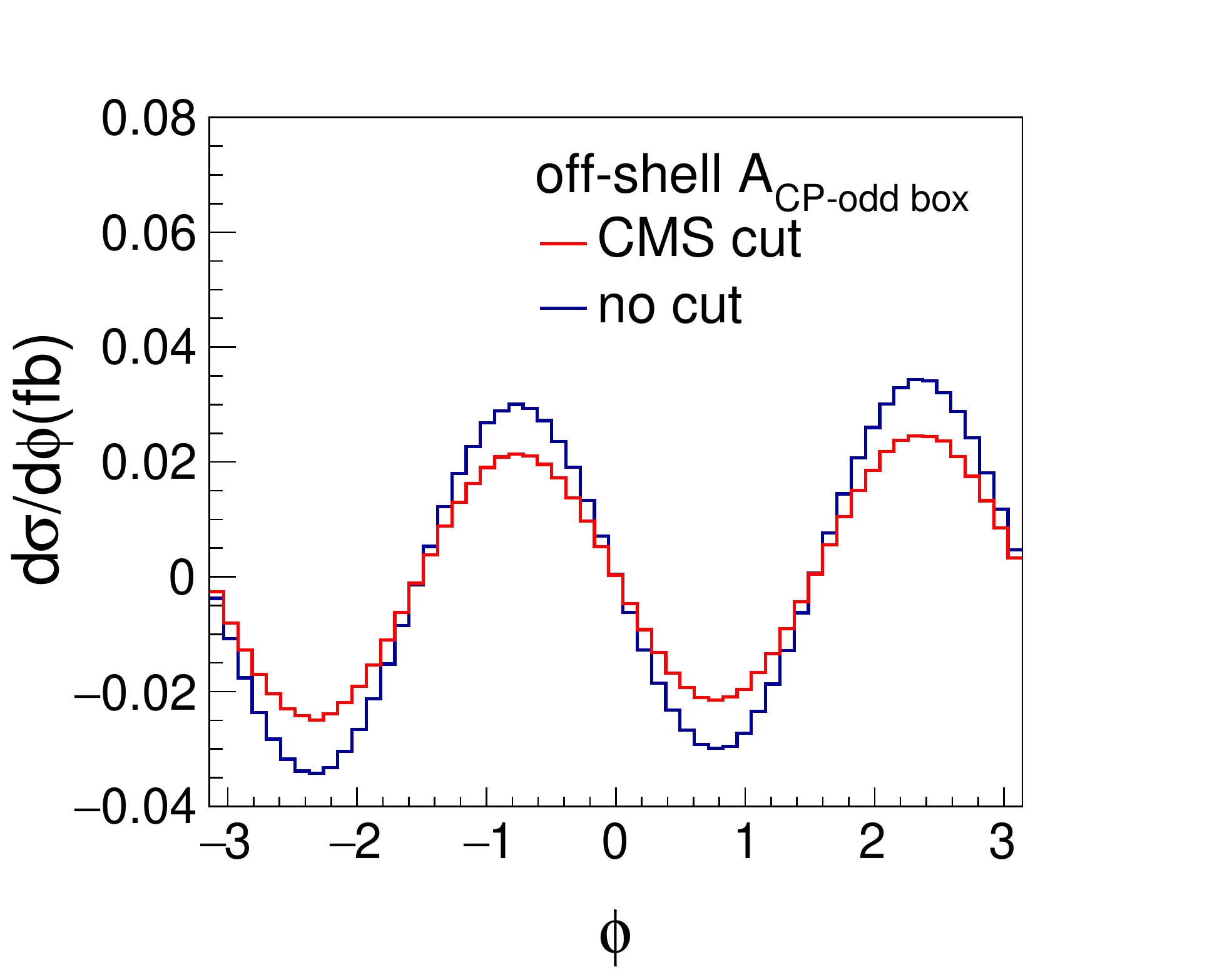}
		\includegraphics[width=3.6cm]{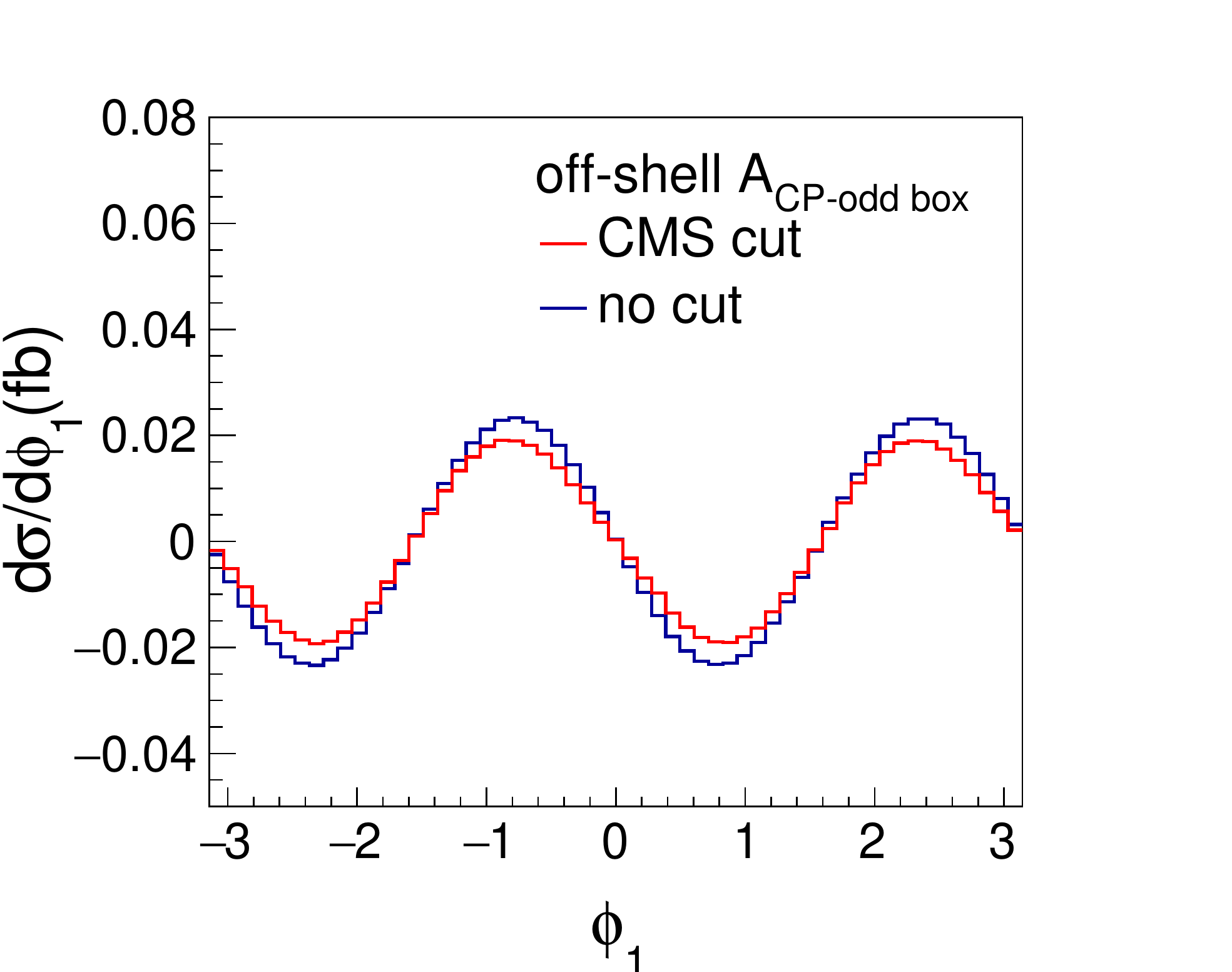}
		\includegraphics[width=3.6cm]{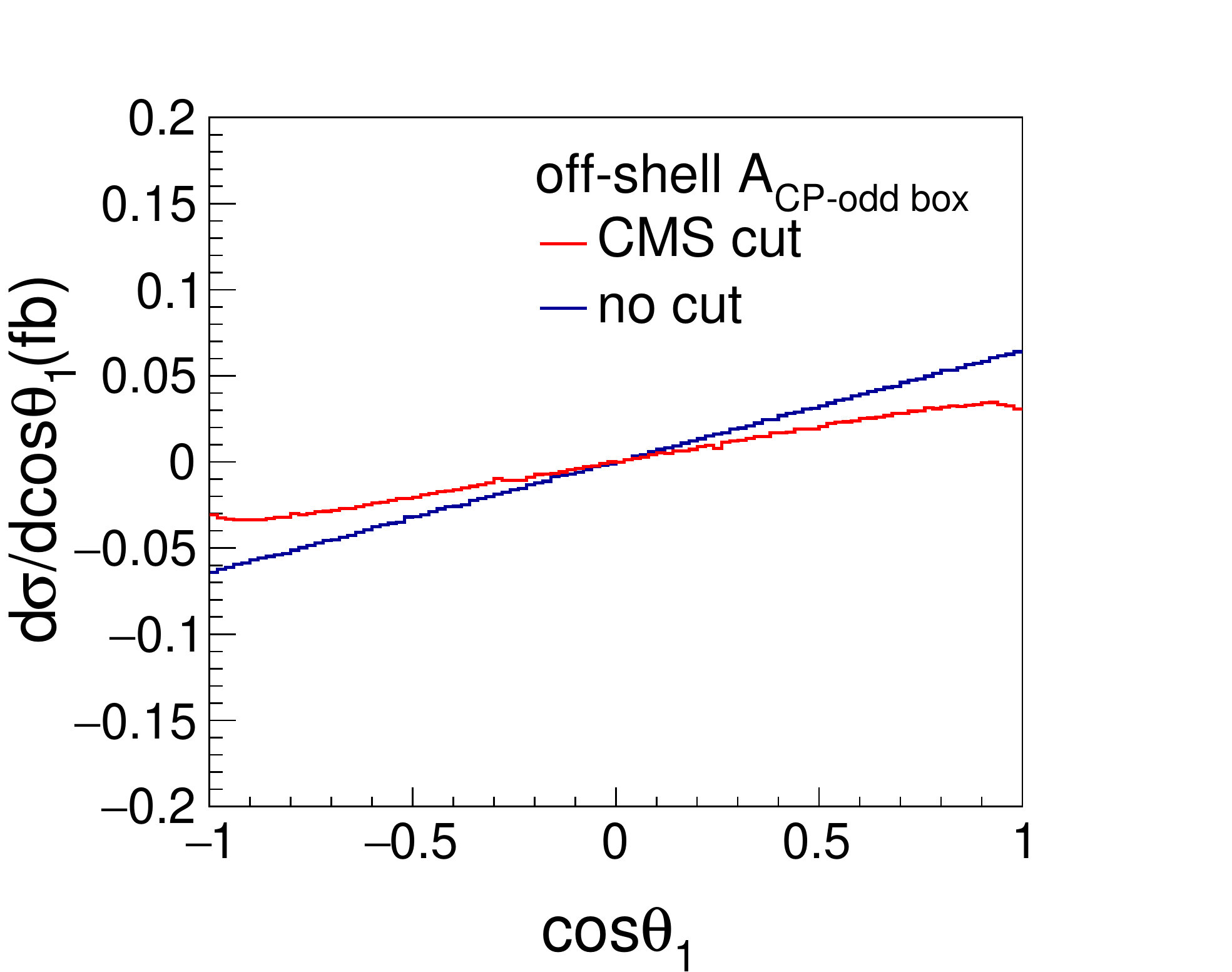}
		\includegraphics[width=3.6cm]{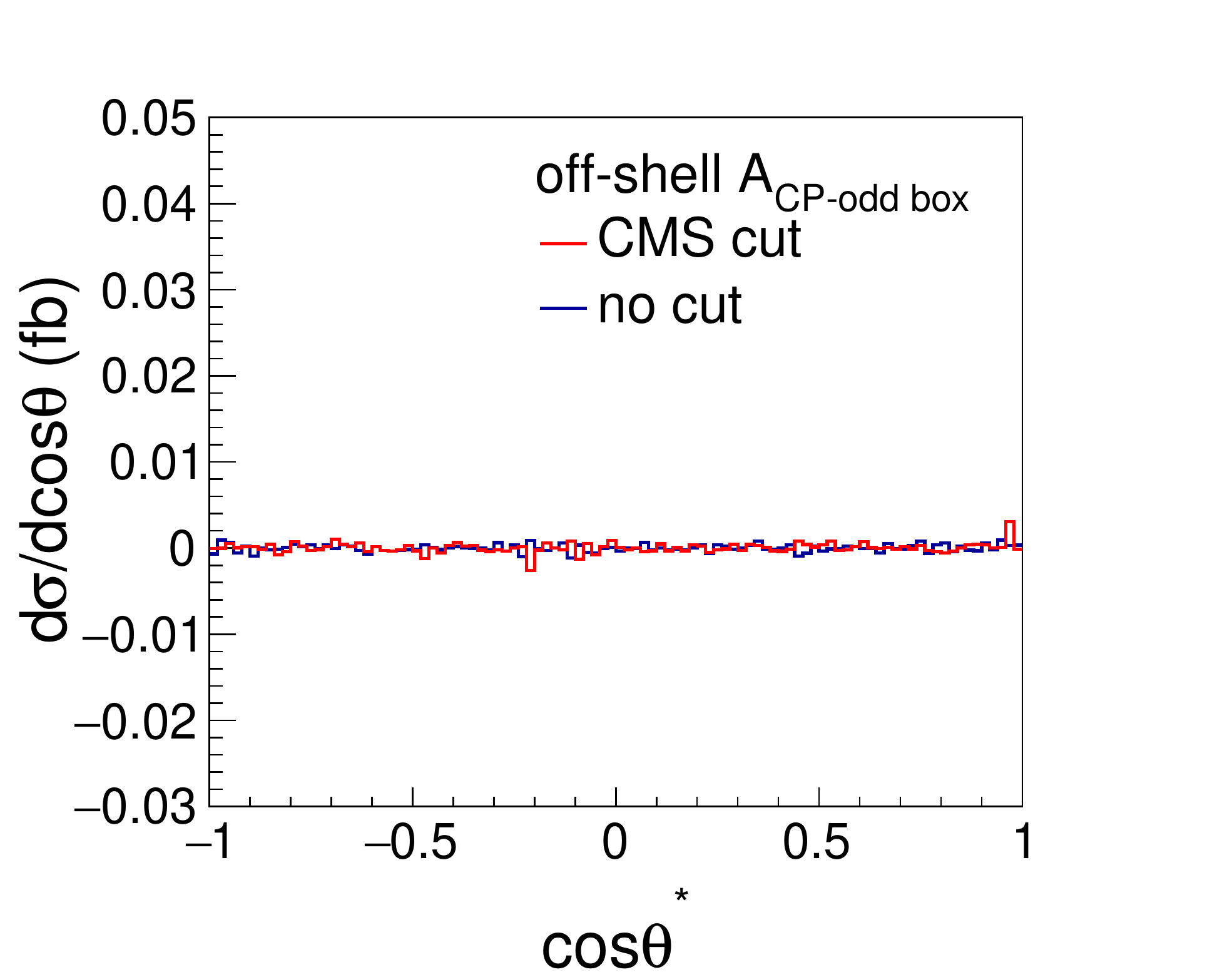}
	
		\caption{The angular differential cross sections from Higgs mediated processes interference with box in Higgs off-shell region. }
		\label{inter-box-off}
	\end{figure}

	\begin{figure}[htbp]
		\centering
		
		\includegraphics[width=3.6cm]{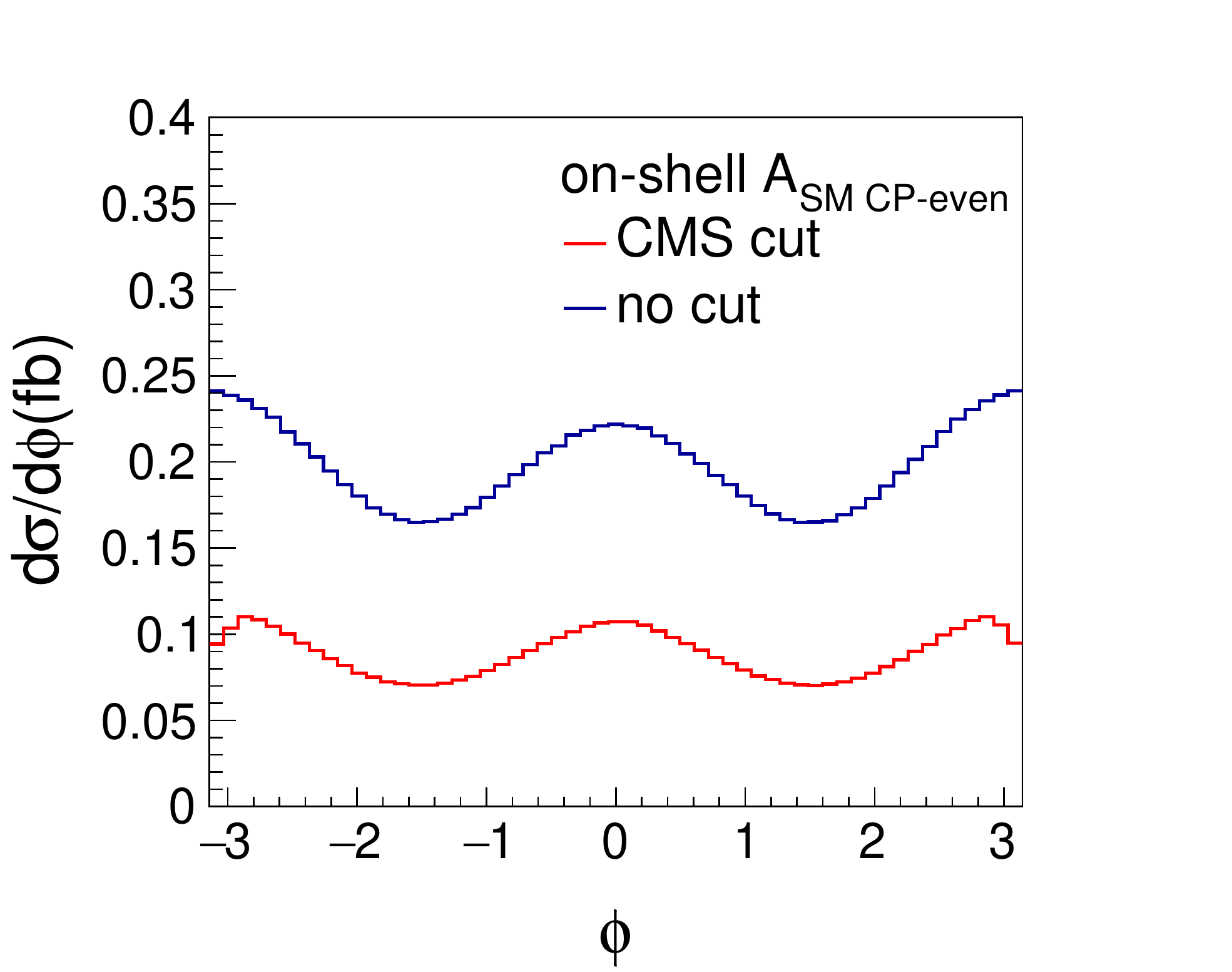}
		\includegraphics[width=3.6cm]{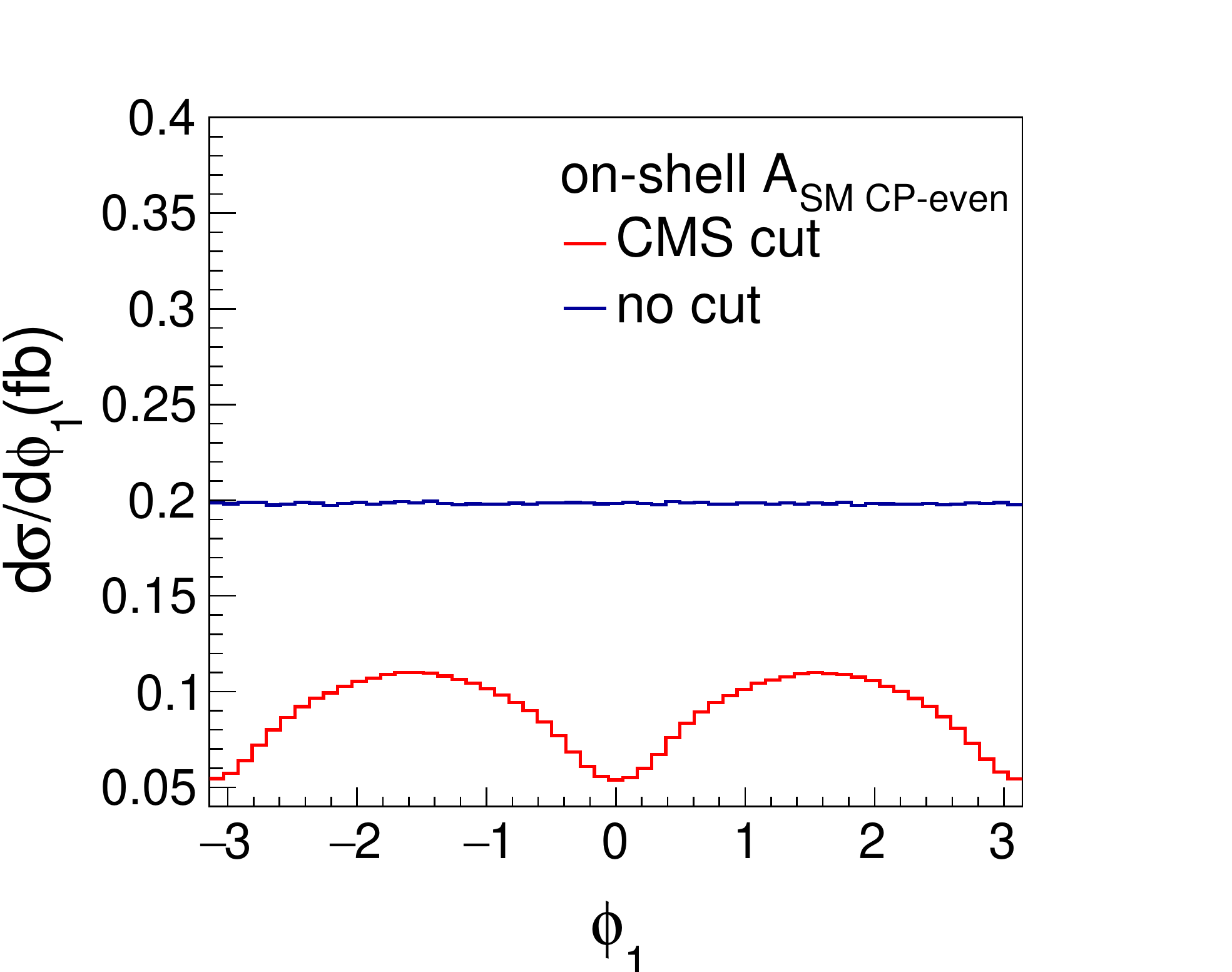}
		\includegraphics[width=3.6cm]{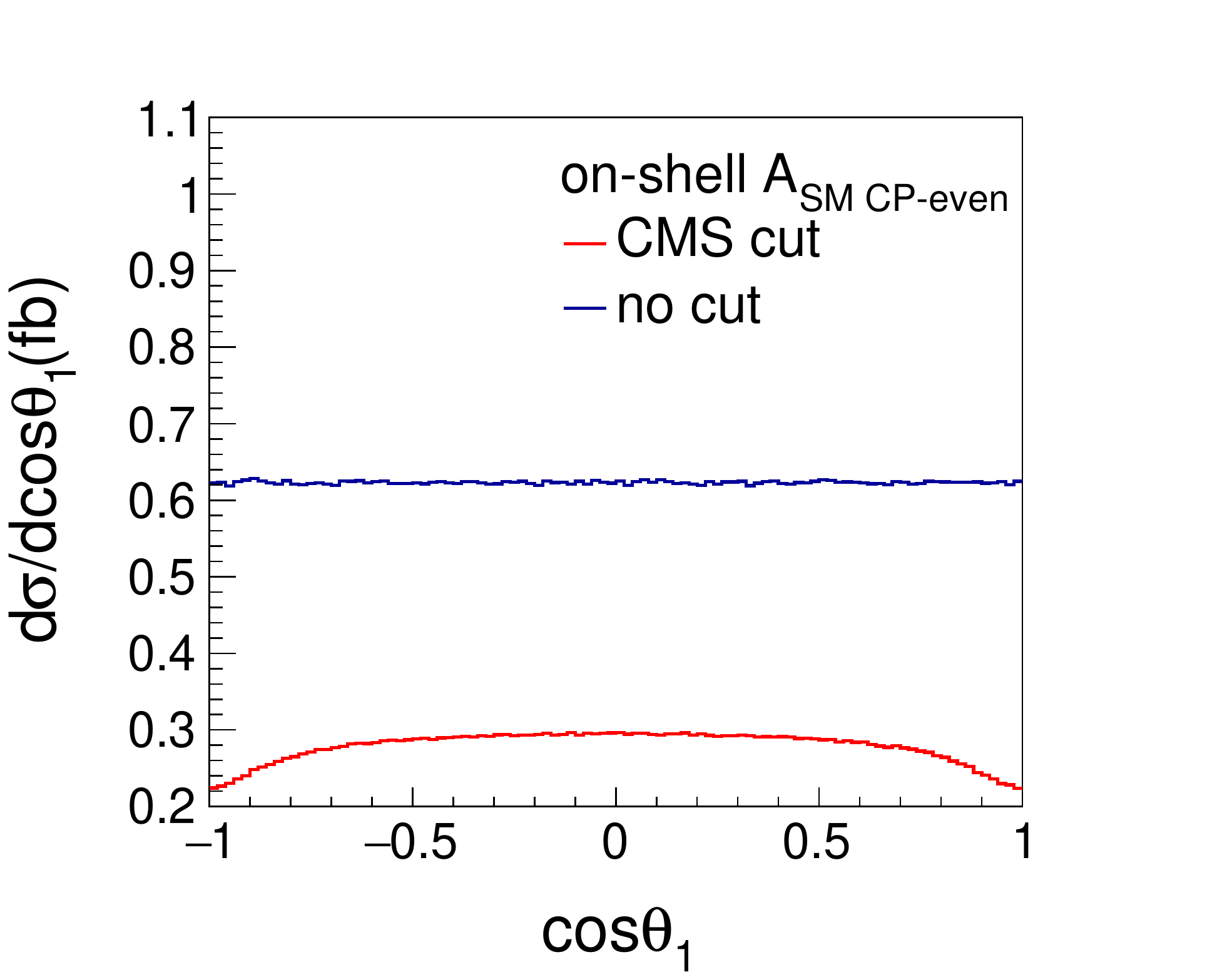}
		\includegraphics[width=3.6cm]{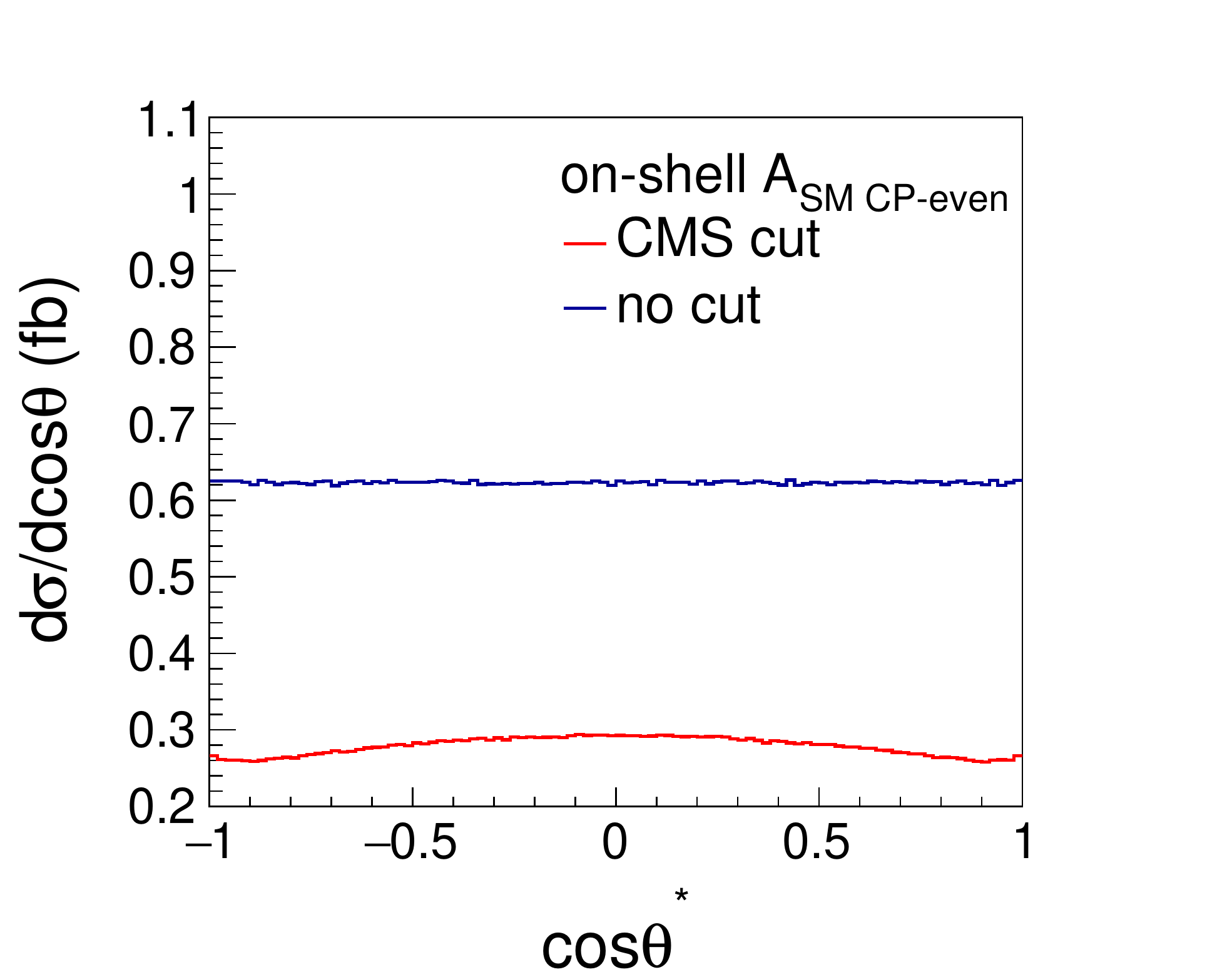}
		\\
		\includegraphics[width=3.6cm]{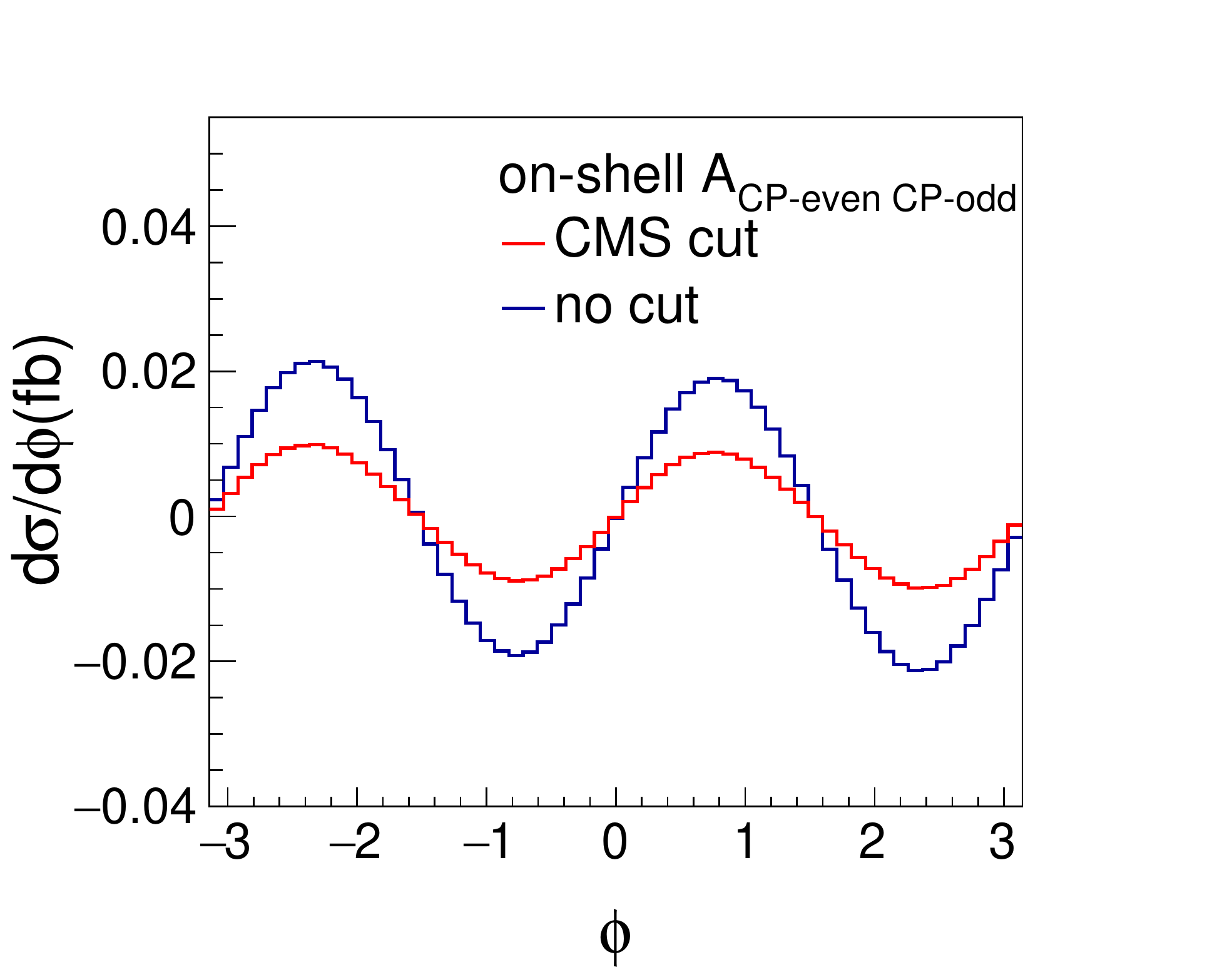}
		\includegraphics[width=3.6cm]{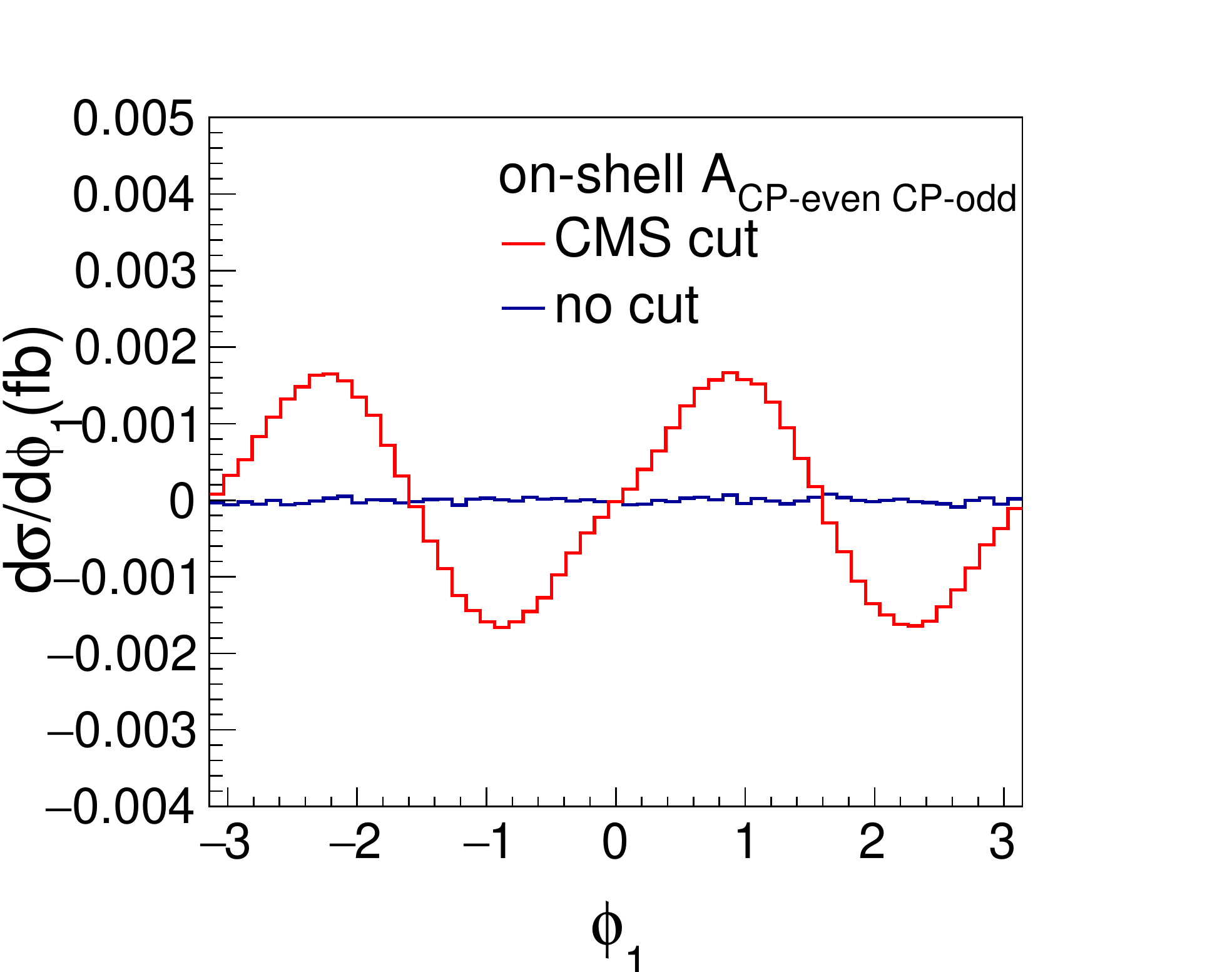}
		\includegraphics[width=3.6cm]{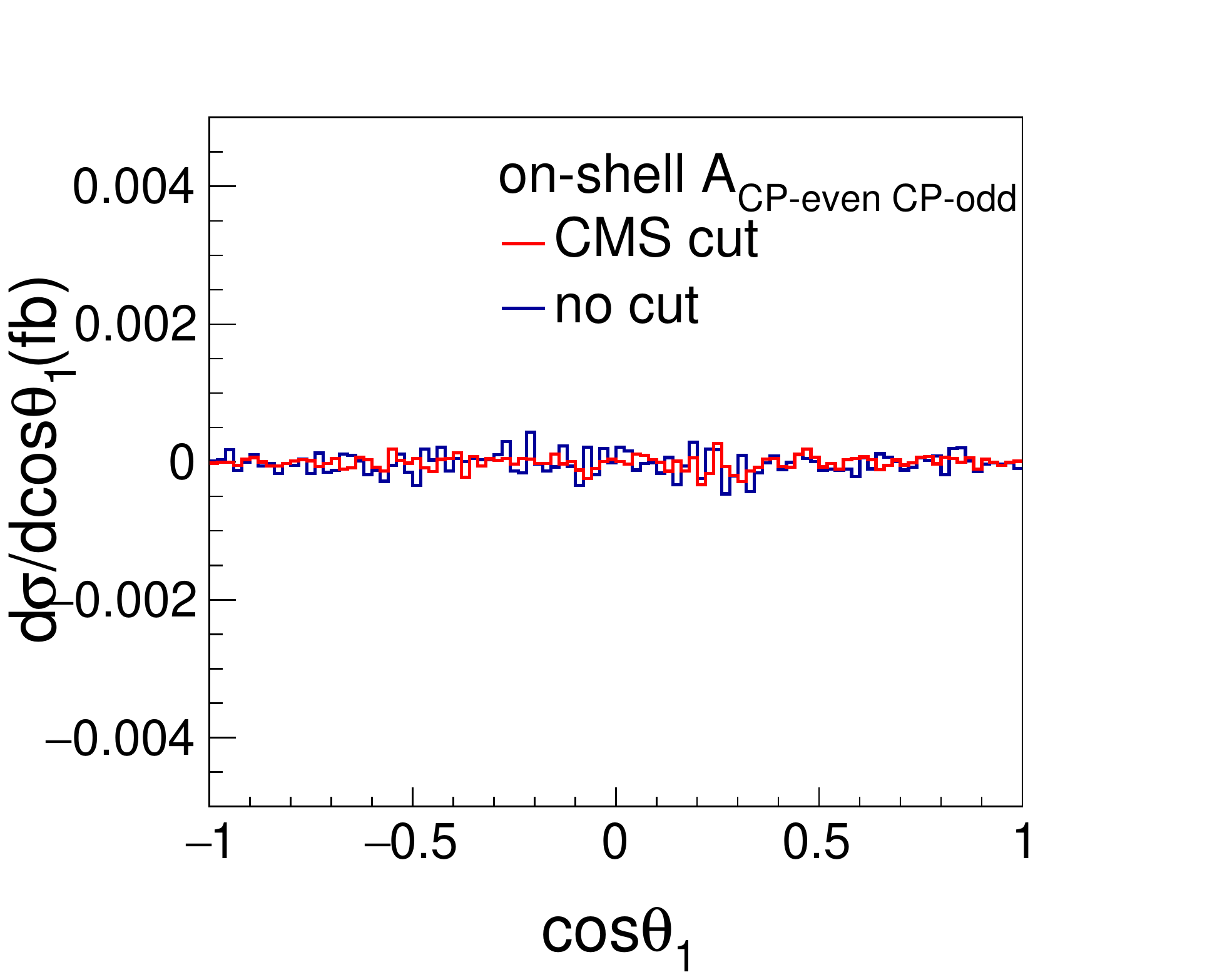}
		\includegraphics[width=3.6cm]{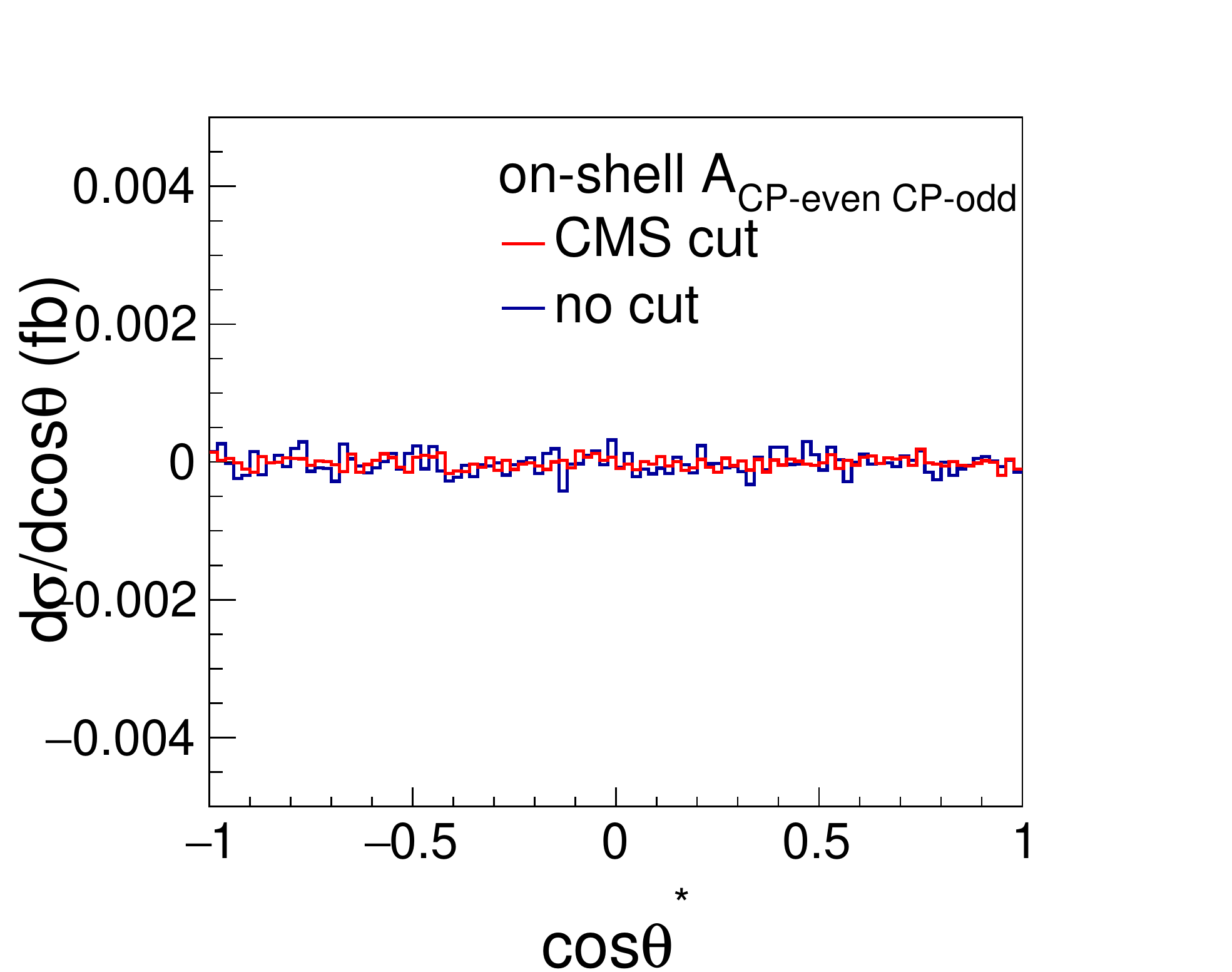}
		\\
		\includegraphics[width=3.6cm]{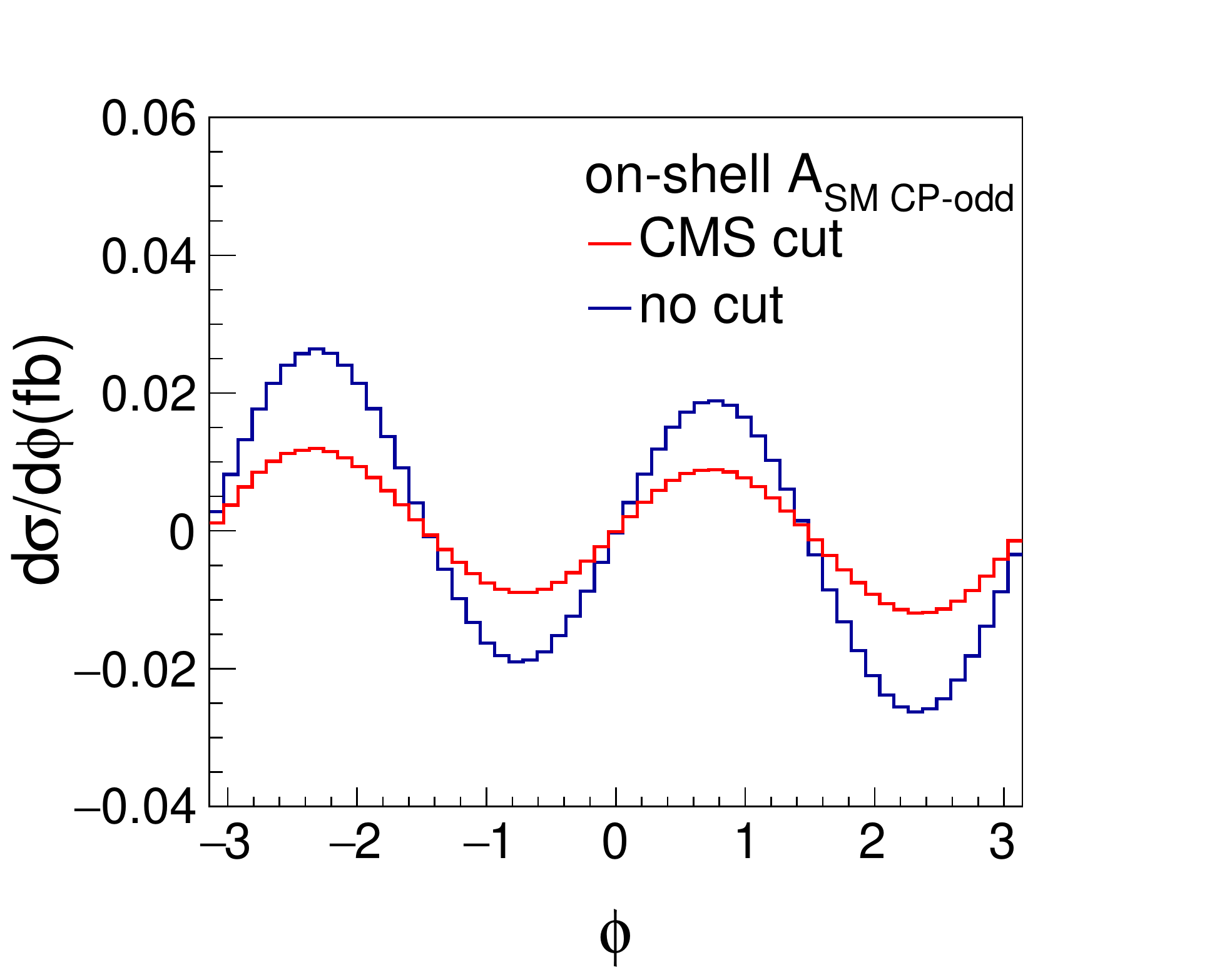}
		\includegraphics[width=3.6cm]{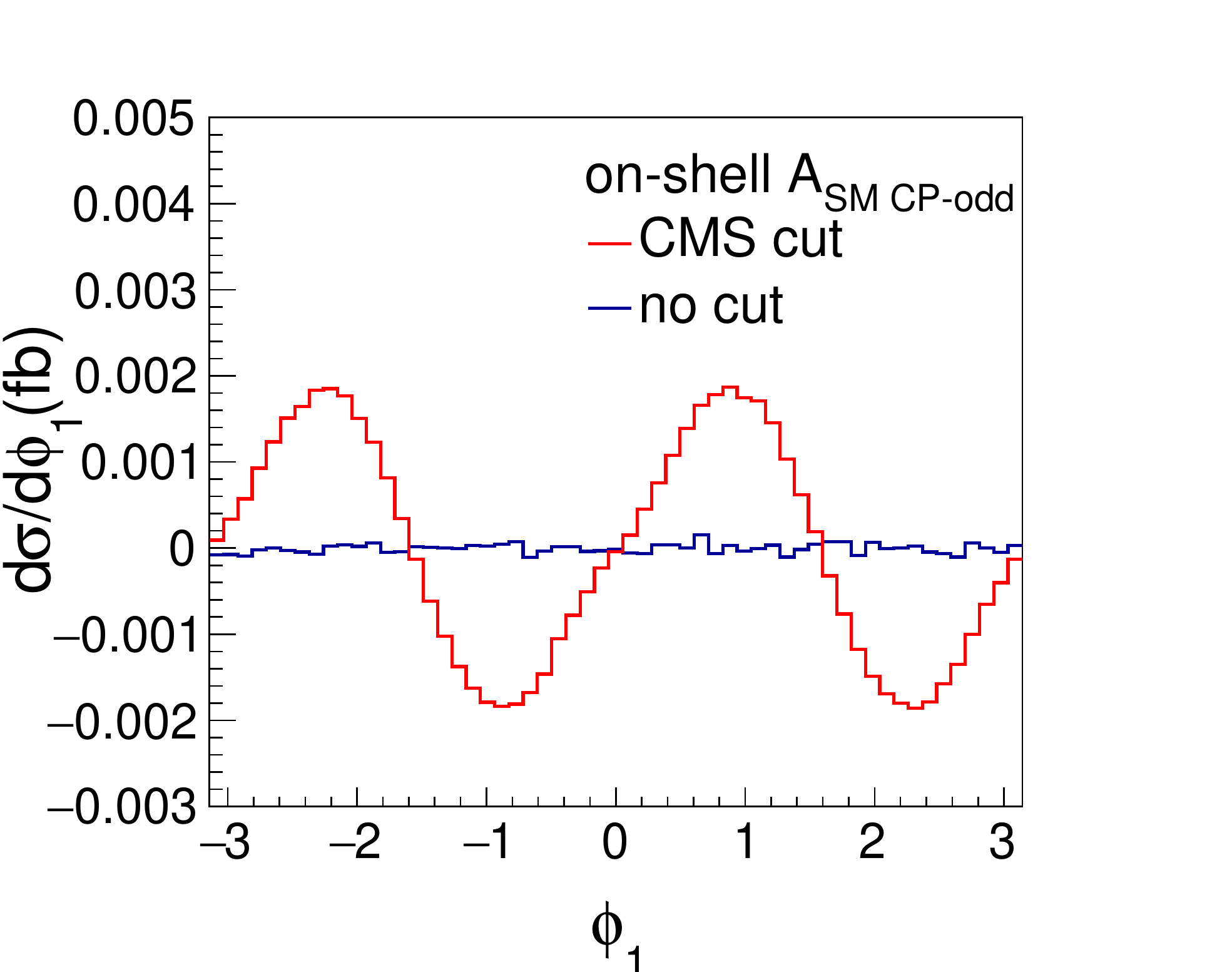}
		\includegraphics[width=3.6cm]{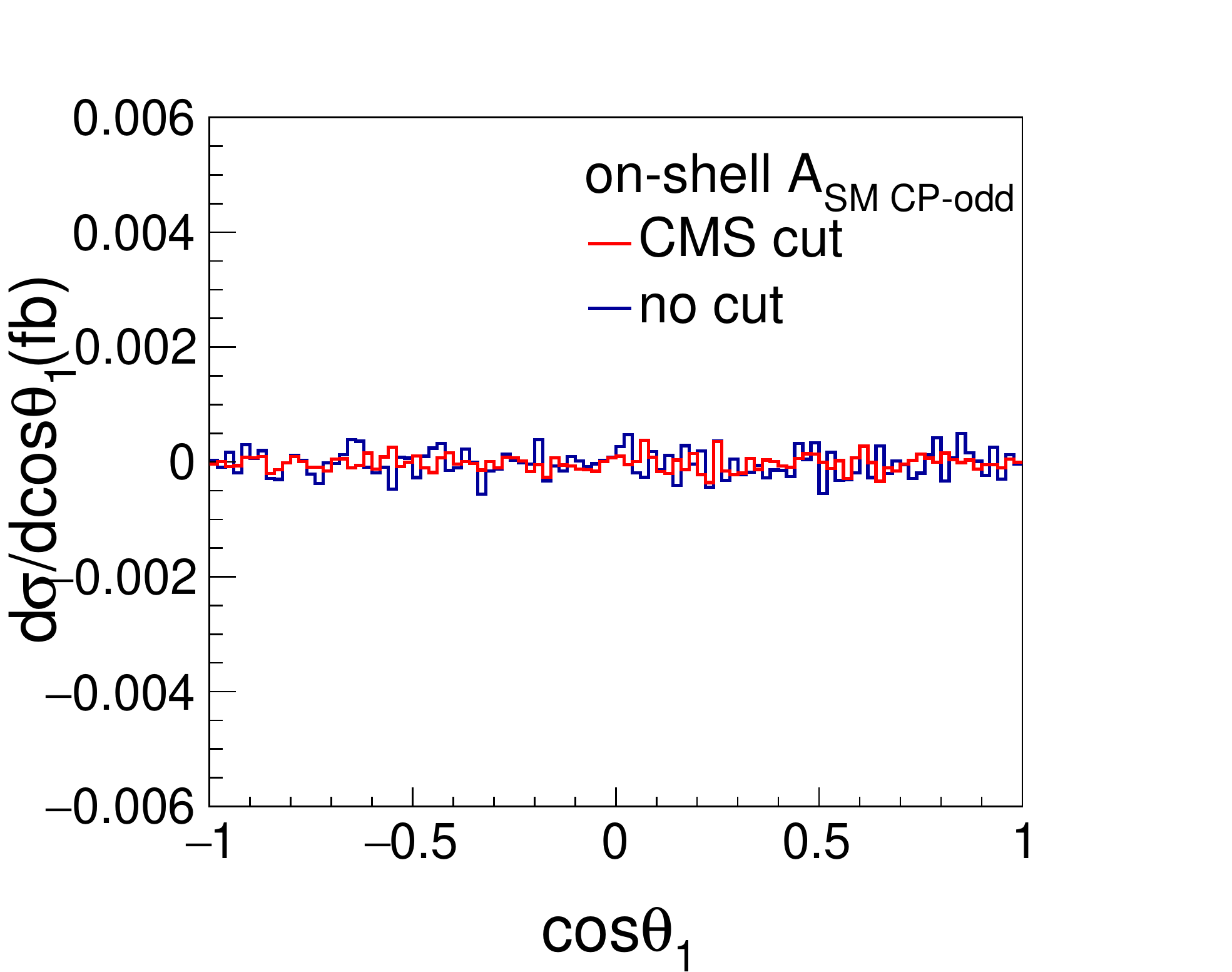}
		\includegraphics[width=3.6cm]{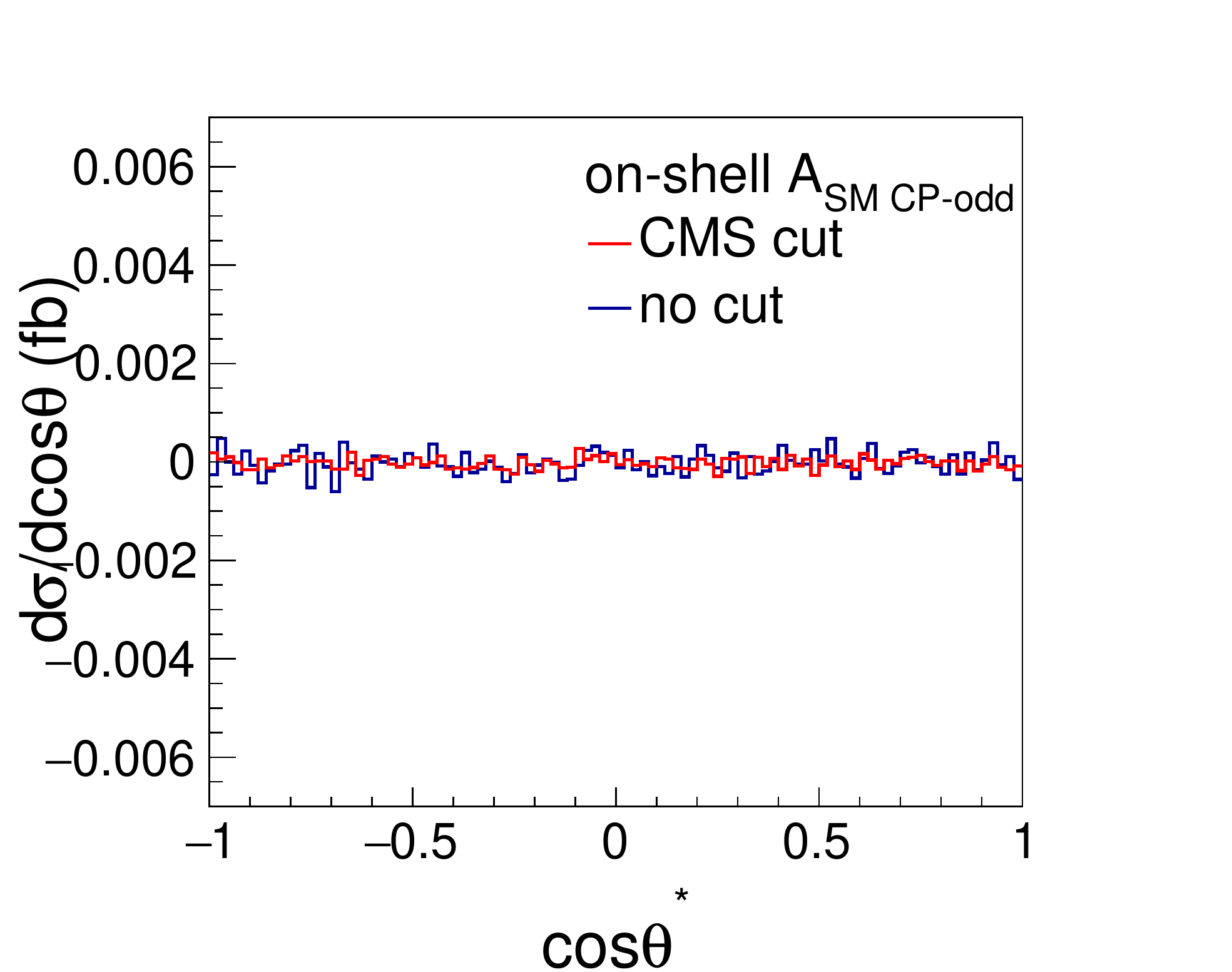}

		\caption{The angular differential cross sections from interference between Higgs mediated processes in Higgs on-shell region. }
		\label{inter-Higgs-on}
	\end{figure}

	\begin{figure}[htbp]
		\centering
		
		\includegraphics[width=3.6cm]{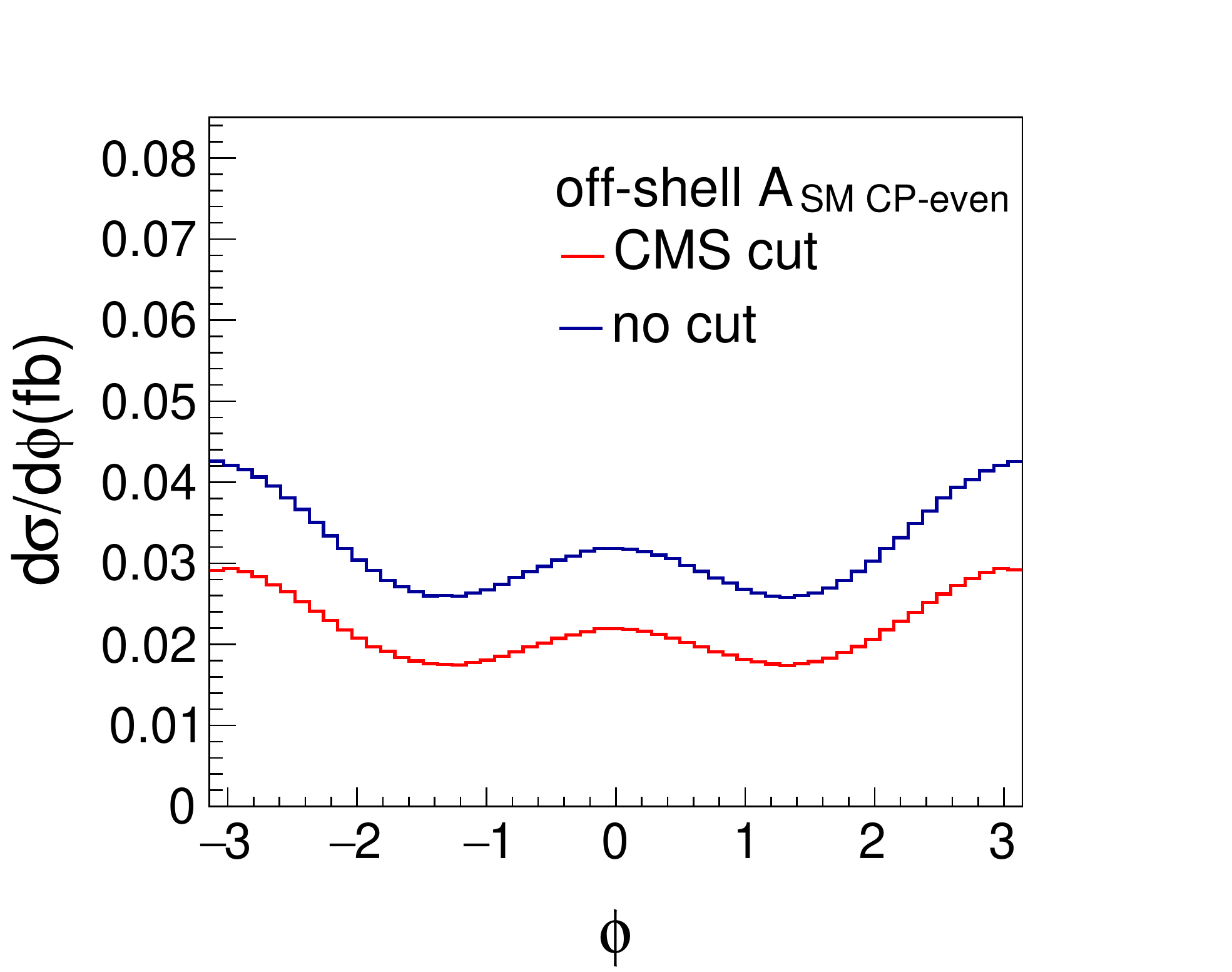}
		\includegraphics[width=3.6cm]{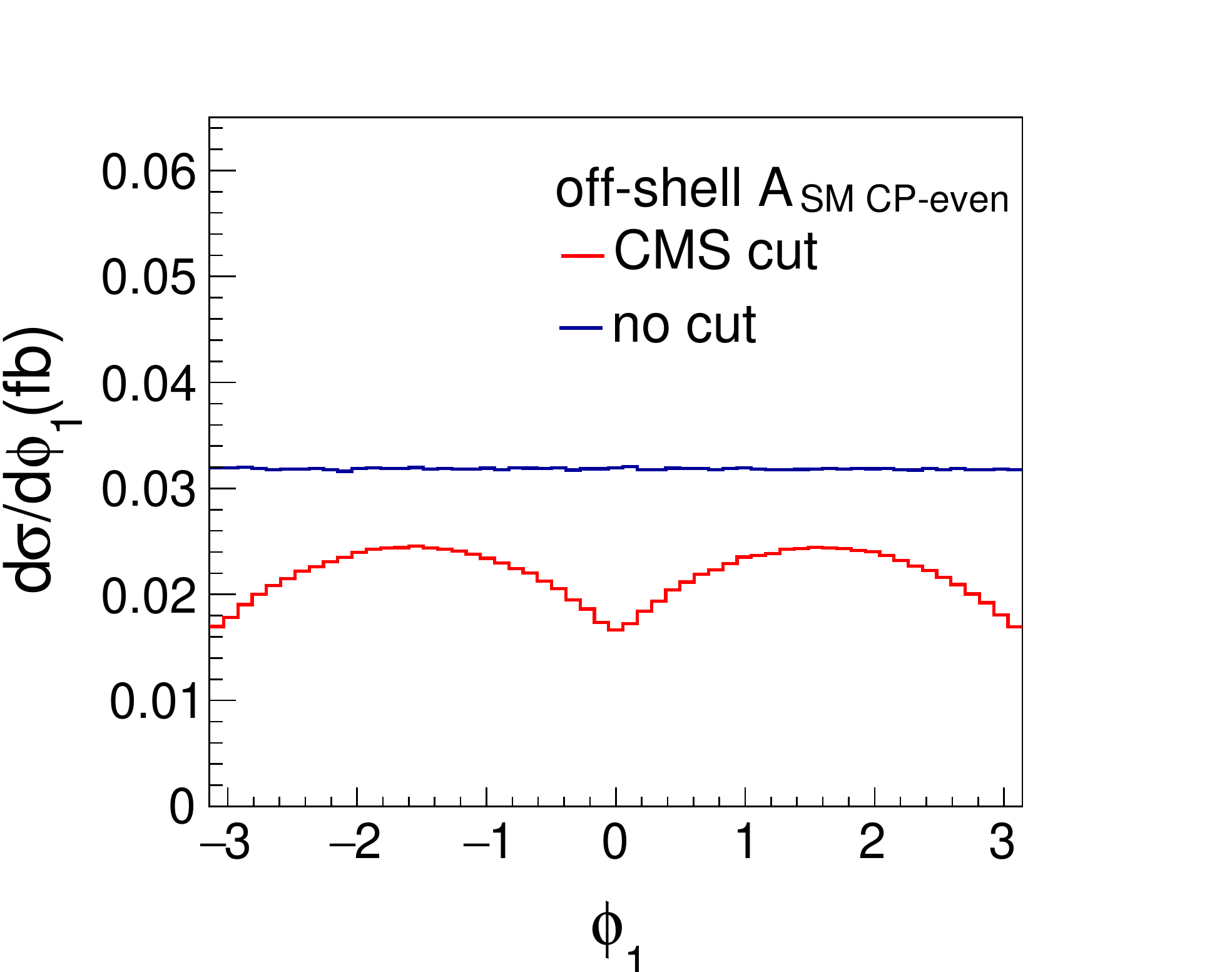}
		\includegraphics[width=3.6cm]{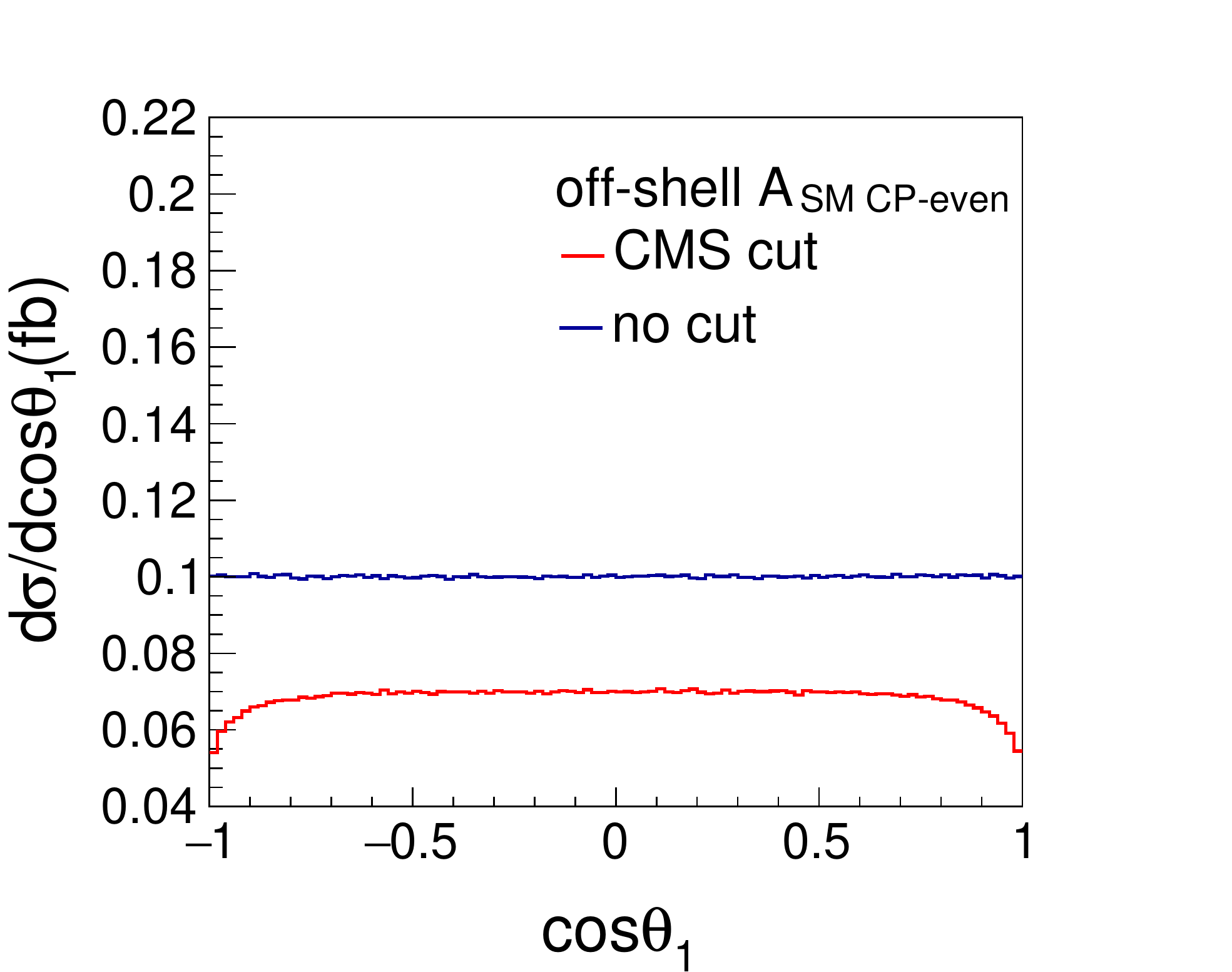}
		\includegraphics[width=3.6cm]{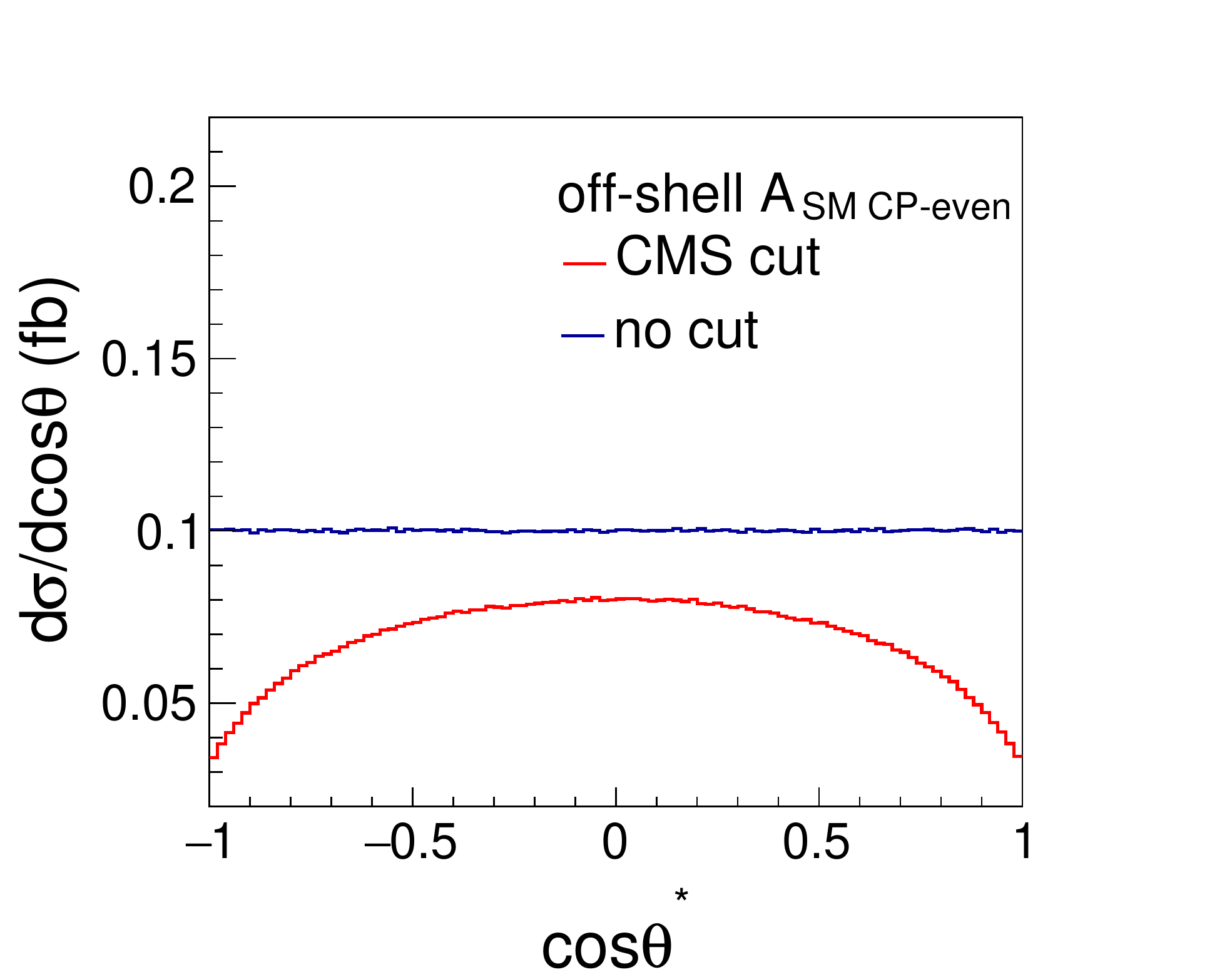}
		\\
		\includegraphics[width=3.6cm]{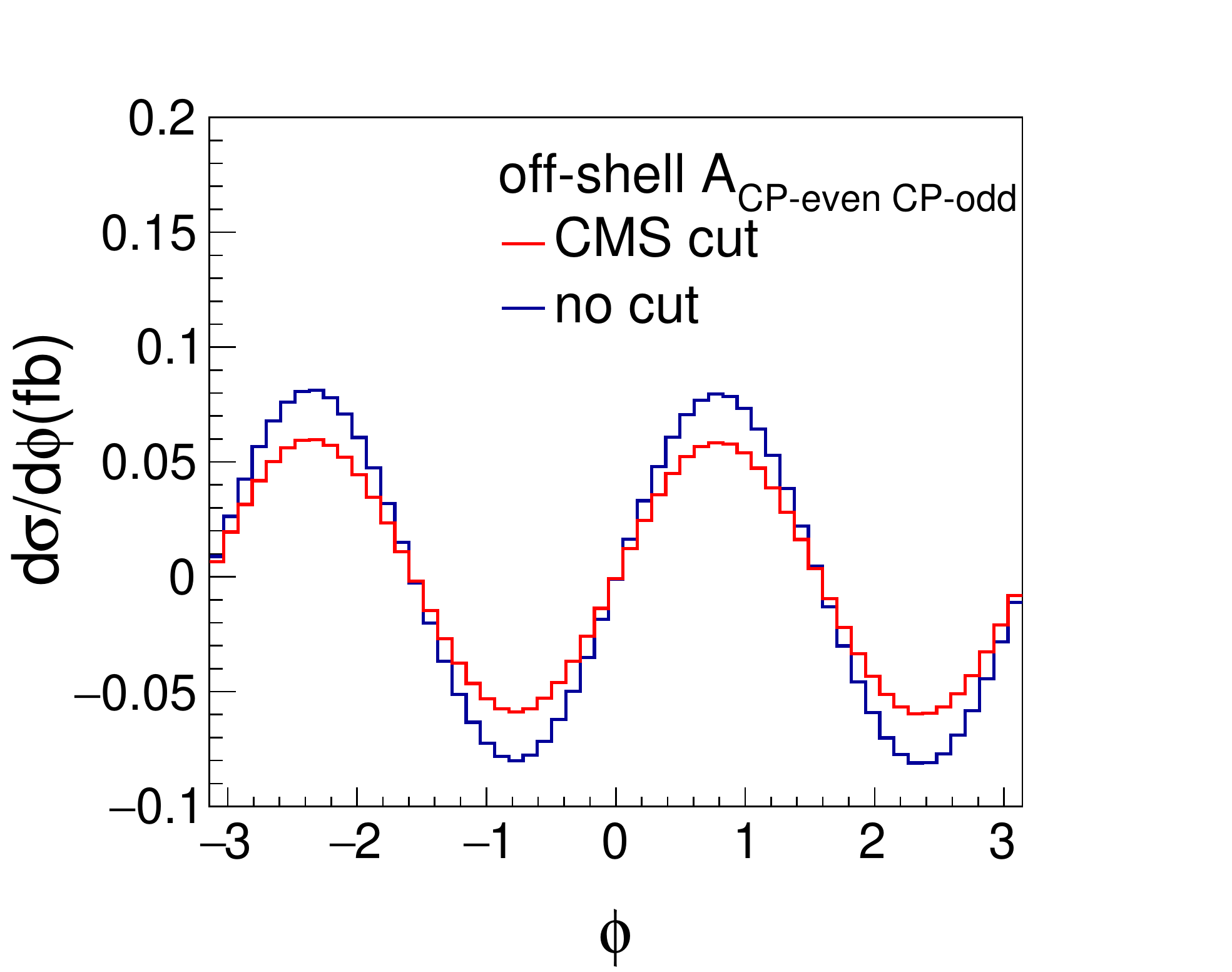}
		\includegraphics[width=3.6cm]{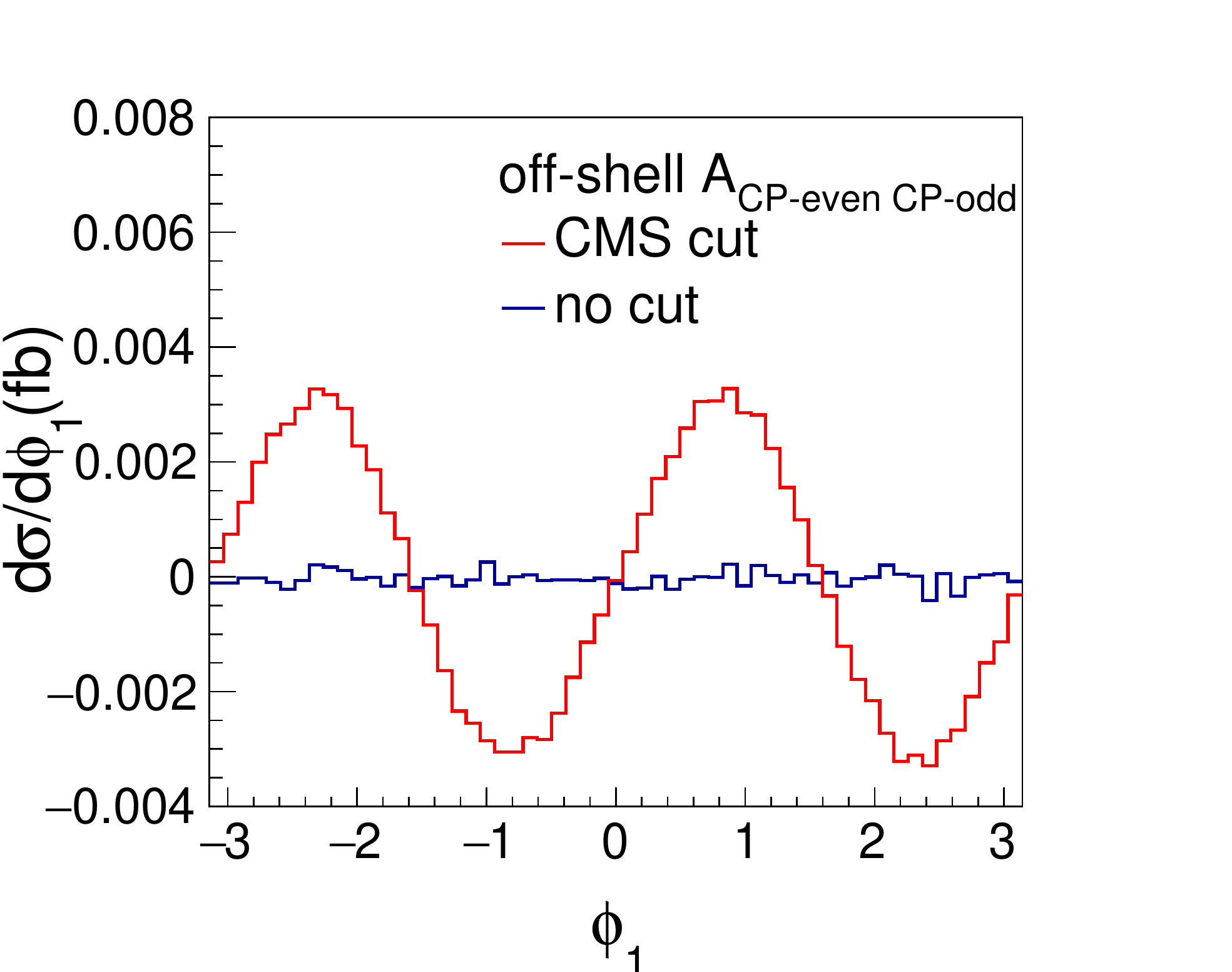}
		\includegraphics[width=3.6cm]{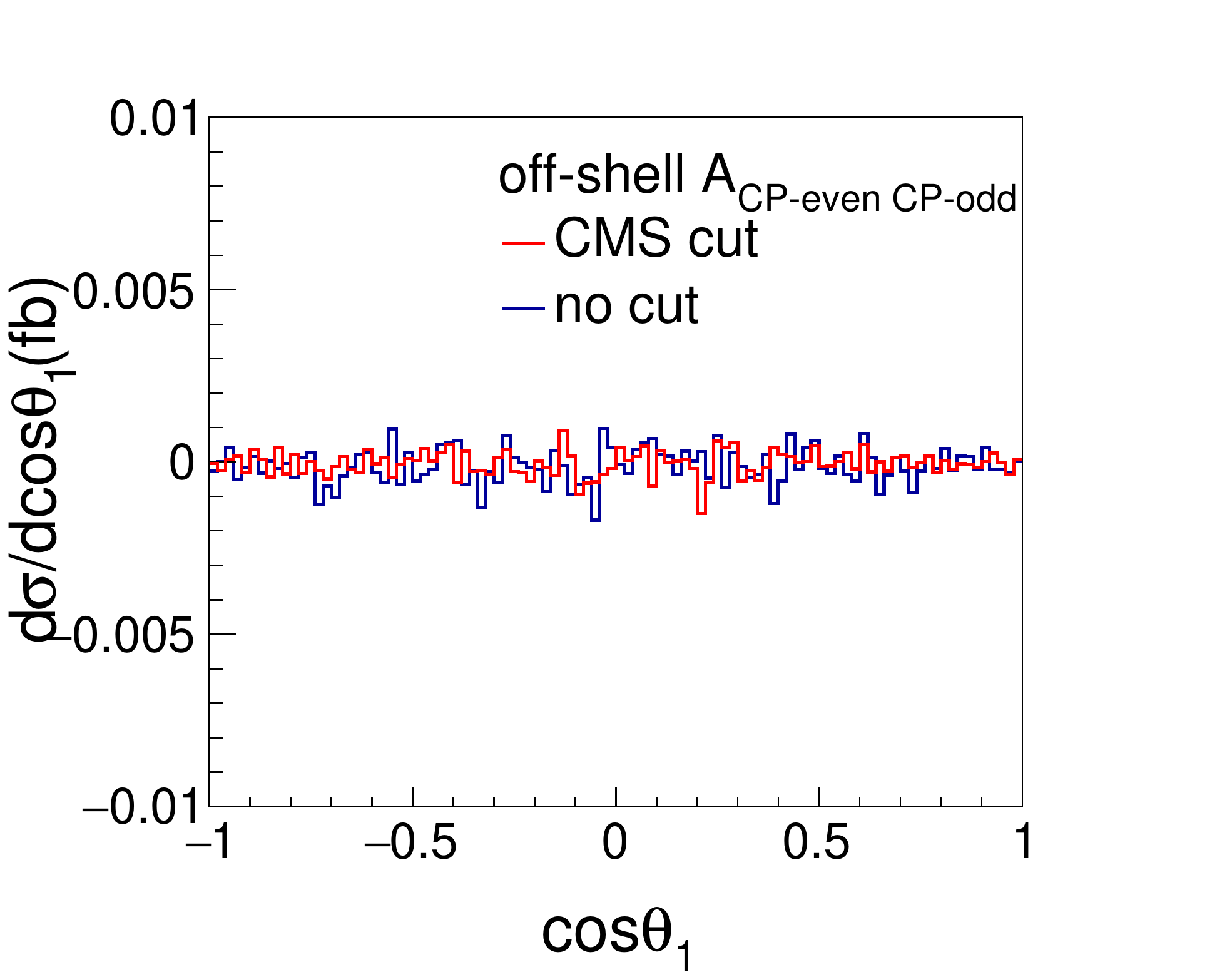}
		\includegraphics[width=3.6cm]{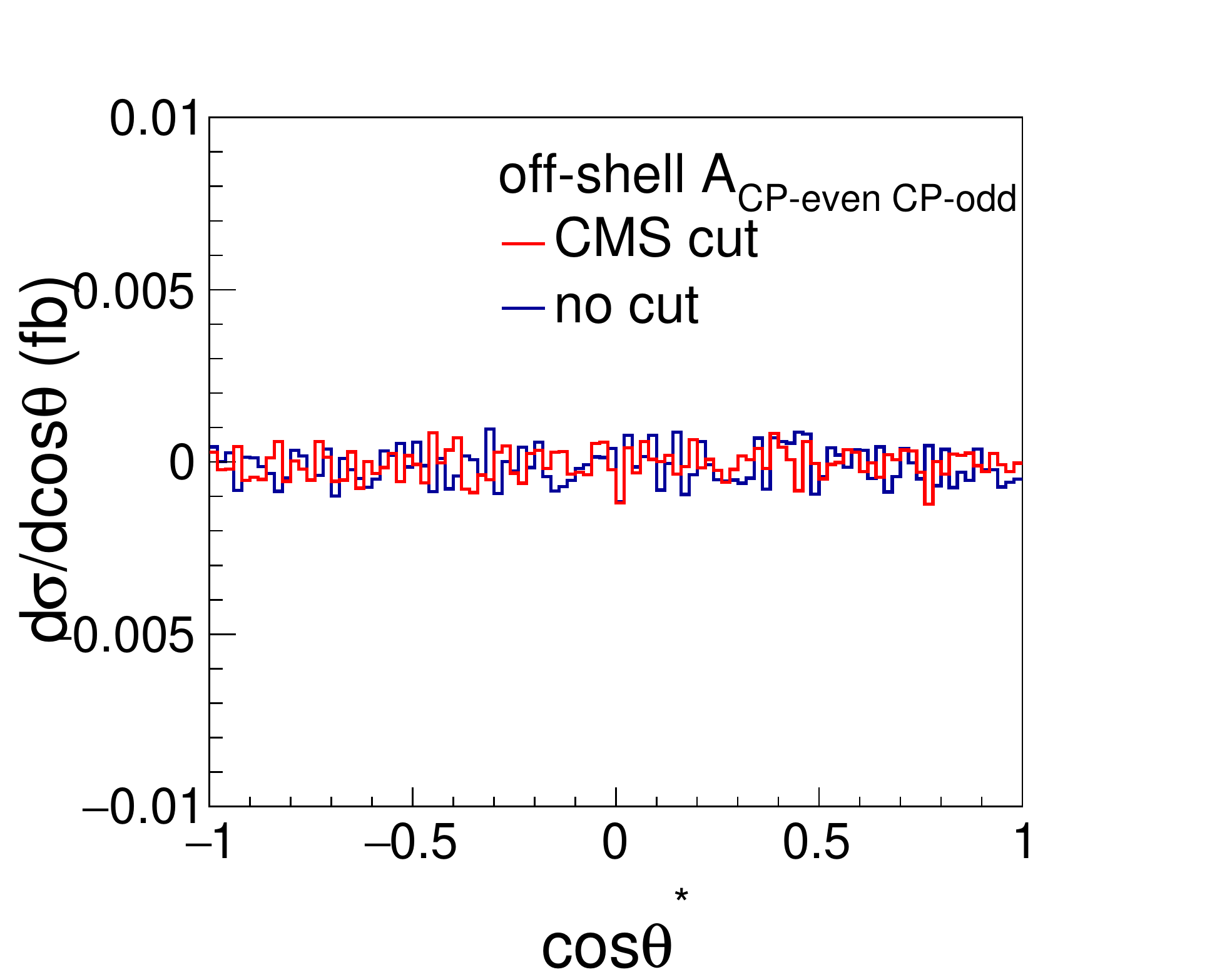}
		\\
		\includegraphics[width=3.6cm]{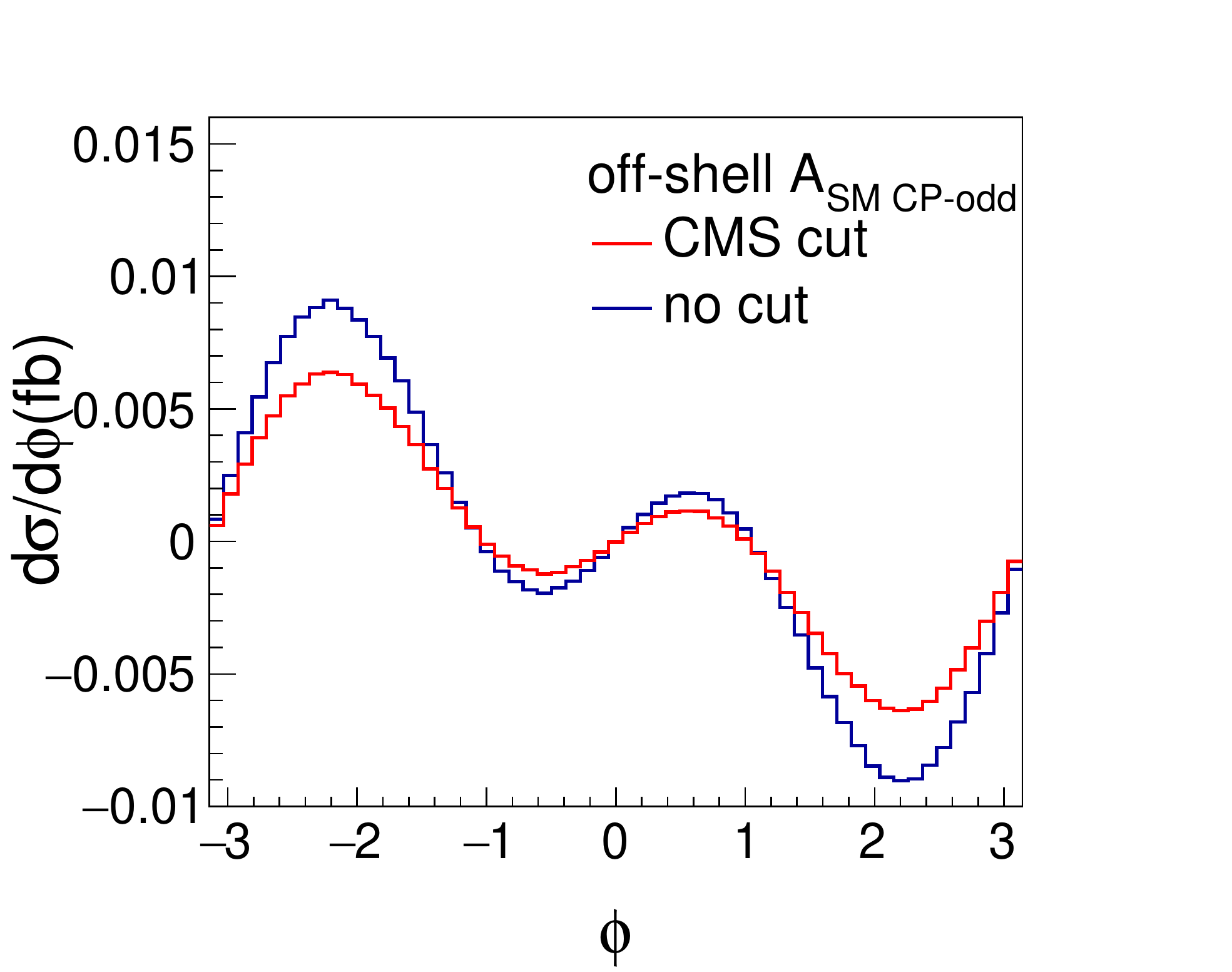}
		\includegraphics[width=3.6cm]{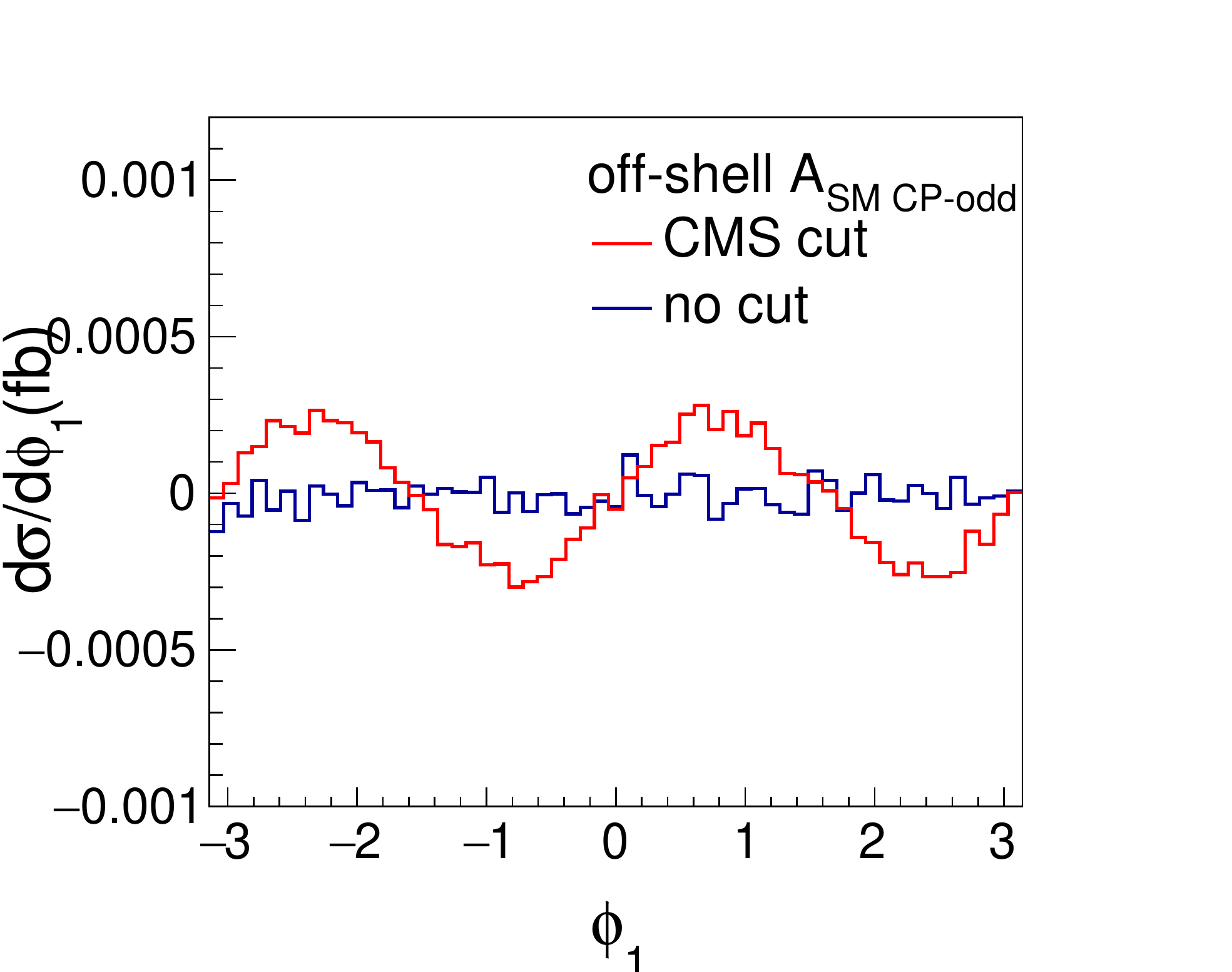}
		\includegraphics[width=3.6cm]{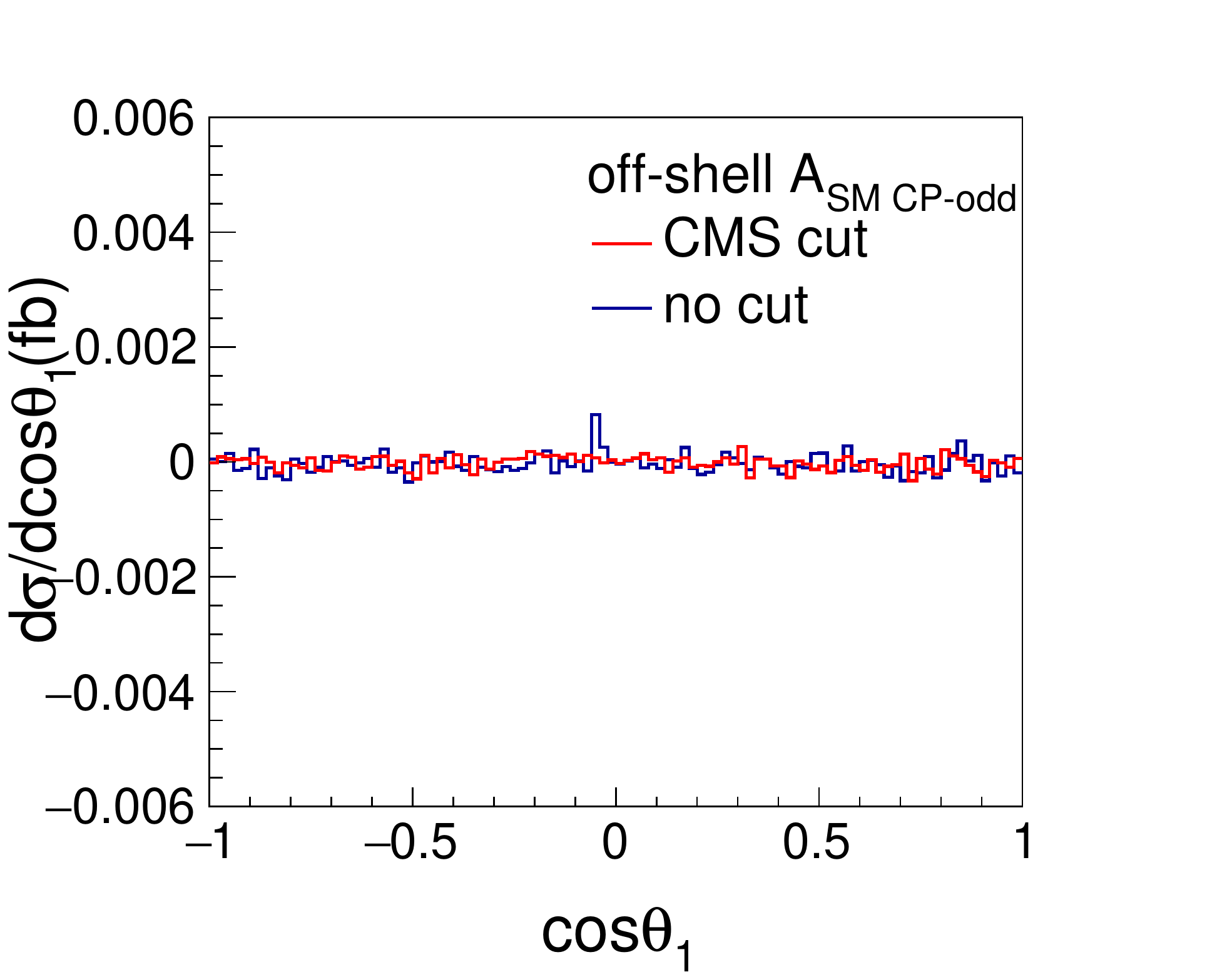}
		\includegraphics[width=3.6cm]{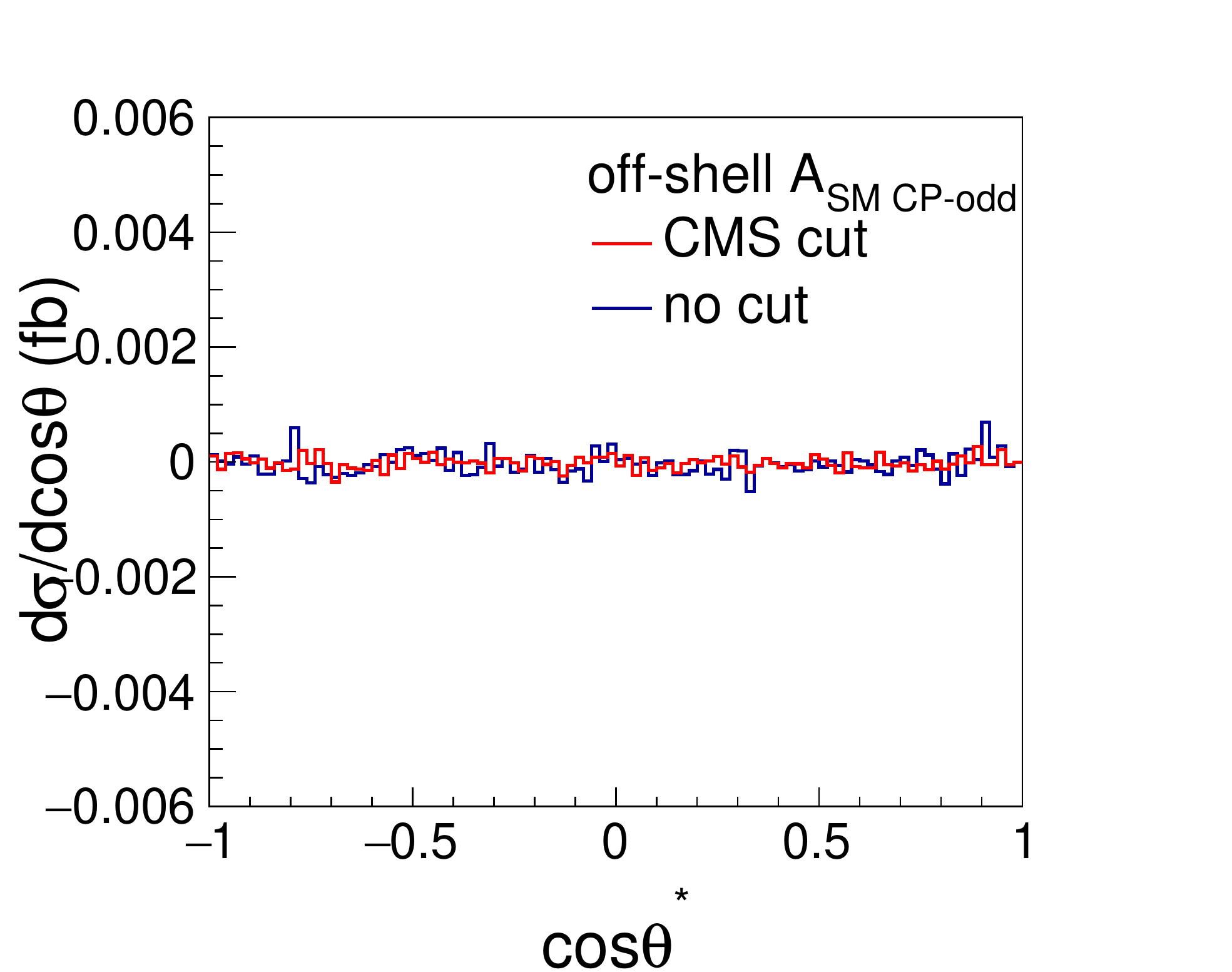}
	
		\caption{The angular differential cross sections from interference between Higgs mediated processes in Higgs off-shell region.}
		\label{inter-Higgs-off}
	\end{figure}

		\begin{figure}
		\centering
		\subfigure{\includegraphics[width=3.6cm]{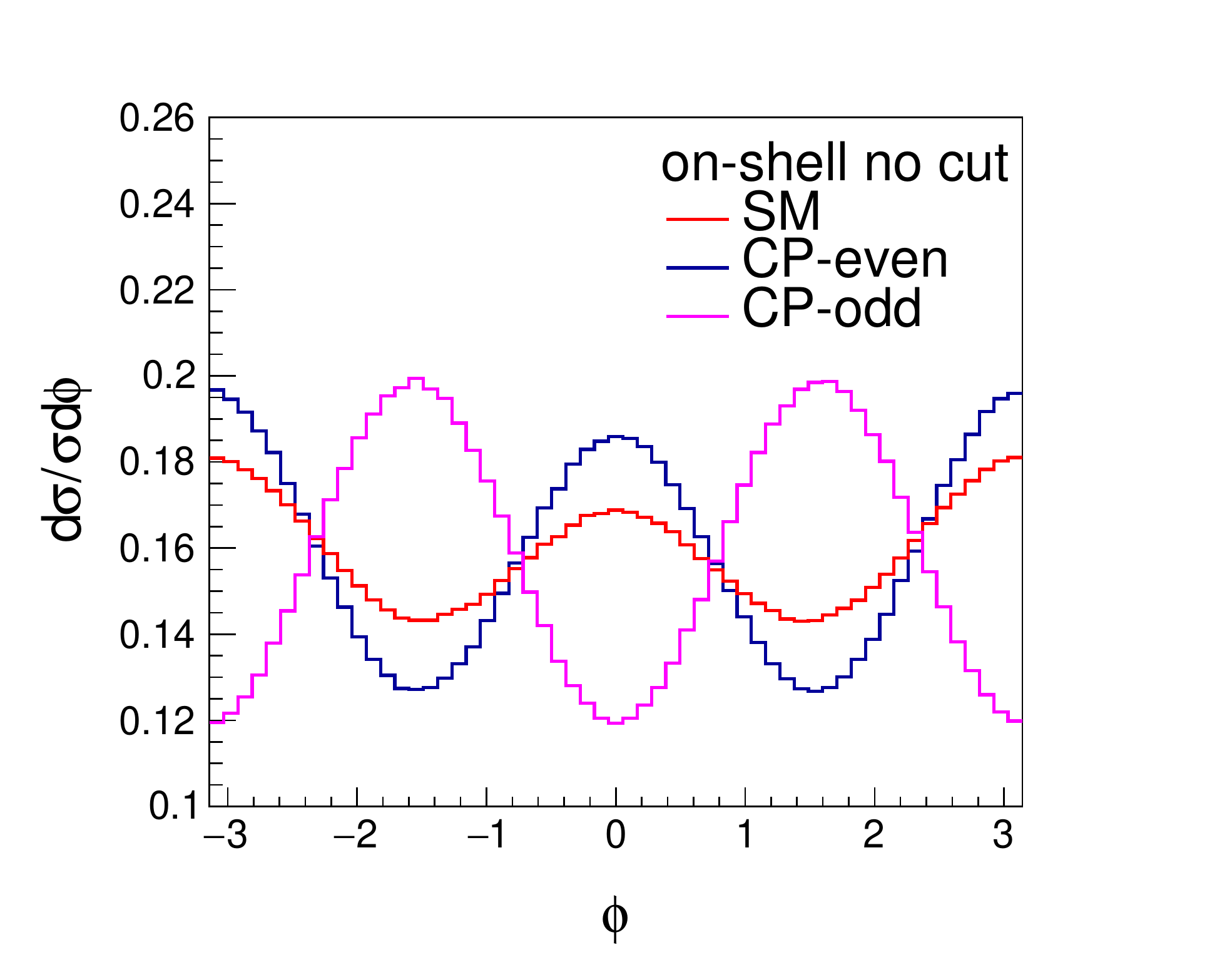}}
		\subfigure{\includegraphics[width=3.6cm]{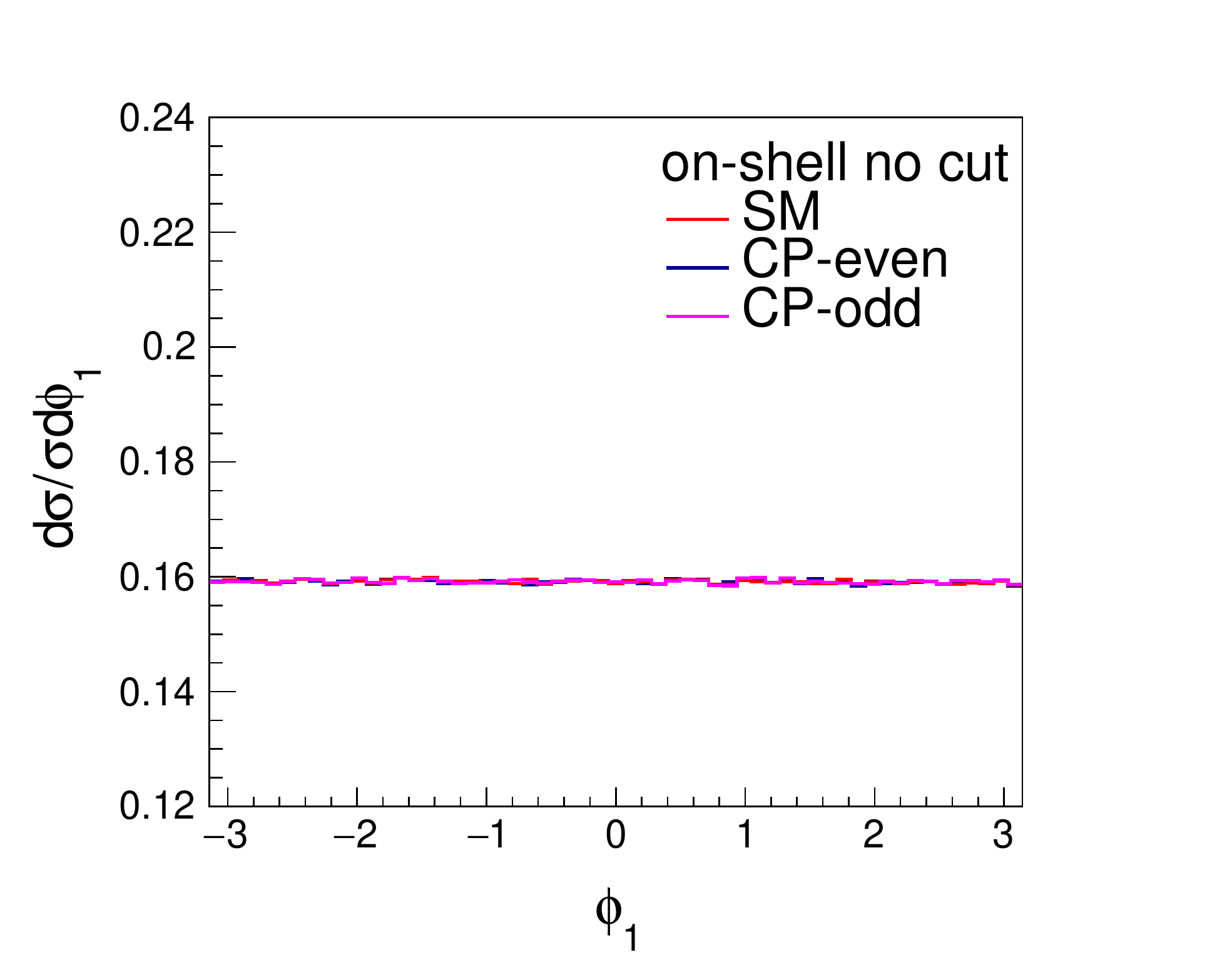}}
		\subfigure{\includegraphics[width=3.6cm]{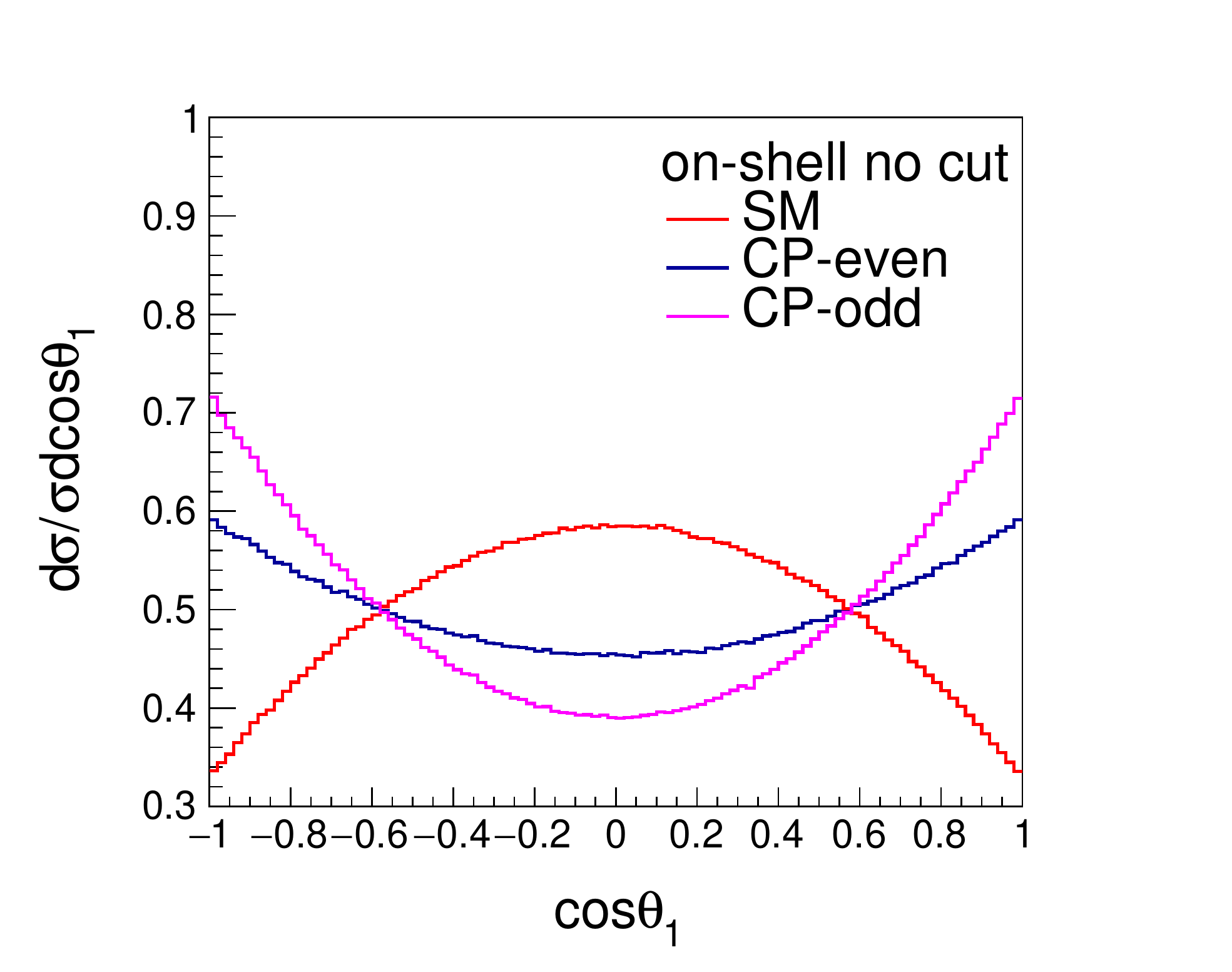}}
		\subfigure{\includegraphics[width=3.6cm]{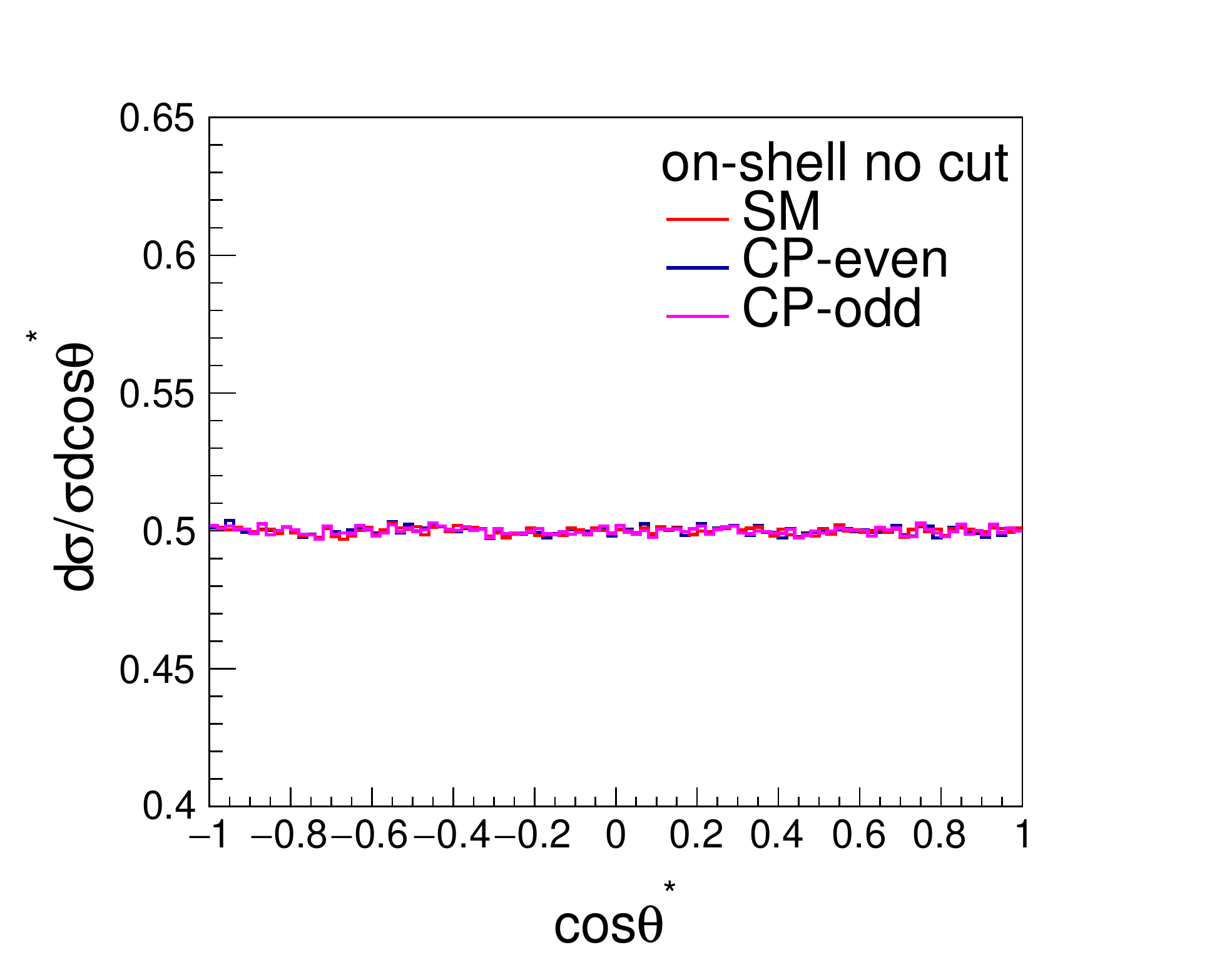}}
		\\
		\subfigure{\includegraphics[width=3.6cm]{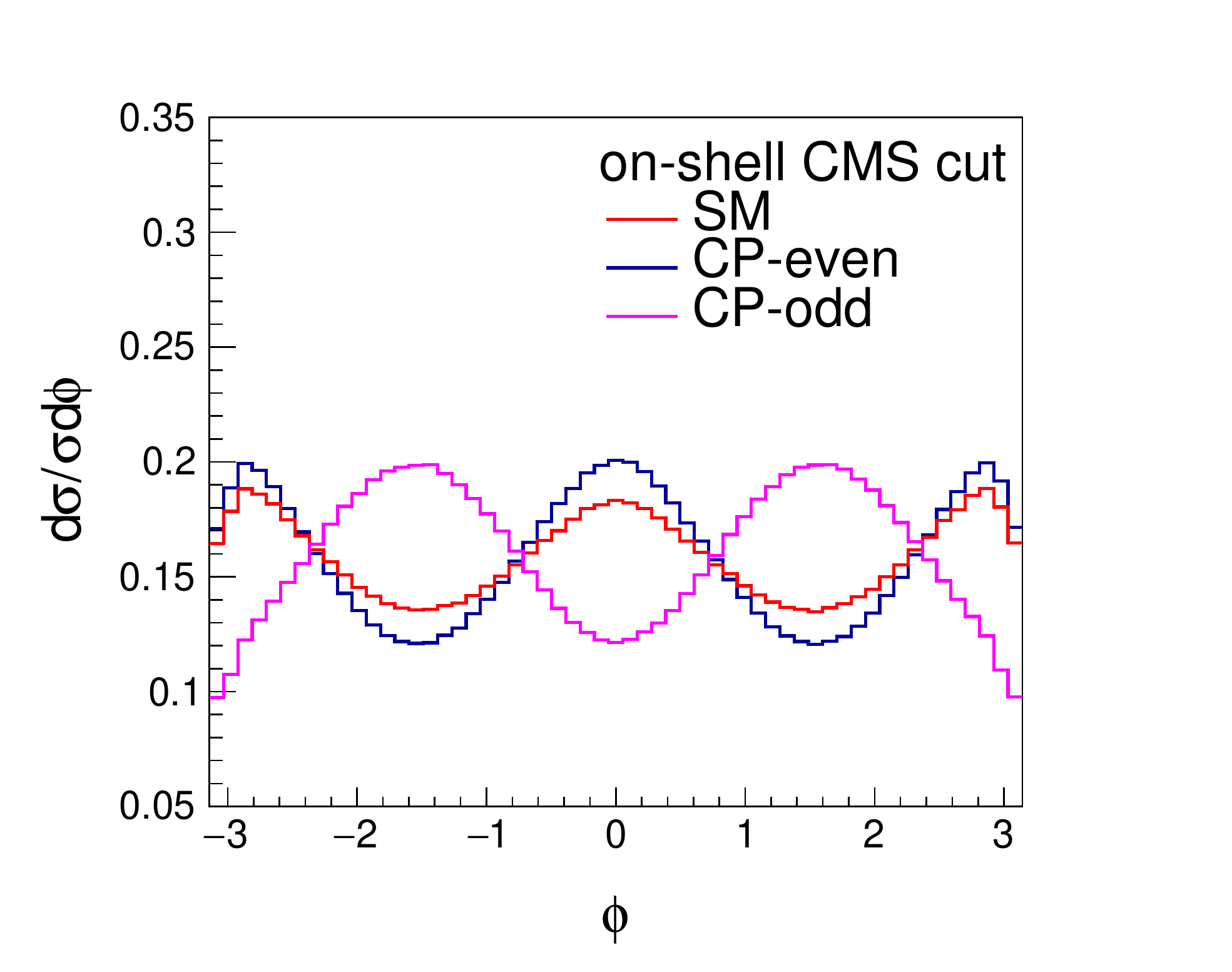}}	
		\subfigure{\includegraphics[width=3.6cm]{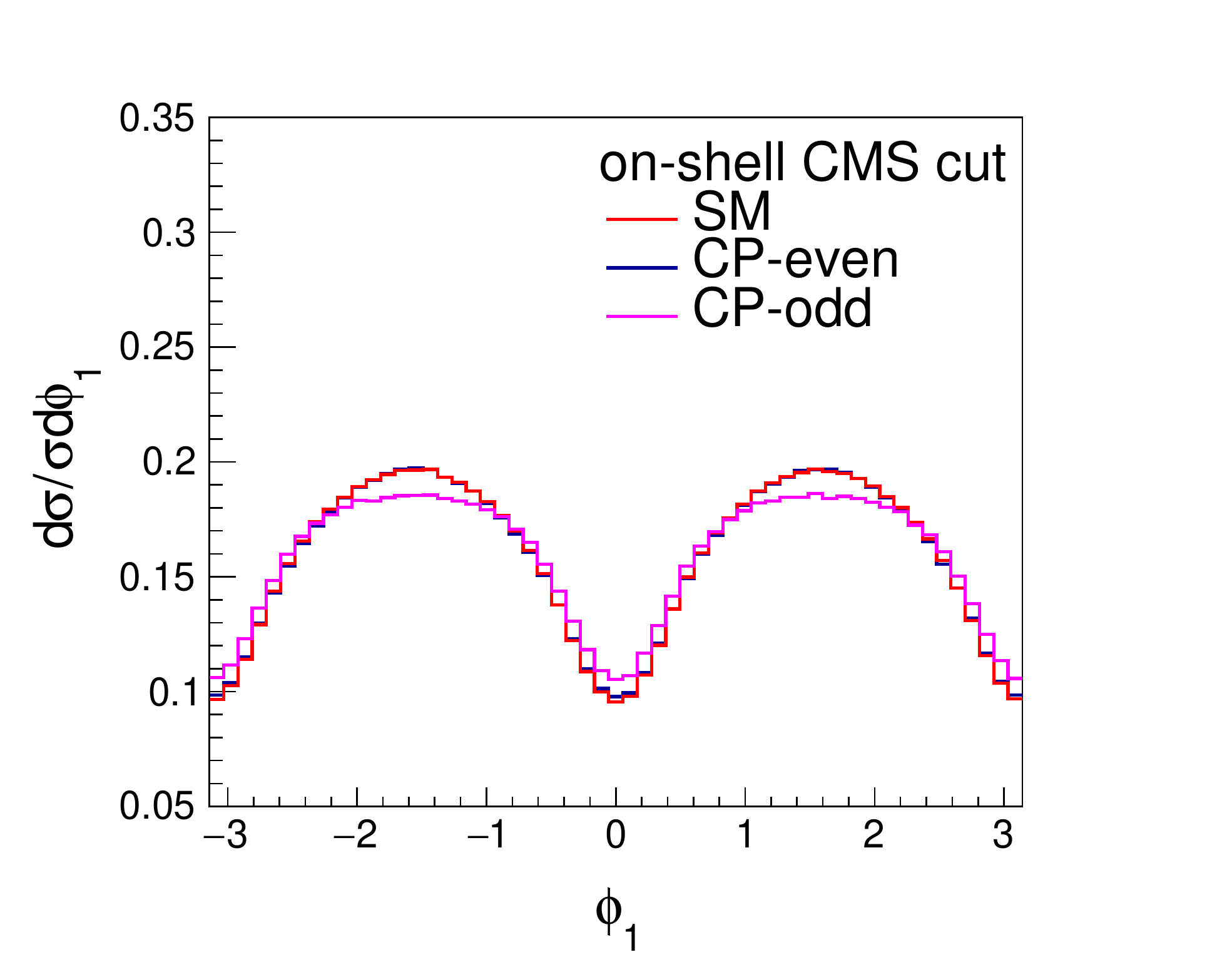}}	
		\subfigure{\includegraphics[width=3.6cm]{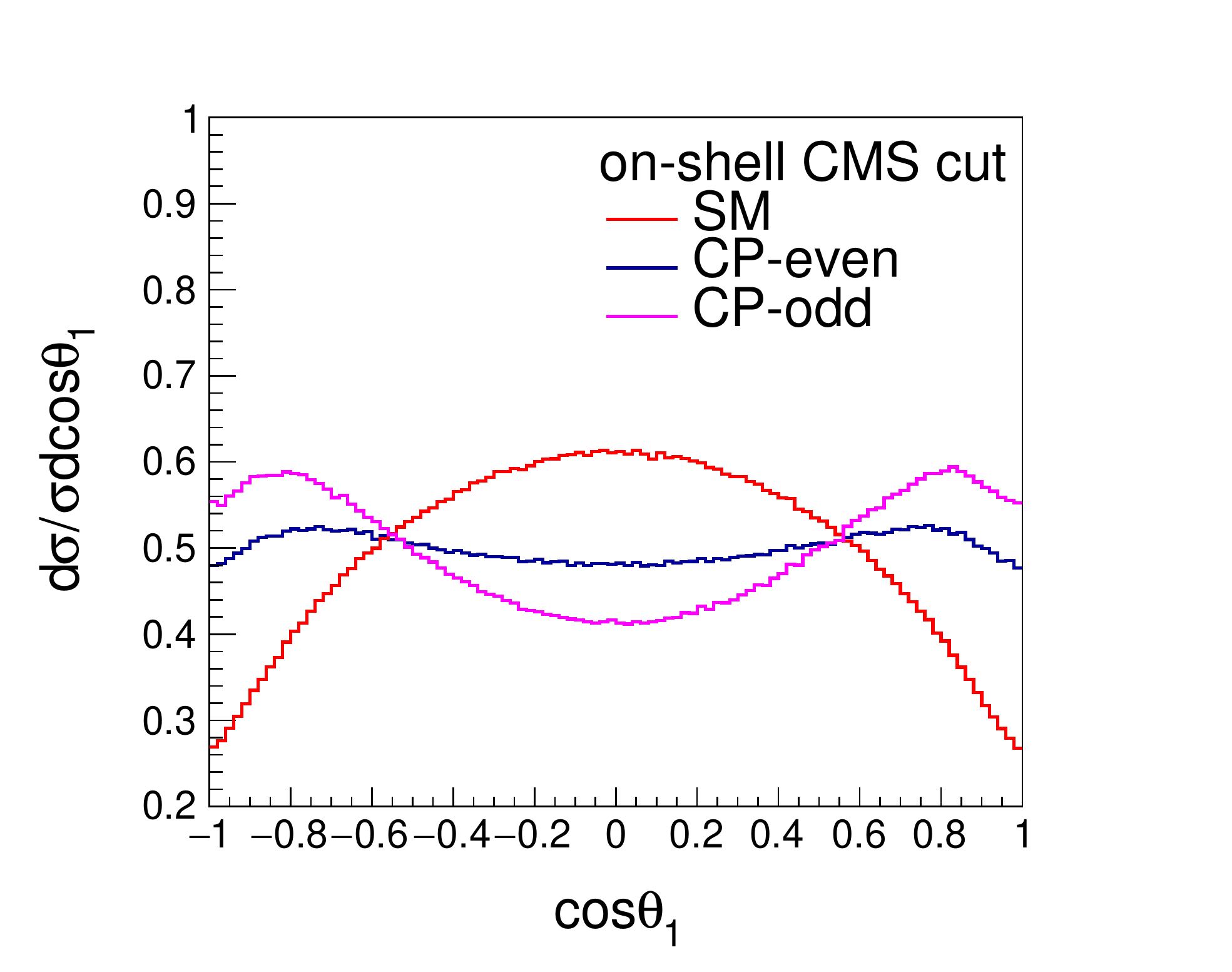}}		
		\subfigure{\includegraphics[width=3.6cm]{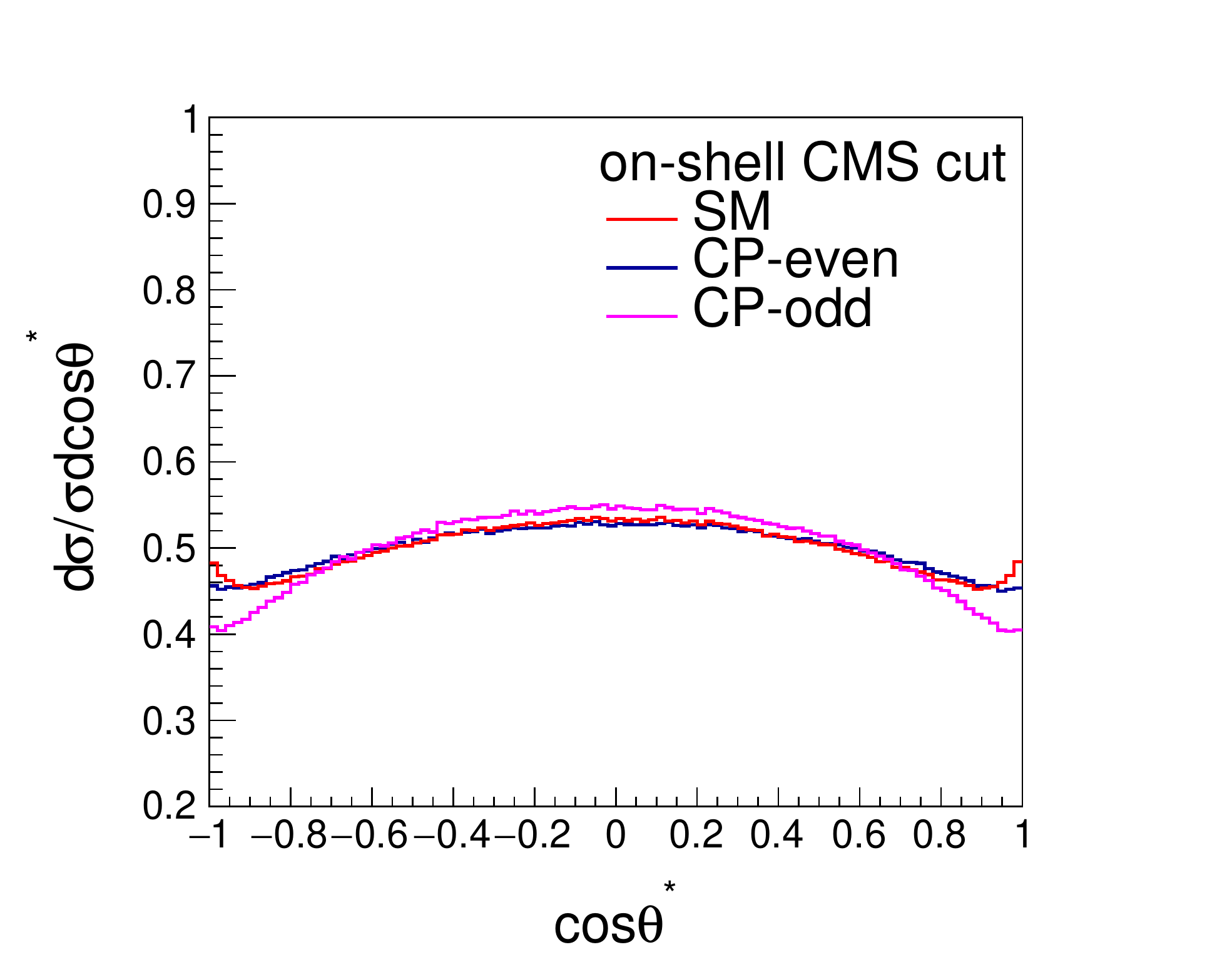}}

		\caption{The angular differential cross sections from angular distribution of signal processes with and without basic CMS cuts in Higgs on-shell region.}
		\label{signal-on}
	\end{figure}

	\begin{figure}
		
		\centering
		\subfigure{\includegraphics[width=3.6cm]{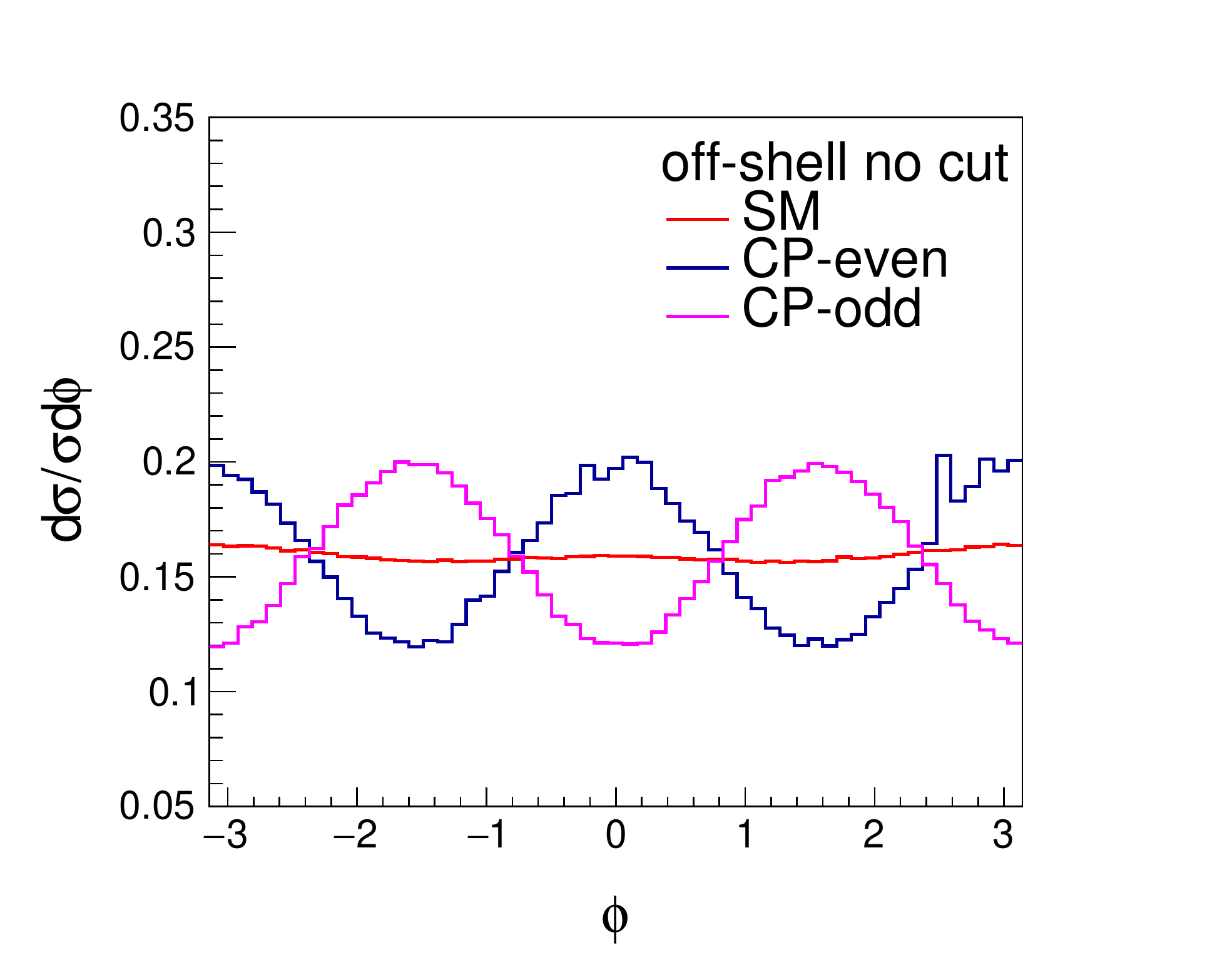}}
		\subfigure{\includegraphics[width=3.6cm]{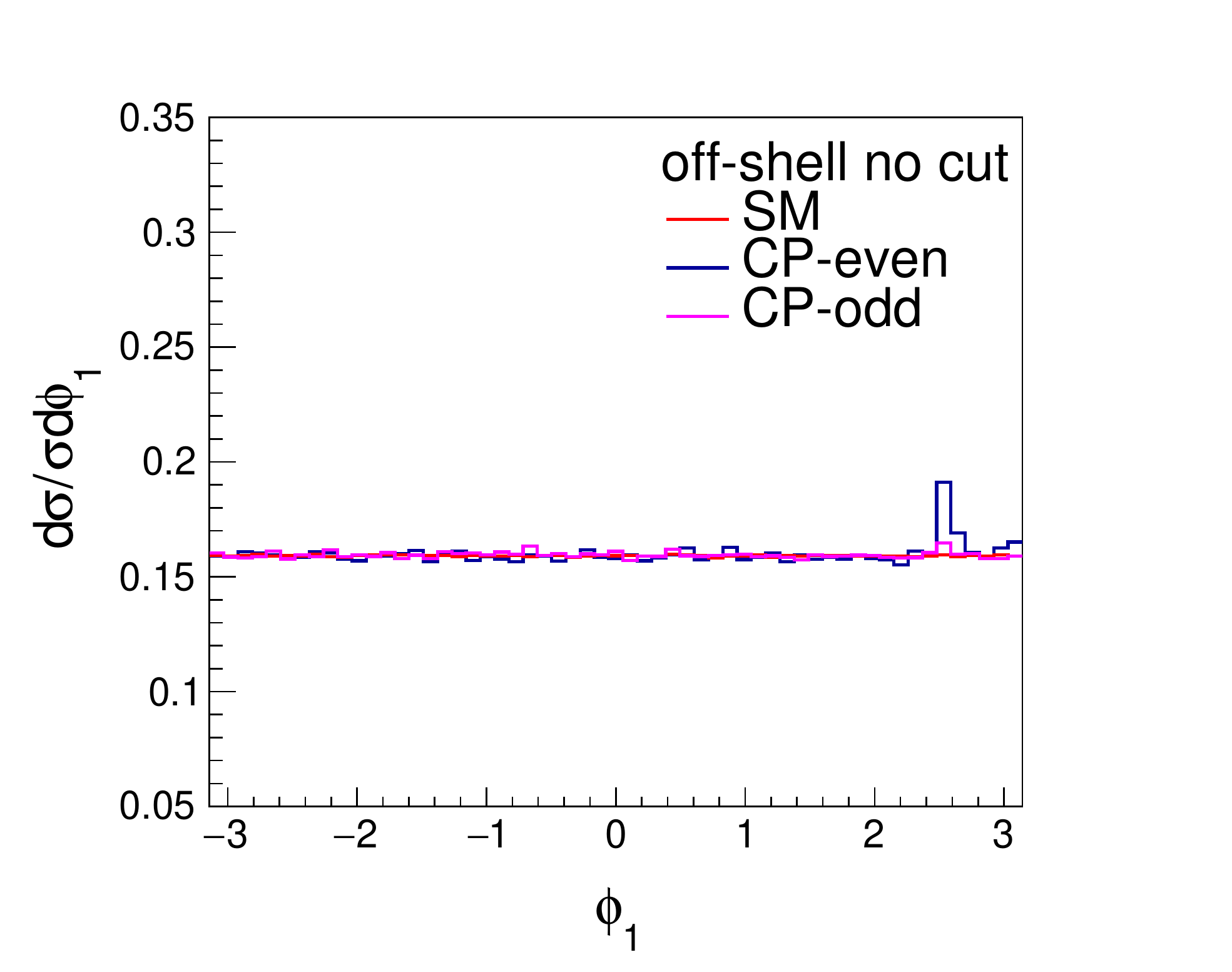}}
		\subfigure{\includegraphics[width=3.6cm]{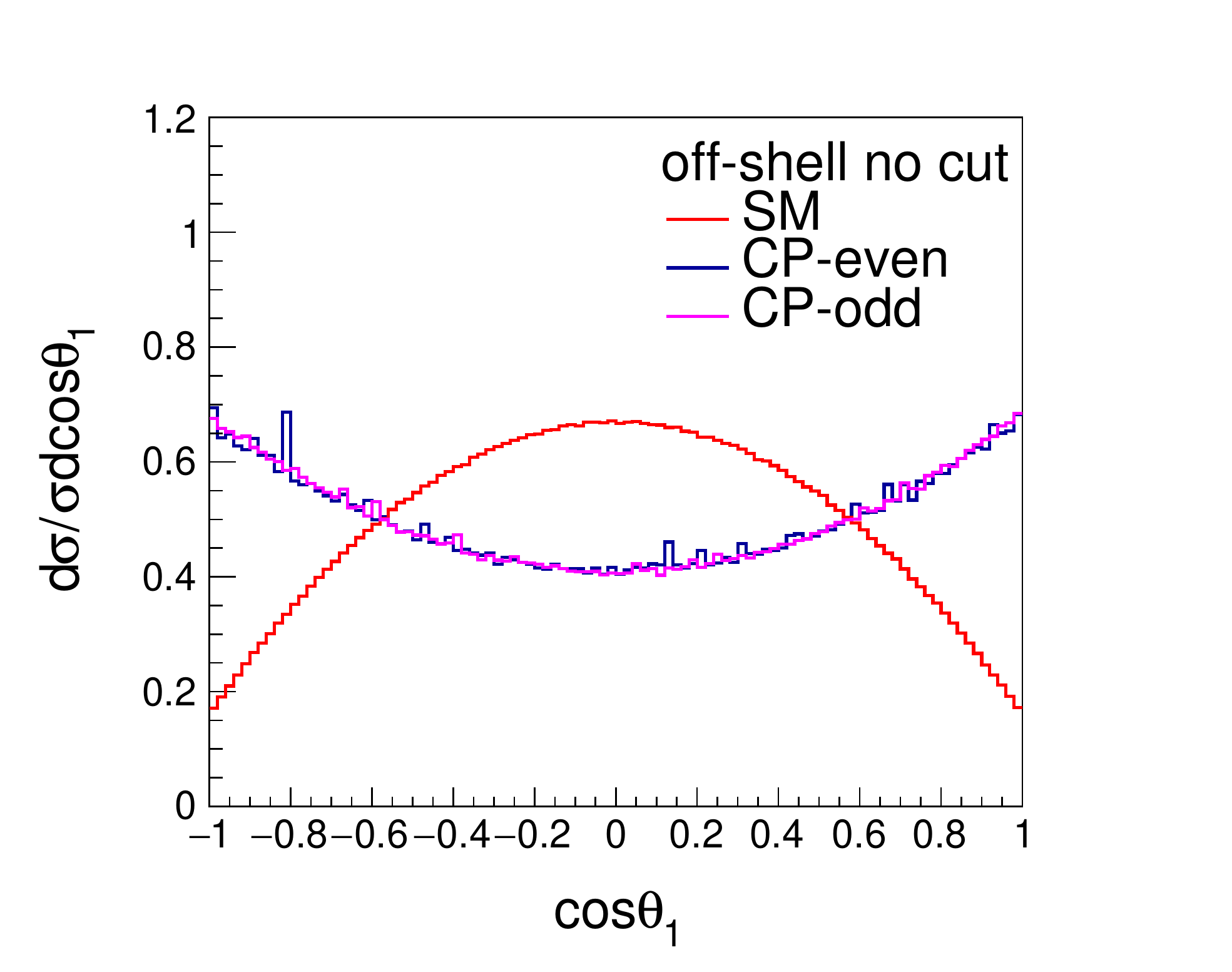}}
		\subfigure{\includegraphics[width=3.6cm]{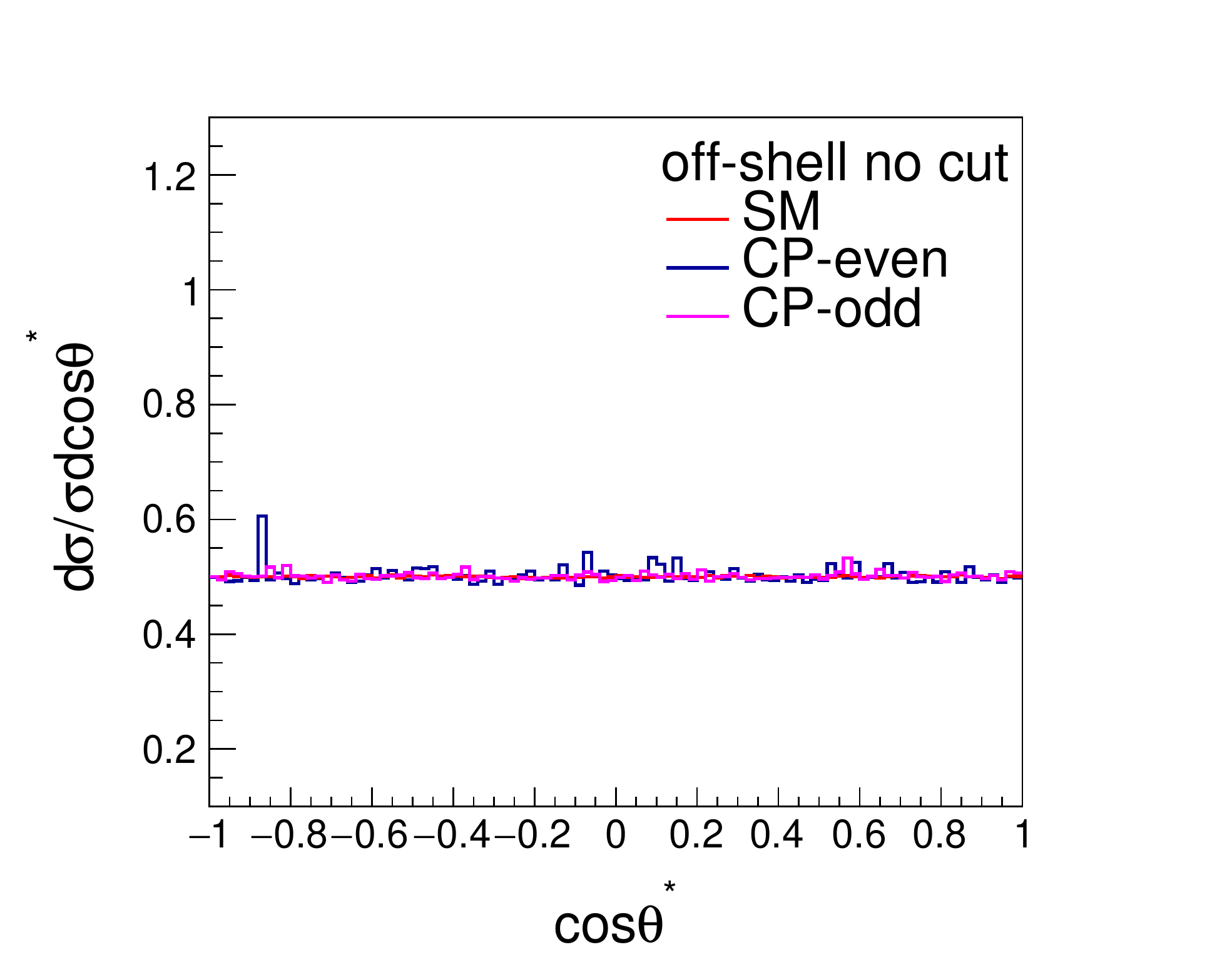}}
		\\
		\subfigure{\includegraphics[width=3.6cm]{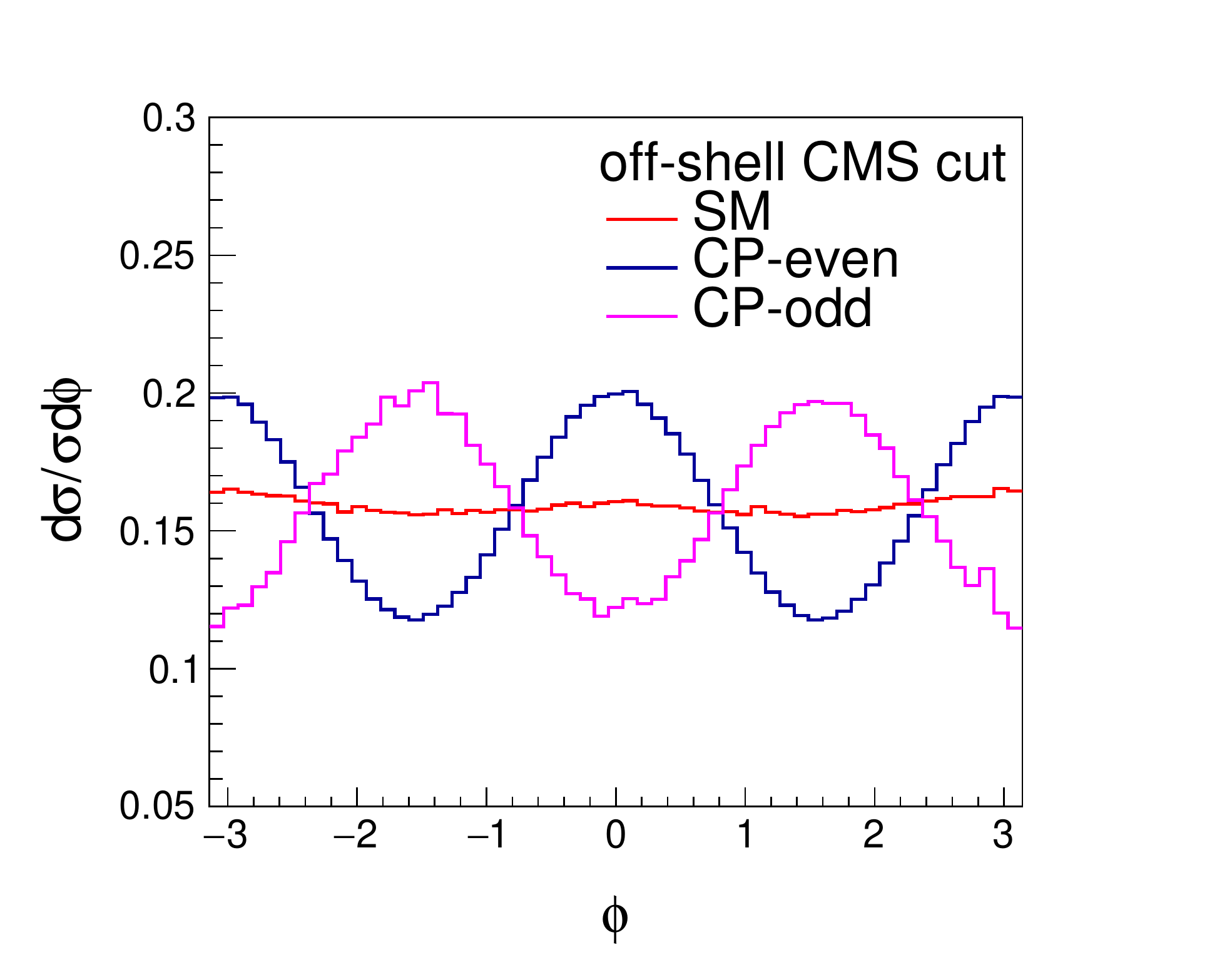}}
		\subfigure{\includegraphics[width=3.6cm]{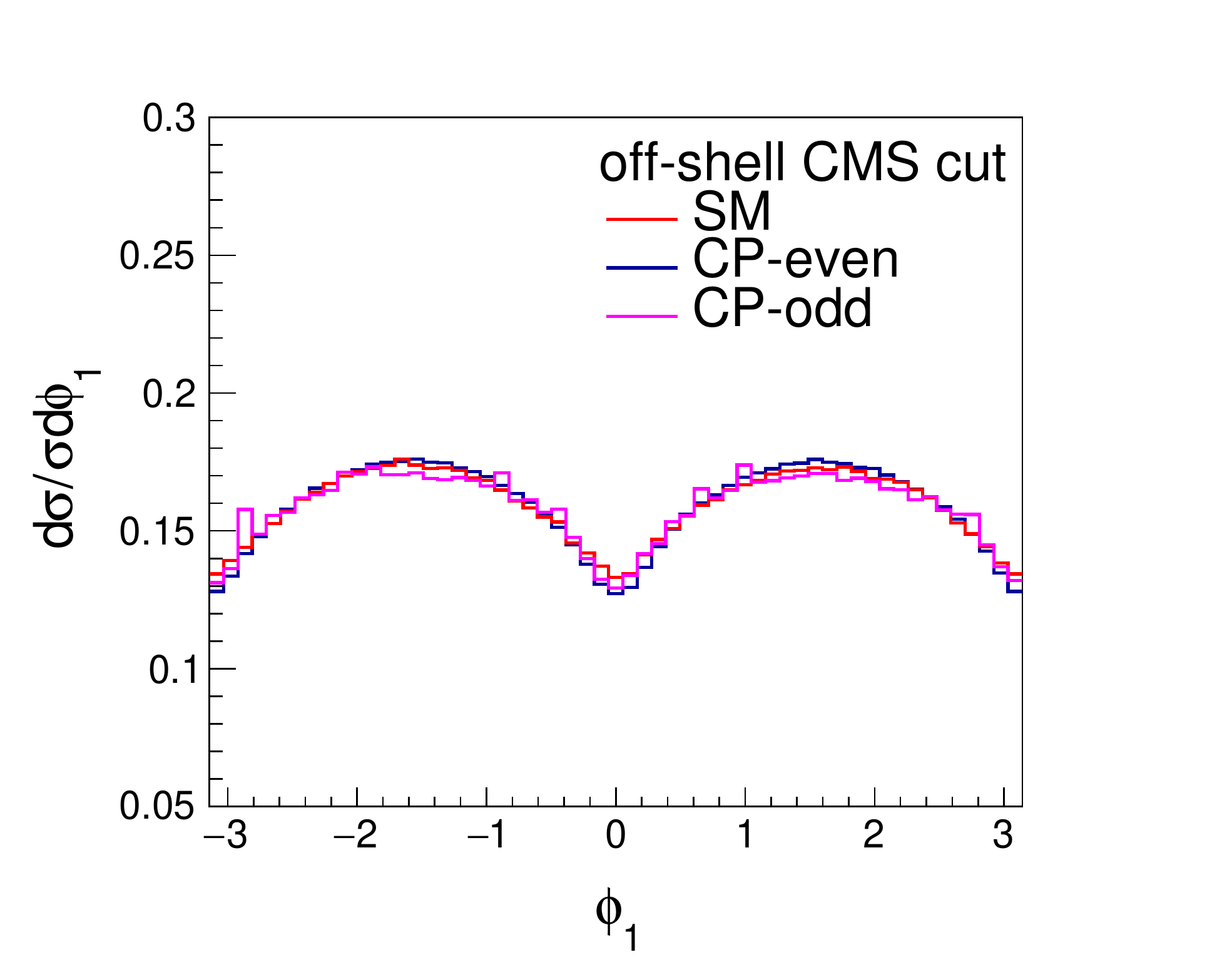}}
		\subfigure{\includegraphics[width=3.6cm]{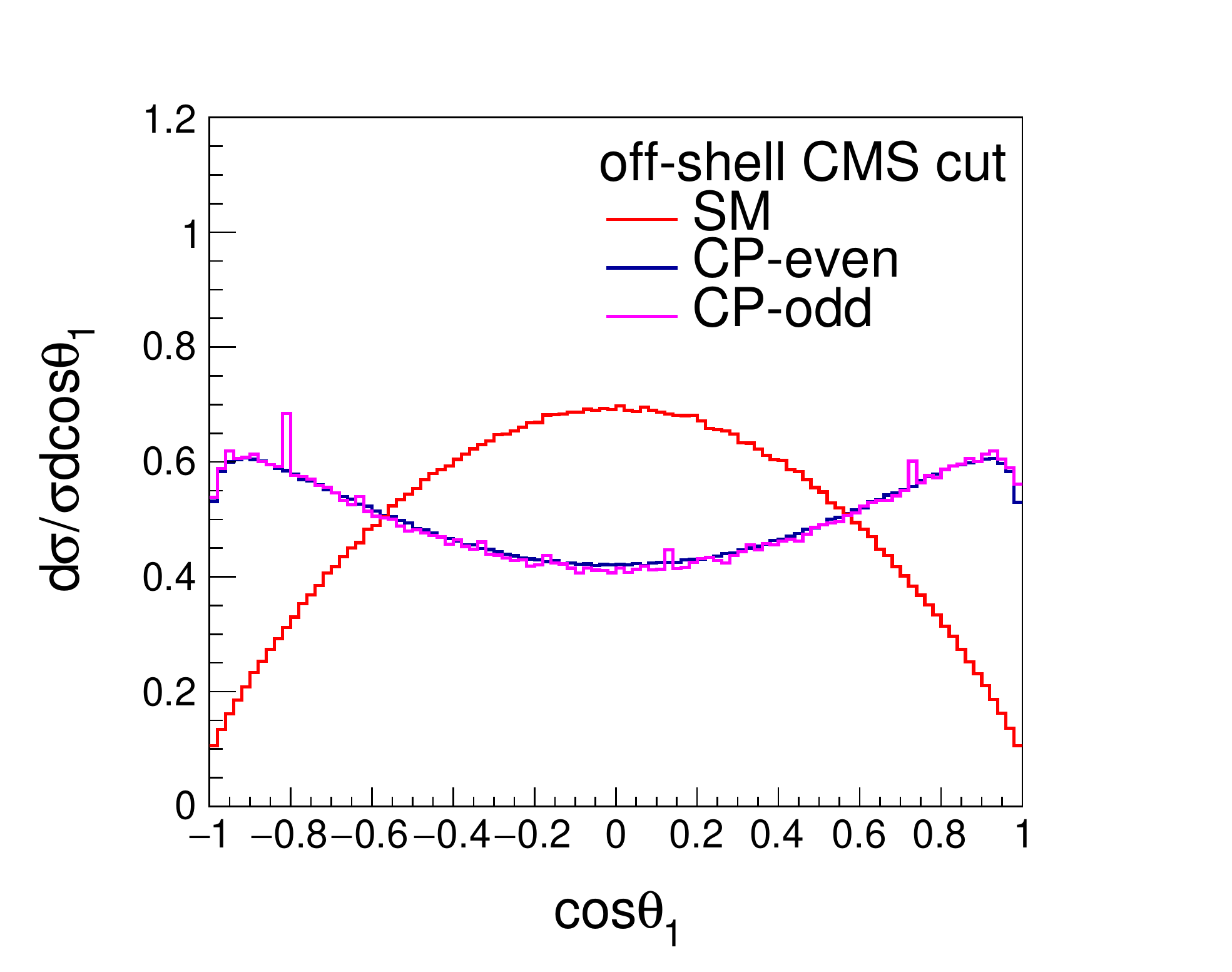}}
		\subfigure{\includegraphics[width=3.6cm]{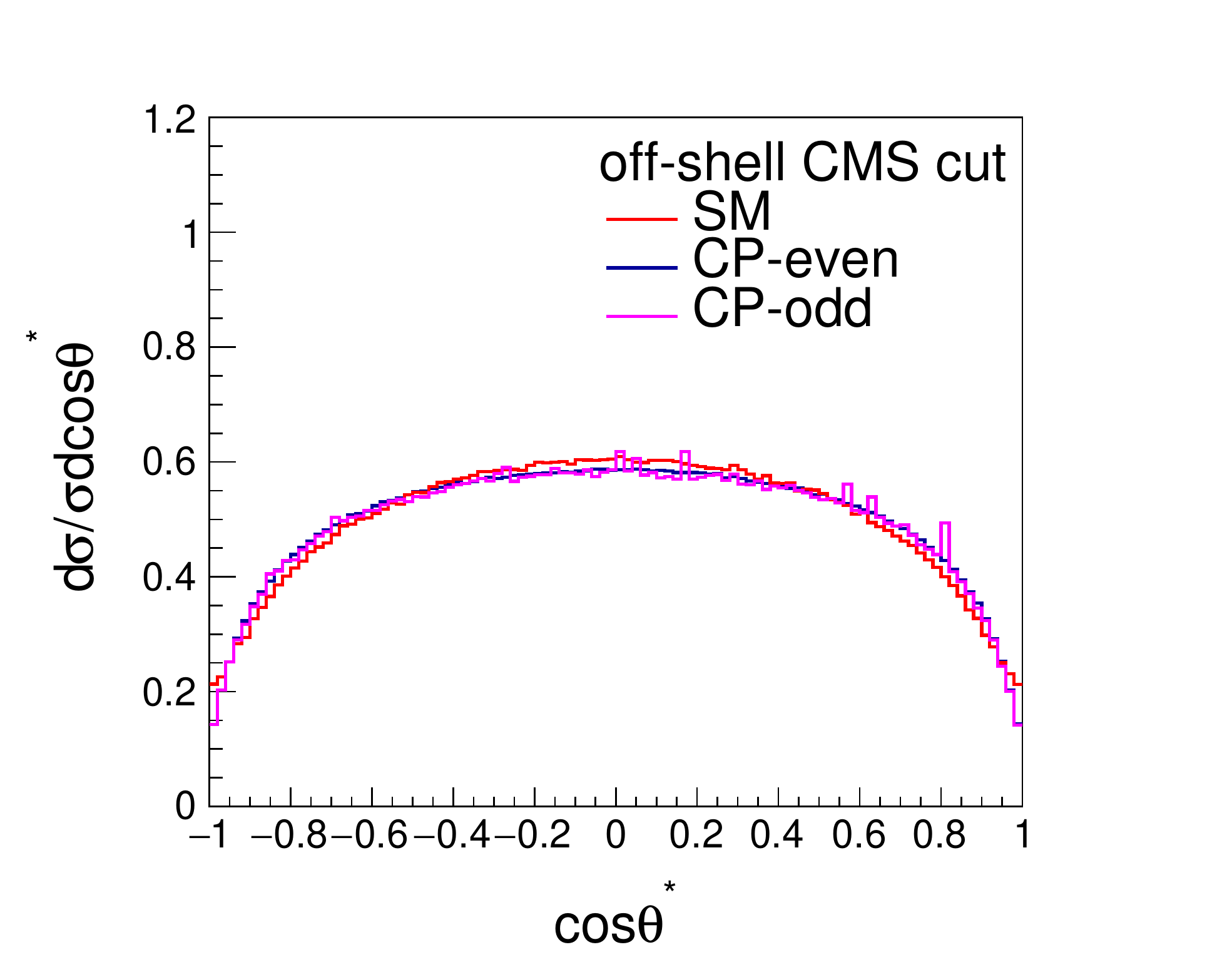}}

		\caption{The angular differential cross sections from angular distribution of signal processes with and without basic CMS cuts in Higgs off-shell region.}
		\label{signal-off}
	\end{figure}

\begin{figure}
	\centering
	\includegraphics[width=3.6cm]{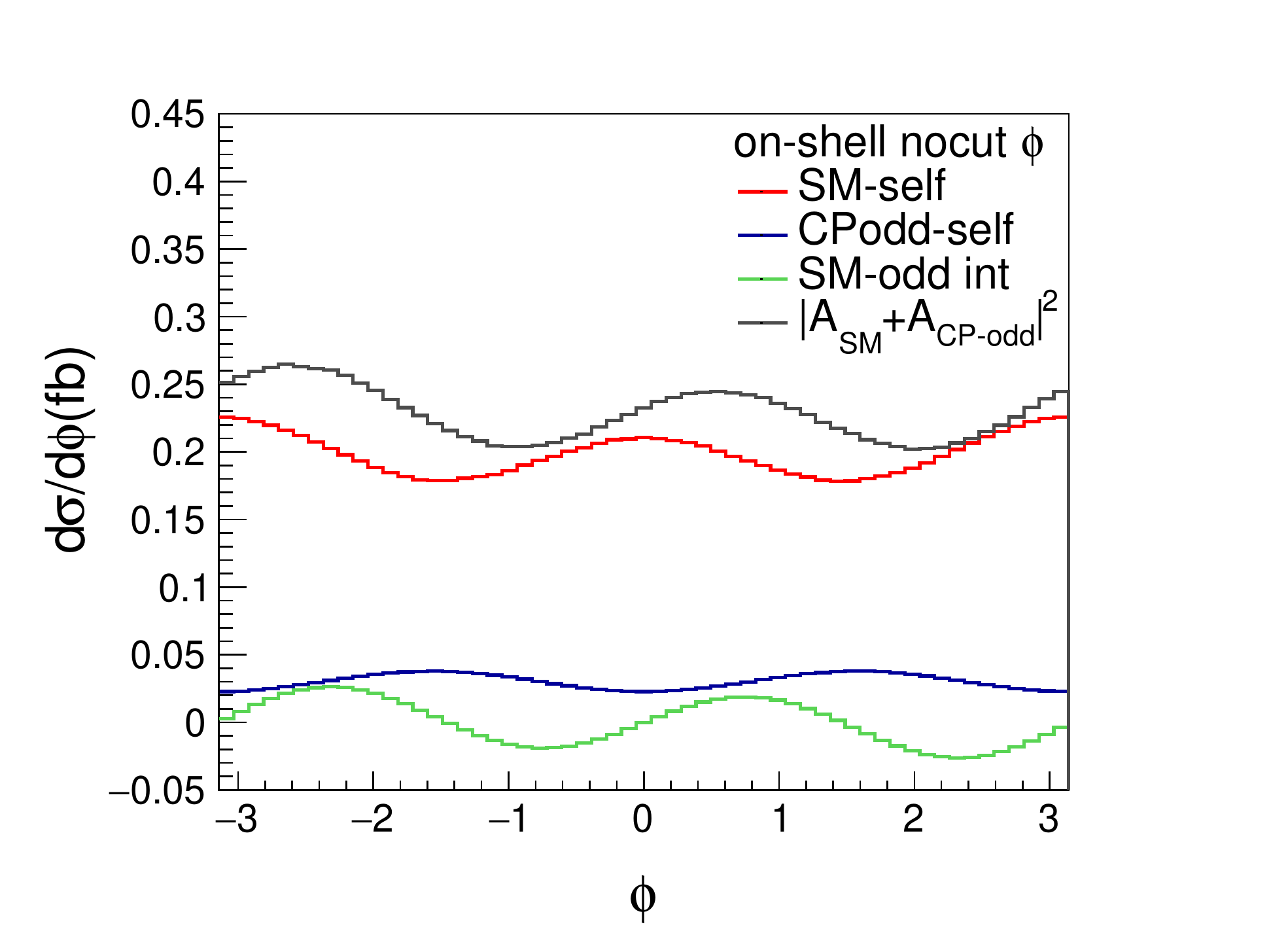}
	\includegraphics[width=3.6cm]{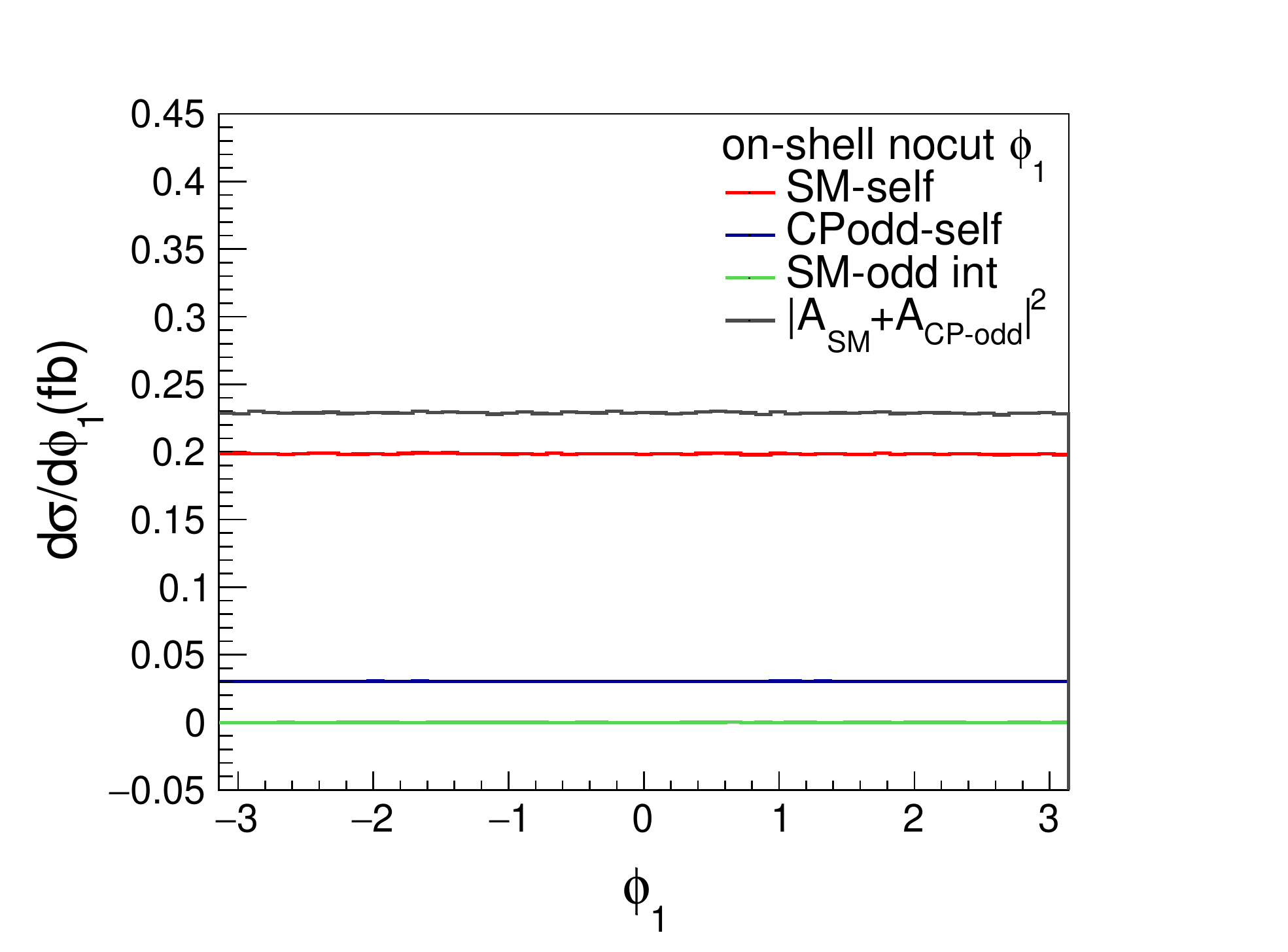}
	\includegraphics[width=3.6cm]{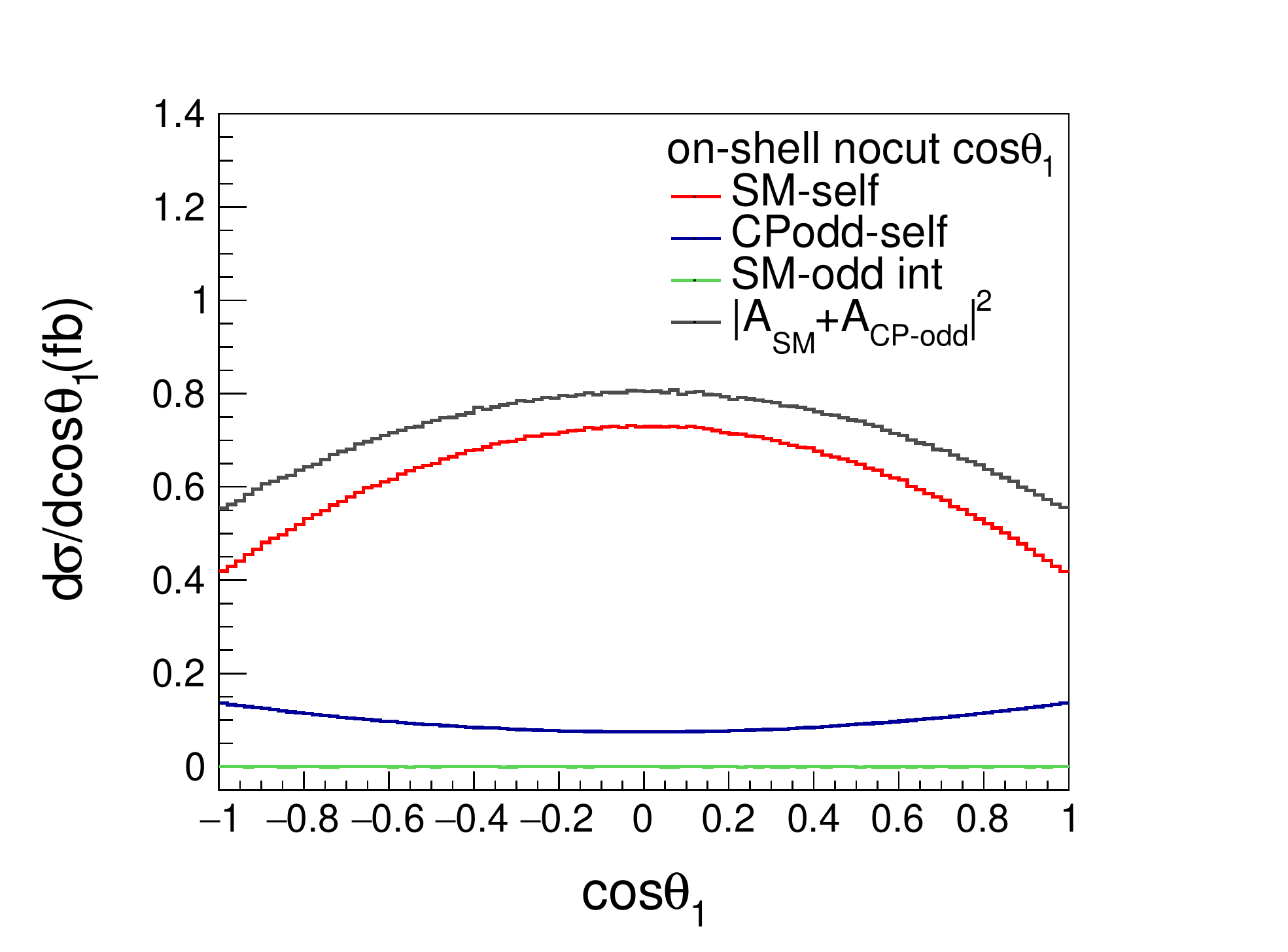}
	\includegraphics[width=3.6cm]{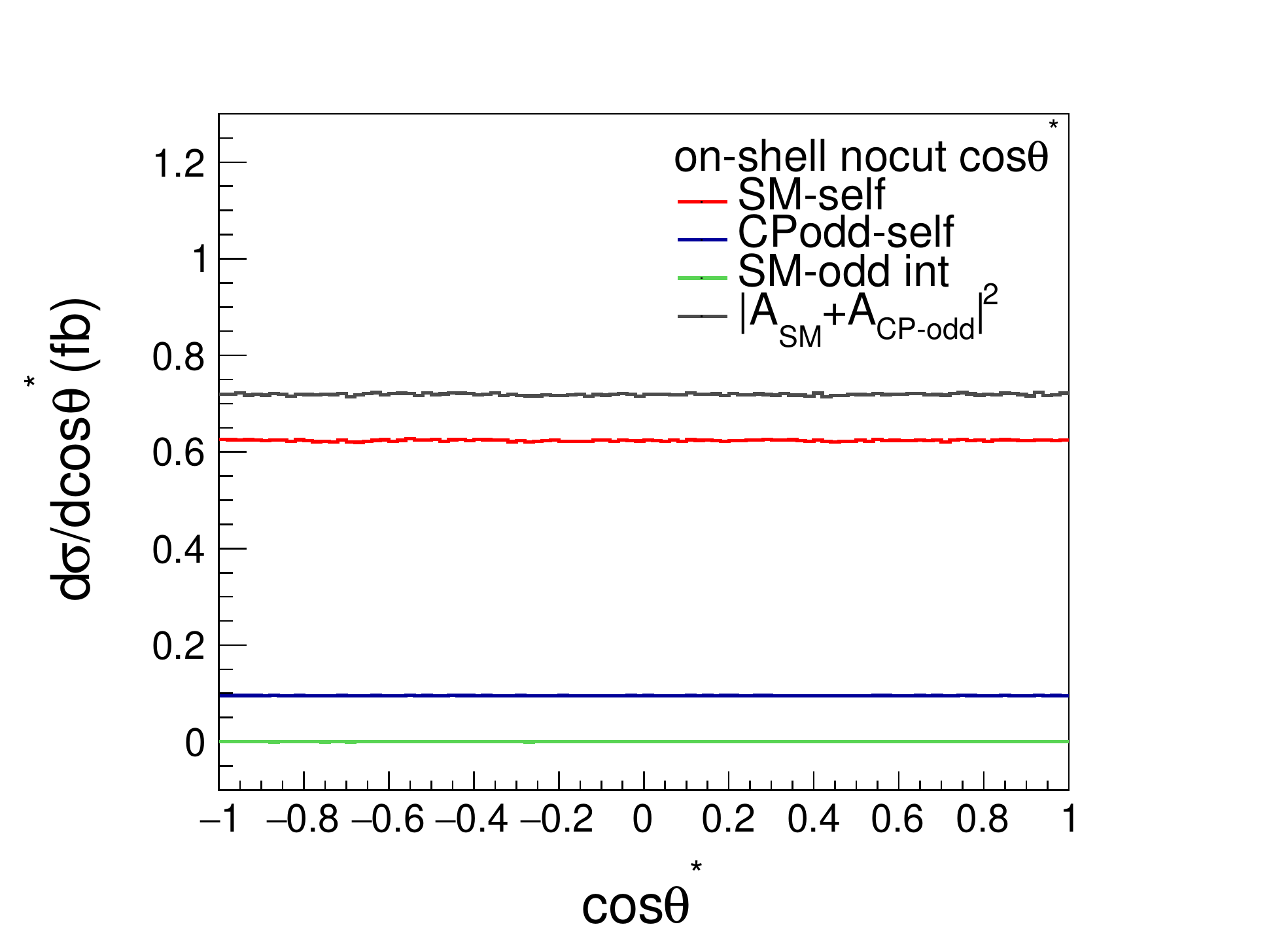}
	\\
    \includegraphics[width=3.6cm]{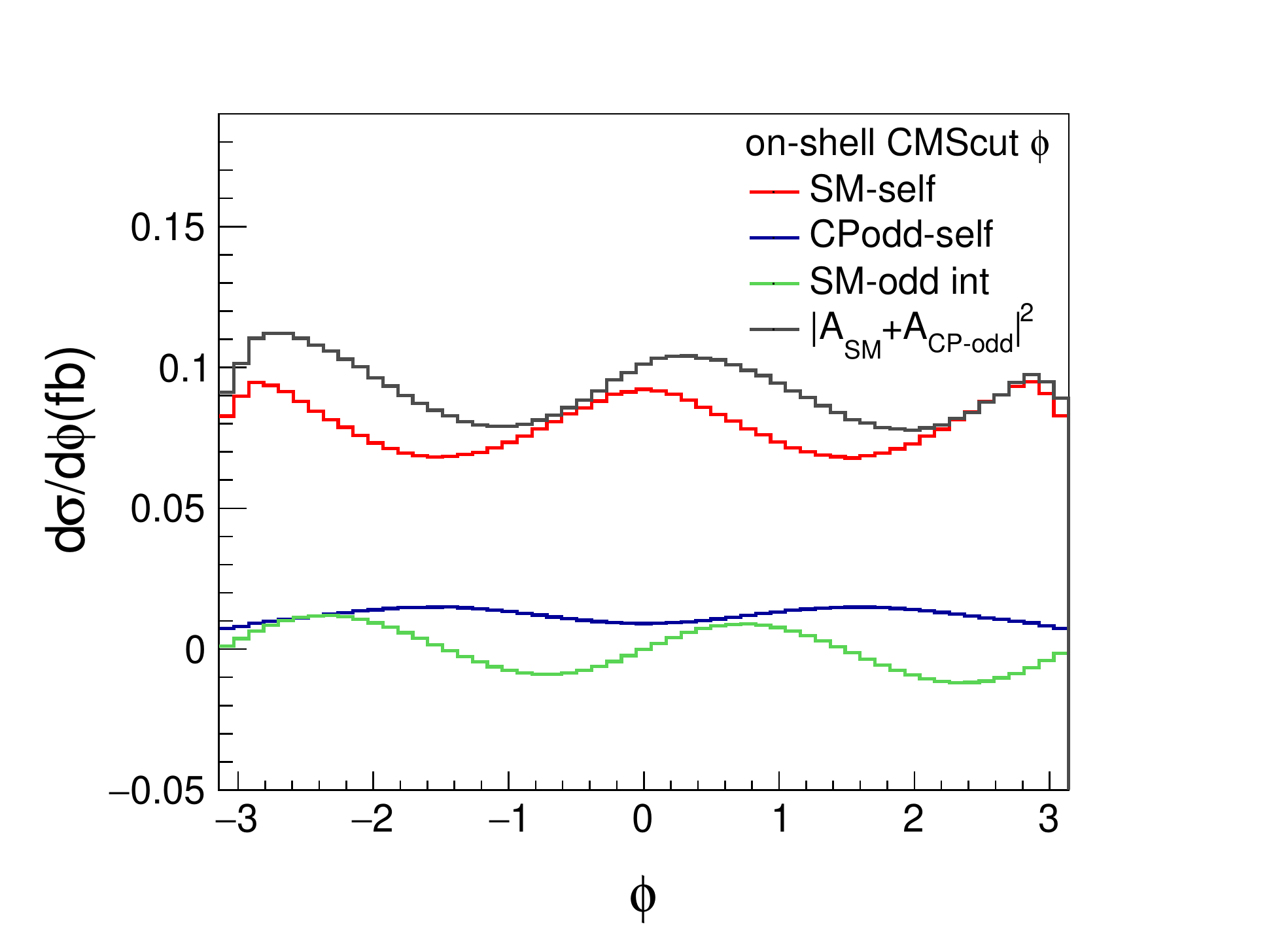}
    \includegraphics[width=3.6cm]{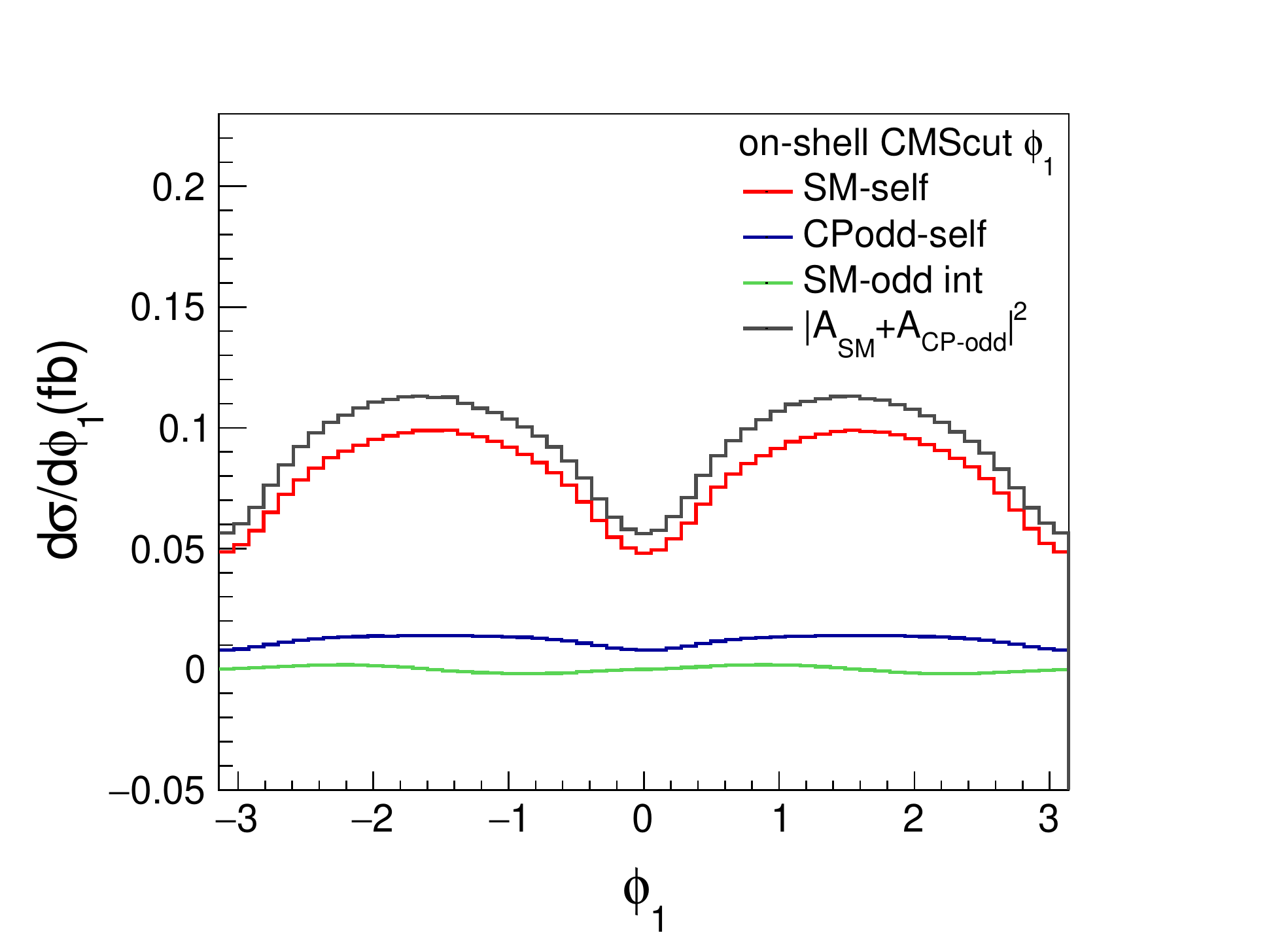}
    \includegraphics[width=3.6cm]{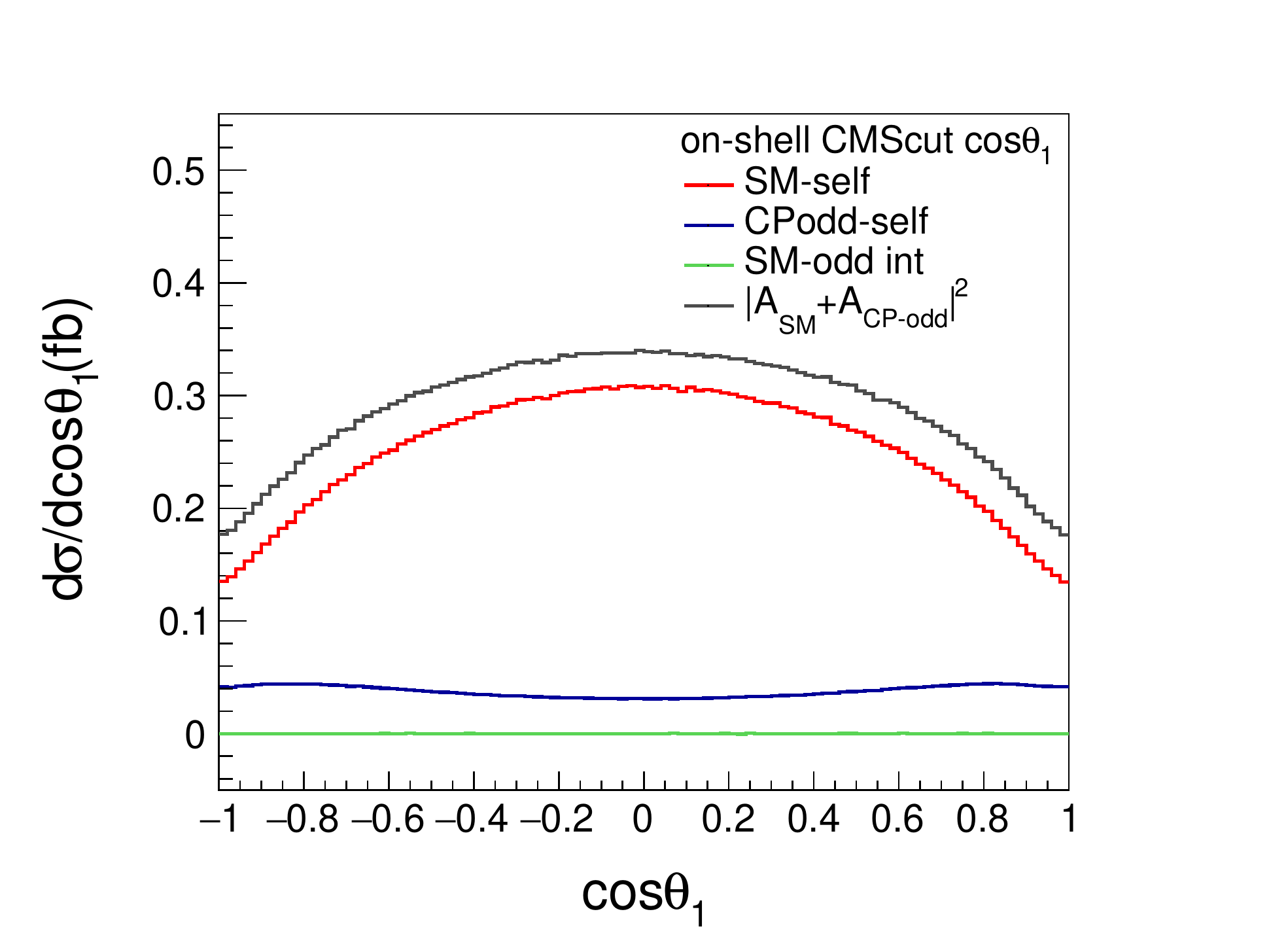}
    \includegraphics[width=3.6cm]{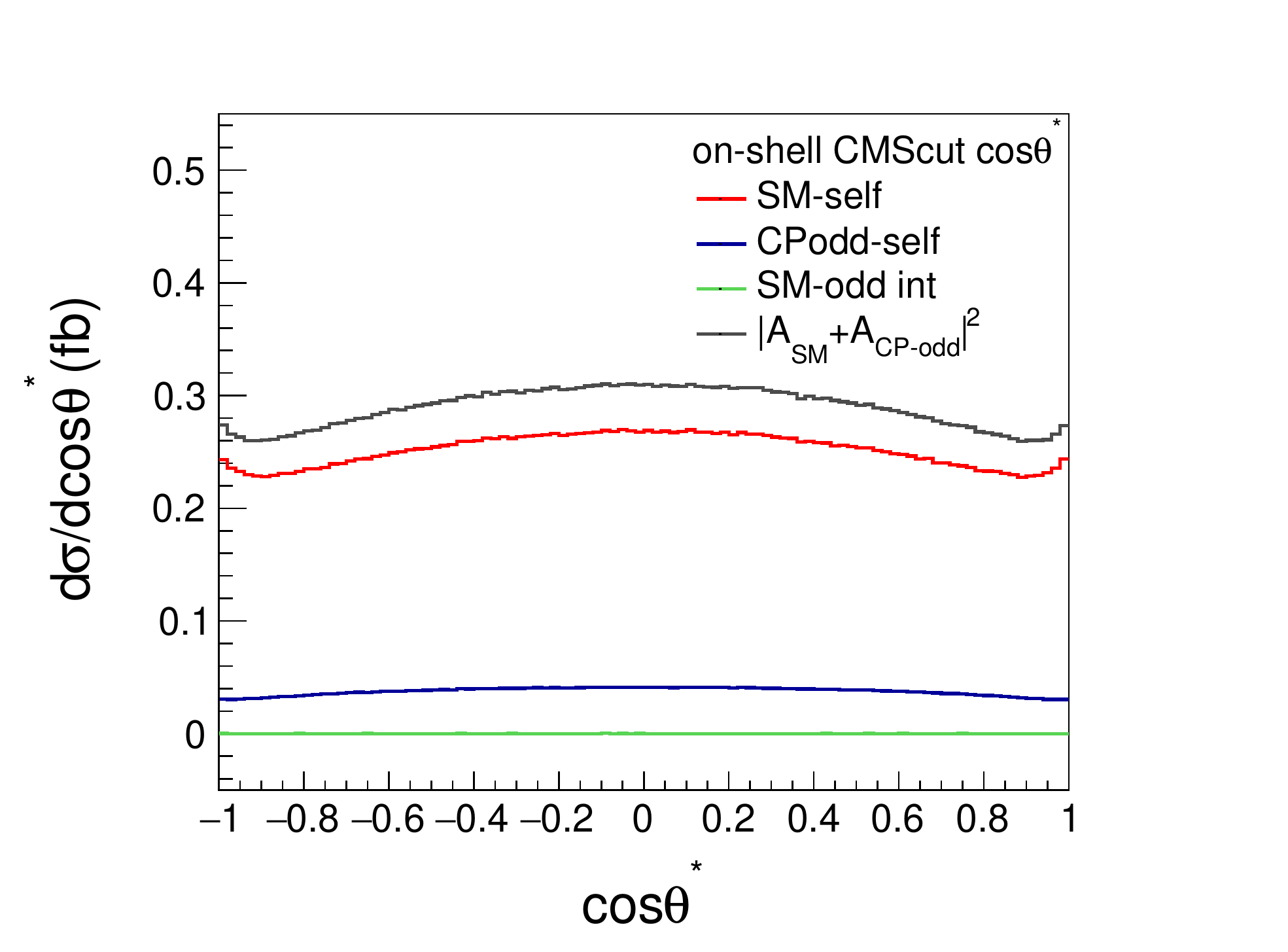}

	\caption{The angular differential cross sections from signal process and its interference between SM and CP-odd in Higgs on-shell region. }\label{signal-inter-SM-CPodd-on}
\end{figure}

\end{appendix}

\bibliographystyle{utphys.bst}
\bibliography{ref}
\end{document}